\documentclass{article}
\usepackage[utf8]{inputenc}

\usepackage{multirow}
\usepackage{fontawesome}

\usepackage{rotating}
\usepackage{geometry}
\usepackage{enumitem}
\setlist[enumerate]{itemsep=0em} % Adjust the value as desired

\usepackage[numbers, round]{natbib}
\setcitestyle{comma} % Separate citation numbers with commas

\usepackage{chngcntr}

\usepackage{booktabs}

\usepackage{xcolor}
\definecolor{iu_crimson}{HTML}{990000}

\usepackage[hidelinks]{hyperref}
\hypersetup{
    colorlinks=true,
    linkcolor=iu_crimson,
    filecolor=black,
    citecolor=iu_crimson,
    urlcolor=blue,
    linktoc=all
}
\usepackage{xurl} % Break long urls correctly

\usepackage{graphicx}

\usepackage{enumitem}

\usepackage{amssymb}
\usepackage{amsmath}

\usepackage{changepage}

\newenvironment{sigstatement}
  {
    \small
    \begin{adjustwidth}{.9cm}{.9cm}
    \bfseries\begin{center}Significance Statement\end{center}\normalfont \vspace{.5em}
  } % Centers title only
  {\end{adjustwidth}}

\usepackage{authblk}

\title{Fact-checking information from large language models can decrease headline discernment}

\author[a,$\dagger$]{Matthew R. DeVerna}
\author[a,b]{Harry Yaojun Yan}
\author[a,c]{Kai-Cheng Yang}
\author[a]{Filippo Menczer}

\affil[a]{\small Observatory on Social Media, Indiana University}
\affil[b]{FSI Cyber Policy Center, Stanford University}
\affil[c]{Network Science Institute, Northeastern University}
\affil[$\dagger$]{Corresponding author. Email: mdeverna@iu.edu}

\date{
\small This preprint is under review at \textit{PNAS}.\\
\today
}

\begin{document}

\maketitle

\begin{abstract}
Fact checking can be an effective strategy against misinformation, but its implementation at scale is impeded by the overwhelming volume of information online.
Recent artificial intelligence (AI) language models have shown impressive ability in fact-checking tasks, but how humans interact with fact-checking information provided by these models is unclear.
Here, we investigate the impact of fact-checking information generated by a popular large language model (LLM) on belief in, and sharing intent of, political news headlines in a preregistered randomized control experiment.
Although the LLM accurately identifies most false headlines~(90\%), we find that this information does not significantly improve participants' ability to discern headline accuracy or share accurate news.
In contrast, viewing human-generated fact checks enhances discernment in both cases.
Subsequent analysis reveals that the AI fact-checker is harmful in specific cases: it decreases beliefs in true headlines that it mislabels as false and increases beliefs in false headlines that it is unsure about. 
On the positive side, AI fact-checking information increases the sharing intent for correctly labeled true headlines. 
When participants are given the option to view LLM fact checks and choose to do so, they are significantly more likely to share both true and false news but only more likely to believe false headlines.
Our findings highlight an important source of potential harm stemming from AI applications and underscore the critical need for policies to prevent or mitigate such unintended consequences.
\end{abstract}

\vspace{4em}

% Use the custom section
\begin{sigstatement}
This study explores how LLMs used for fact-checking affect the perception and dissemination of political news headlines.
Despite the growing adoption of AI and tests of its ability to counter online misinformation, little is known about how people respond to LLM-driven fact-checking.
This randomized control experiment reveals that even LLMs that accurately identify false headlines do not necessarily enhance users' abilities to discern headline accuracy or promote accurate news sharing.
LLM fact checks can actually reduce belief in true news wrongly labeled as false and increase belief in dubious headlines when the AI is unsure about an article's veracity.
These findings underscore the need for research on AI fact-checking's unintended consequences, informing policies to enhance information integrity in the digital age.
\end{sigstatement}

\clearpage

Digital misinformation has rapidly become a critical issue of modern society~\cite{lewandowsky2017beyond, Lazer2018Mar}.
Recent work suggests that misinformation can erode support for climate change~\cite{van2017inoculating,biddlestone2022climate}, contribute to vaccine hesitancy~\cite{pierri2022online,rathje2022social,loomba2021measuring}, exacerbate political polarization~\cite{tucker2018social}, and even undermine democracy~\cite{van2021political}.
As a mitigation strategy, fact checking has proved effective at reducing people's belief in~\cite{walter2020fact,brashier2021timing,nyhan2020taking} and intention to share~\cite{yaqub2020effects} misinformation in various cultural settings~\cite{Porter2021Sep}.
However, this approach is not scalable, greatly limiting its applications~\cite{pennycook2021psychology}.

To tackle this challenge, researchers and social media platforms have been exploring automated methods~\cite{lee_etal_2020_language} to detect misinformation~\cite{shu2017fake,zhou2020survey} and fact-check claims~\cite{ijcai2021p619, Yang2021Dec, lee_etal_2020_language, graves2018understanding, 2017NaeemulClaimBuster, Ciampaglia2015Jun}.
A robust fact-checking system must possess the ability to detect claims, retrieve relevant evidence, assess the veracity of each claim, and yield justifications for the provided conclusions~\cite{zeng2021automated,guo2022survey}.
Previous work attempting to meet these goals typically adopts cutting-edge artificial intelligence (AI) methods, specifically natural language processing.
Nevertheless, building a functional system that can handle the vast volume of digital information on the internet, spanning various contexts and languages, remains a daunting task.

Recent advances in large language models (LLMs) may appear to provide a feasible path forward. 
Trained on massive datasets of text from the internet, including news articles, books, and websites~\cite{brown2020language}, these models are knowledgeable about a wide range of topics and have shown impressive performance on tasks such as text summarization and named entity recognition~\cite{ye2023comprehensive,qin2023chatgpt}. 
Outside the laboratory, LLMs have demonstrated remarkable abilities, even passing challenging exams designed for humans~\cite{Katz2023GPTpassbar, openai2023gpt4}.

Analyses of ChatGPT, a prominent LLM, suggest it can rate the credibility of news outlets~\cite{Yang2023LLMRating} and has great potential to fact-check claims~\cite{Quelle2023Oct, Hoes2023GPTrateClaims, Kuznetsova2023generative}, especially when augmented with additional data~\cite{Zhou2024Muse}.
Messages provided by LLMs to correct social media misinformation can be better than corrective messages generated by humans~\cite{He2023Mar}.
These models can generate convincing justifications for the information they provide and even engage in conversations with users to provide additional context and facilitate understanding in multiple languages.
Such capabilities of LLMs, coupled with open-sourcing efforts~\cite{standfordAlpaca2023, databricks2023Dolly}, create a favorable environment for the development of scalable and reliable AI systems that can verify substantially more claims on the internet than is currently possible.

However, realizing this potential requires humans to integrate LLMs into the digital information ecosystem effectively.
Unfortunately, human-AI interaction is notoriously complex~\cite{Hancock2020Mar}.
Prior work has shown that AI is often seen as objective~\cite{sundar2008main, sundar_machine_2019, sundar2020rise, DeAndrea2014May}, yet trust in AI depends on various factors such as individual expectations~\cite{Luger2016May, Meurisch2020Dec}, system interactivity~\cite{Shi2022Mar, smith-renner_no_2020}, and whether the AI provides information about its recommendations~\cite{Zhang2020Jan, bansal_does_2021}.

In the present context, it remains unclear how humans would interact with fact-checking information provided by state-of-the-art LLMs.
Therefore, a thorough analysis of this misinformation intervention is necessary before deploying models in the wild.
To this end, we conduct a preregistered~\cite{DeVerna2022cgptOSF}, randomized controlled experiment to examine the causal effects of viewing fact-checking information provided by ChatGPT 3.5 on individual beliefs in and intention to share political news headlines.
We selected ChatGPT for our study despite it not being specifically tailored for fact-checking.
This decision was driven by its widespread public availability and use as well as the promising results emerging from tests of its claim verification capabilities at the time~\cite{Kuznetsova2023generative, Quelle2023Oct, Hoes2023GPTrateClaims}.

\section*{Results}
% Including the \section(s) with the "*" removes them from the PDF outline. This adds them back. Form: \pdfbookmark[level][text][label]
\pdfbookmark[0]{Results}{sec:results}
\label{sec:results}

We recruited a representative sample of $N = 2{,}159$ U.S. participants (see \nameref{sec:methods} for more information).
All participants were presented with the same 40 real political news stories, which included a headline, lede sentence (if present), and image.
Half of these headlines were true and the other half were false. 
Half were favorable to Democrats and the other half were favorable to Republicans (see \nameref{sec:methods} for details).

Participants were separated into ``belief'' and ``sharing'' groups in which they were asked to indicate, respectively, whether they believed headlines to be accurate or would be willing to share them on social media.
The response options for both questions were ``Yes'' or ``No.''
These questions were asked separately as priming individuals to think about headline veracity can alter sharing behavior~\cite{Pennycook2021Shifting, pennycook2021psychology}.
Each group included four conditions: a control group and three treatment conditions. 
In the \textit{human fact check} condition, participants were presented with traditional fact checks generated by humans.
The other two conditions emulated hypothetical scenarios for an automated fact-checking system on a social media platform: treated subjects were either forced to view fact-checking information provided by ChatGPT (\textit{LLM-forced}) or given the option to reveal that information by clicking a button (\textit{LLM-optional}). 
ChatGPT fact-checking information was identical for all treated subjects and presented directly below the corresponding headline.
Participants in the LLM treatment conditions were informed that the fact checks were generated by ChatGPT, while those in the human fact checks condition were only informed that they would receive fact-checking information.
Subjects in the control condition were only shown headlines and asked the belief/sharing question without being exposed to any fact-checking information. 
The experimental design is illustrated in Fig.~\ref{fig:main_effects}a (see \nameref{sec:methods} for more details). 

Unless otherwise stated, all $P$ values presented here are generated with two-tailed Mann-Whitney U tests and adjusted with Bonferroni correction for multiple comparisons.
In the Supplementary Information, we also report on linear regression for all results, employing robust standard errors clustered on participant and headline.

\subsection*{Accuracy of LLM fact-checking information}
\pdfbookmark[1]{Accuracy of LLM fact-checking information}{sec:res:chatGPTaccuracy}
\label{sec:res:chatGPTaccuracy}

\begin{figure}
    \centering
    \includegraphics[width=.95\linewidth]{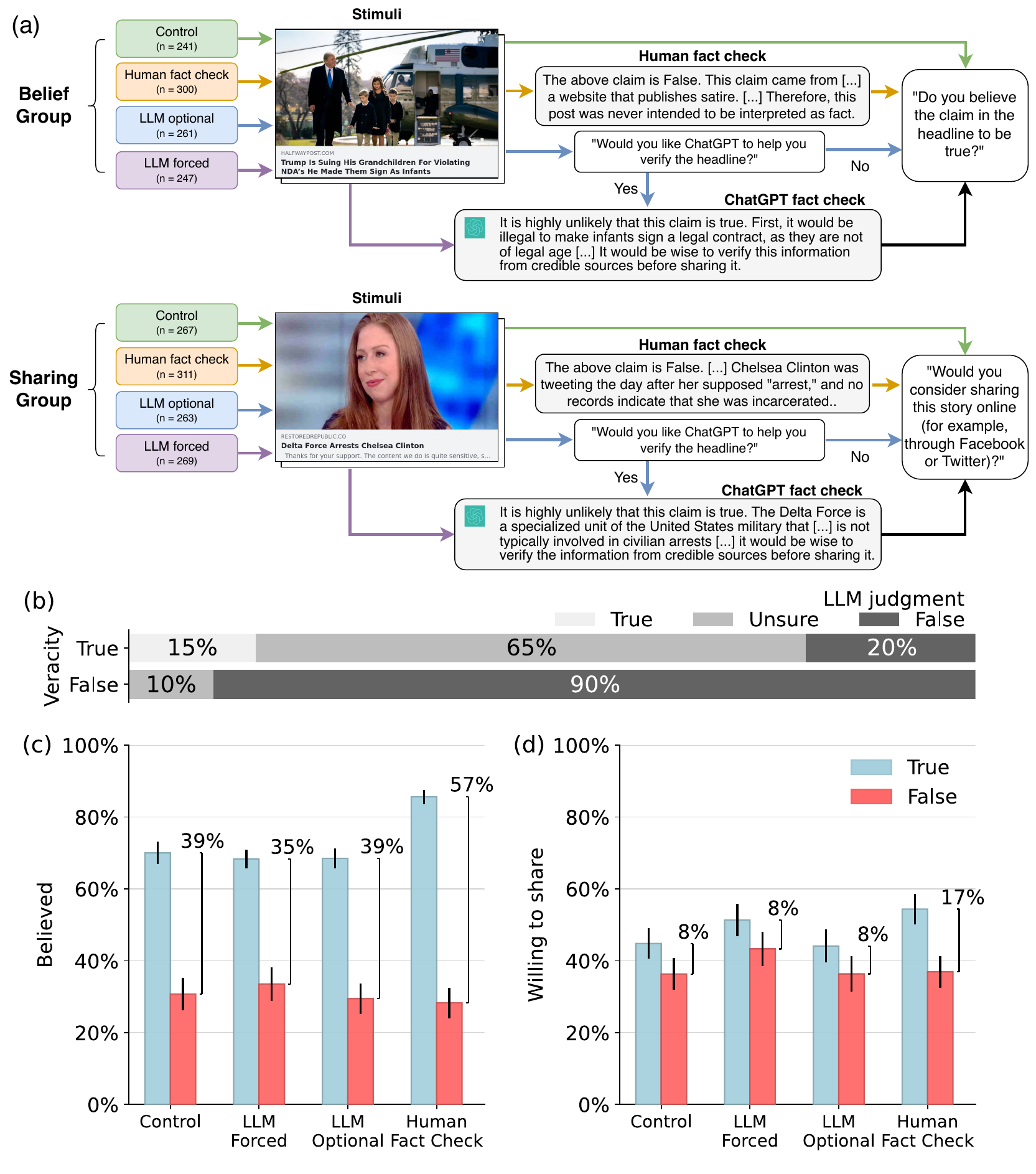}
    \caption{
    Experimental design, accuracy, and main effects of the LLM fact-checking intervention. 
    (a) Graphical representation of the experimental design and participant flow. Although two different false claims are shown as examples along with their respective ChatGPT fact-checking information, both belief and sharing groups are exposed to the same set of stimuli and fact checks. 
    (b) ChatGPT's judgment (shade) based on headline veracity.
    The bottom two panels show the proportion of headlines that participants indicated they (c) believed or (d) were willing to share on social media.
    The x-axes indicate the experimental conditions and the colors of the bars represent headline veracity.
    Error bars represent 95\% confidence intervals, calculated using a bootstrapping method with 5,000 resamples.
    Mean group discernment (rounded to whole percentages) is annotated for each condition, calculated as the mean difference between the proportion of true and false headlines believed (or willing to be shared).
    }
    \label{fig:main_effects}
\end{figure}

To contextualize our results, we first illustrate in Fig.~\ref{fig:main_effects}b the accuracy of ChatGPT's fact-checking information.
True headlines were accurately fact-checked 15\% of the time (3/20) whereas 20\% (4/20) were erroneously reported as ``false.'' 
For the remaining 65\% of true headlines (13/20), ChatGPT expressed some degree of uncertainty (labeled as ``unsure'').
These responses often contained language such as ``It is possible that ... but I don't have any information on whether this has happened or not.'' 
For false headlines, ChatGPT was unsure in 10\% (2/20) of cases; the remaining 90\% (18/20) were accurately judged as ``false.''

Although limited in size, our set of headlines provides us with a balanced representation of political biases and factual accuracy to evaluate the LLM.
Overall, this analysis suggests that the LLM is an accurate fact-checker for false content.
For true headlines, it is less accurate but can generally identify and explain when it cannot provide accurate fact-checking information.
These results align with earlier studies that delve into the accuracy of LLM fact-checking utilizing much larger datasets~\cite{Quelle2023Oct, Hoes2023GPTrateClaims, Kuznetsova2023generative}.

An additional analysis of various prompt engineering methods and their performance, measured using standard binary classification metrics, is available in the Supplementary Information.
Techniques that forced ChatGPT to produce only ``true'' or ``false'' judgments did not enhance the model's overall accuracy.

\subsection*{Ineffectiveness of LLM intervention}
\pdfbookmark[1]{Ineffectiveness of LLM intervention}{sec:res:ineffectiveness}
\label{sec:res:ineffectiveness}

To evaluate the effectiveness of a misinformation intervention, it is crucial to measure its impact on belief in and sharing of both true and false headlines~\cite{Guay2023Mar, Pennycook2021Shifting}.
Although the veracity of headlines may not always fit neatly into true and false categories, as in the case of rumors with unclear veracity that are later clarified~\cite{DeVerna2022Rumors}, this framework defines the desired outcome: an effective misinformation intervention should enhance individuals' ability to distinguish between true and false headlines such that they believe/share more accurate news.

To capture the causal effects of LLM-generated fact-checking information, we compare the average discernment of participants in the treatment conditions (LLM-forced and LLM-optional) to those in the control condition.
Discernment is defined as the difference between the proportion of true and false headlines that participants believe (or are willing to share), capturing the intervention's impact on both news categories.
The inclusion of the human fact check condition allows us to differentiate AI-related effects from those associated with traditional fact checking.

Figure~\ref{fig:main_effects} (panels c,d) illustrates the effects of fact checking on belief in and intent to share true and false headlines under each condition, including the mean group discernment as an annotation.
In contrast with our preregistered expectations, discernment within both the belief and sharing groups was unaffected by the LLM treatment, regardless of condition.
In the belief group (Fig.~\ref{fig:main_effects}c), participants who were forced to view AI fact checks displayed a slight mean reduction ($-4.50$\%) in discernment when compared to the control group ($U = 31{,}993$, $P = 0.61$, $d = -0.15$, 95\% CI: $[-10.04\%, 0.90\%]$).
The discernment of those given the option to view fact checks in this group was virtually unaffected, decreasing on average by only $-0.27$\% ($U = 31{,}265$, $P = 1$, $d = -0.01$, 95\% CI: $[-5.52\%, 5.03\%]$).

We observe similar results regarding sharing behavior.
Participants in the LLM-forced and LLM-optional conditions of the sharing group displayed a mean reduction of $-0.43$\% ($U = 35{,}785$, $P = 1$, $d = -0.02$, 95\% CI: $[-3.93\%, 3.05\%]$) and $-0.67$\% ($U = 34{,}860$, $P = 1$, $d = -0.03$, 95\% CI: $[-4.18\%, 2.92\%]$), respectively, when compared with the control group.

In contrast to the above, we observe a significant increase in discernment for participants who viewed human fact checks within both the belief and sharing groups.
On average, belief discernment increased by 18.06\% ($U = 25{,}210$, $P < 0.001$, $d = 0.50$, 95\% CI: $[12.00\%, 24.00\%]$), and sharing discernment increased by 8.98\% ($U = 34{,}224$, $P = 0.001$, $d = 0.35$, 95\% CI: $[4.89\%, 13.33\%]$).
These effects are primarily due to an increased belief in and willingness to share true headlines---which increased by 15.63\% and 9.58\%, respectively---while the impact on false headlines remained largely unchanged.

In summary, these results indicate that human fact checks served as an effective misinformation intervention, while those generated by the LLM did not.
This is unexpected, considering that the AI provides participants with useful information, particularly for false headlines.
However, this analysis does not account for the accuracy of the AI's responses, nor does it examine how behaviors vary when individuals choose to view or not view this information.
To delve deeper into these dynamics, we have supplemented our preregistered design with two additional exploratory analyses in the sections that follow.

\subsection*{Accounting for LLM accuracy}
\pdfbookmark[1]{Accounting for LLM accuracy}{sec:res:chatgpt_judgement}
\label{sec:res:chatgpt_judgement}
% {'False_False': 18,
%  'True_Unsure': 13,
%  'True_False': 4,
%  'True_True': 3,
%  'False_Unsure': 2}
Here, we explore the causal effects of viewing LLM fact-checking information when accounting for model accuracy.
The judgments made by ChatGPT for both true and false headlines fall into one of three categories: correct, incorrect, or unsure.
This results in six different scenarios (True/False $\times$ Correct/Incorrect/Unsure) in which effects may be observed.
However, our data contain no false headlines judged by ChatGPT to be true, resulting in five scenarios for each previously considered comparison  (Belief/Sharing $\times$ Control/LLM-optional/LLM-forced).

\begin{figure*}
    \centering
    \includegraphics[width=\linewidth]{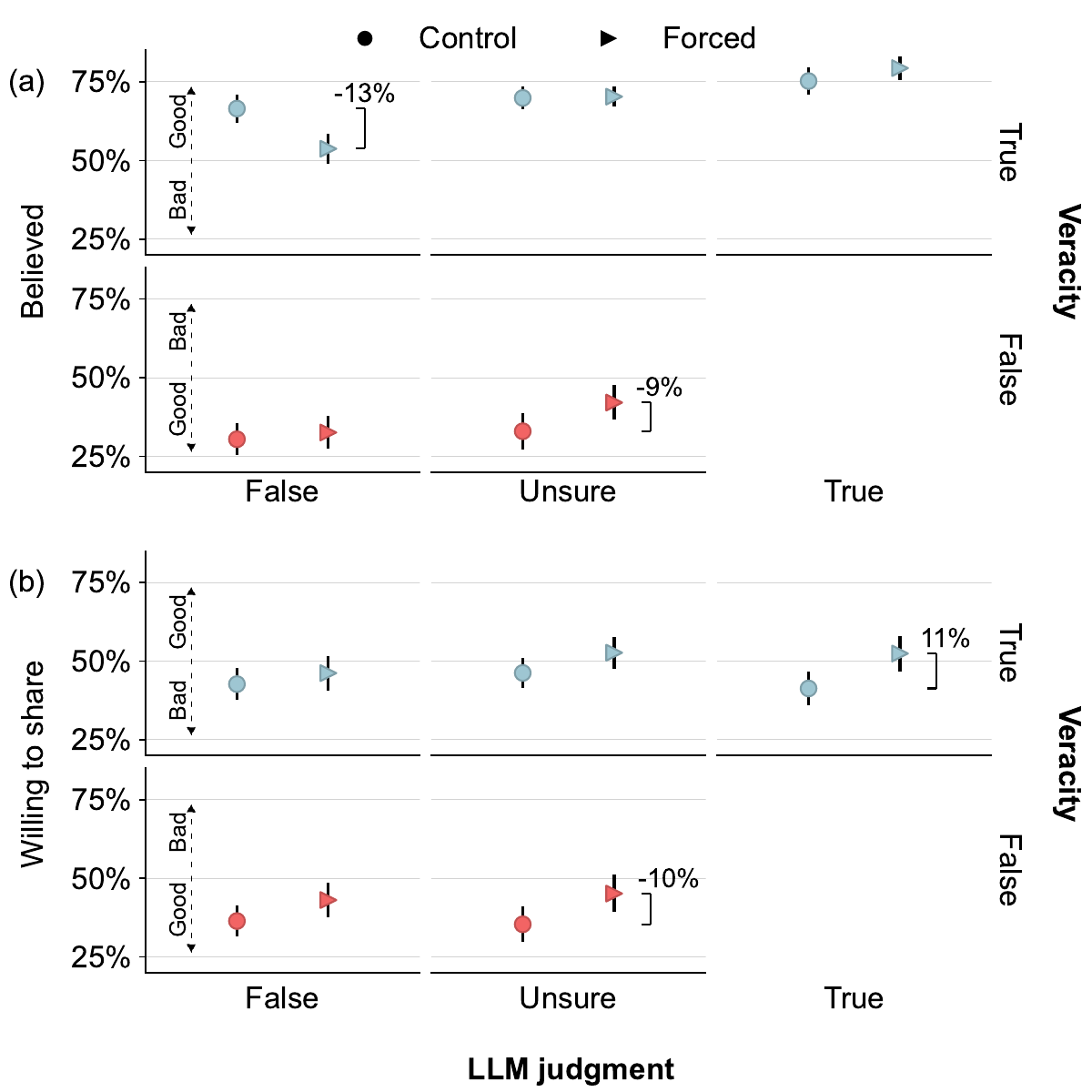}
    \caption{
    Effects of LLM fact-checking information on headline belief and sharing intent, contingent on headline veracity and fact check judgment.
    Each panel shows the proportion of participants in the control (circles) and forced (triangles) conditions who (a) believed or (b) were willing to share a specific group of headlines.
    Headlines are grouped by the combination of veracity and LLM judgment, e.g., the top left panel indicates the proportion of participants who believed true headlines that ChatGPT judged as false.
    As no false headlines were judged to be true by ChatGPT, this panel is left empty.
    A visual guide on the left (dashed arrows) helps the reader understand the desired directional effect of a misinformation intervention, given the veracity of a headline.
    Mean group differences (rounded to whole percentages) are annotated for panels that illustrate effects discussed in the main text---positive (negative) annotations illustrate desirable (undesirable) changes. 
    Error bars represent 95\% confidence intervals, calculated using a bootstrapping method with 5,000 resamples.
    }
    \label{fig:five_way}
\end{figure*}

To evaluate the potential impact of LLM-generated fact checks, we compare the LLM-forced and control conditions in these five scenarios, as illustrated in Fig.~\ref{fig:five_way}.
Annotations indicate mean group differences and highlight the significant effects identified through Bonferroni-adjusted Mann-Whitney U tests.

In the belief group, we found significant undesirable effects showing that LLM fact checks decreased participants' discernment.
Specifically, there was a 12.75\% decrease in the belief of true headlines incorrectly judged as false by ChatGPT ($U = 35{,}937$, $P < 0.001$, $d = -0.38$, 95\% CI: $[-18.67\%, -6.89\%]$) and a 9.12\% increase in the belief of false headlines where the AI expressed uncertainty ($U = 25{,}931$, $P = 0.03$, $d = 0.22$, 95\% CI: $[1.69\%, 16.35\%]$).
Both cases demonstrate a behavioral change that is counter to the ideal outcomes of any misinformation intervention.

Regarding the sharing group, we observed mixed results.
While there was an 11.09\% increase in participants' intention to share true headlines correctly judged by ChatGPT ($U = 30{,}897$, $P = 0.02$, $d = 0.26$, 95\% CI: $[4.02\%, 18.06\%]$), there was also a 9.77\% increase in the intention to share false headlines where ChatGPT expressed uncertainty ($U = 31{,}856$, $P = 0.05$, $d = 0.22$, 95\% CI: $[2.31\%, 17.25\%]$).
The former increases sharing discernment, while the latter reduces it by a similar amount.

These results indicate that LLMs can affect belief in and intent to share both true and false news, depending on how they judge a headline.
While most effects are small, some reflect harmful outcomes in the sense of reduced discernment. 

\subsection*{Opt in versus opt out}
\pdfbookmark[1]{Opt in versus opt out}{sec:res:in_vs_out}
\label{sec:res:in_vs_out}

We next analyze participants' behavior in the LLM-optional conditions, comparing those who opt in to see LLM fact-checking information versus those who opt out.

On average, participants chose to view fact checks for slightly more than half of the headlines. 
The mean number of fact checks viewed for the belief and sharing groups was 21.6 (SD = 15.8) and 23.8 (SD = 15.7), respectively.
However, the distribution was bimodal: about half the participants viewed fact checks for most headlines, while the other half viewed them for only a handful.
Participants who viewed fact checks for more than half of the 40 headlines (52.1\%) averaged viewing 36.7 (SD = 5.5). 
In contrast, those who viewed less than half averaged 7.5 views (SD = 6.4).
A Mann-Whitney U test revealed no significant difference in opt-in behavior between true and false headlines ($P = 0.1$).
See the Supplementary Information for more details.

Figure~\ref{fig:in_vs_out} illustrates the belief in and intention to share headlines for which subjects chose to see versus not see LLM fact checks, for both true and false headlines.
Each subject's contribution to the group mean values and confidence intervals are weighted by the number of times they chose to see (or not see) each type of headline.
Figure~\ref{fig:in_vs_out}a shows that participants who chose to see LLM fact checks were significantly more likely to believe false headlines accurately identified by the LLM, with a 29.35\% increase ($U = 23{,}480$, $P = 0.005$, $d = 0.63$, 95\% CI: $[20.81\%, 37.93\%]$), as well as those the model was unsure about, with a 28.12\% increase ($U = 14{,}260$, $P < 0.001$, $d = 0.64$, 95\% CI: $[18.43\%, 38.12\%]$).
There was no significant difference in belief for true headlines that the model could not classify, with a 5.51\% increase ($U = 20{,}452$, $P = 1$, $d = 0.12$, 95\% CI: $[-3.70\%, 14.47\%]$), and we only observe a 7.46\% increase for accurately identified true headlines ($U = 10{,}941$, $P = 0.04$, $d = 0.18$, 95\% CI: $[-1.50\%, 16.35\%]$).
However, a significant decrease for those misjudged as false (16.59\% decrease; $U = 10{,}299$, $P < 0.001$, $d = -0.35$, 95\% CI: $[-26.34\%, -7.24\%]$) was found.
Figure~\ref{fig:in_vs_out}b shows that participants who chose to see LLM fact checks were significantly more likely to share headlines in all scenarios. 
These increases ranged from 29\% for true headlines judged as false to 39\% for true headlines judged as true (see the Supplementary Information for statistics).

\begin{figure*}[t]
    \centering
    \includegraphics[width=\linewidth]{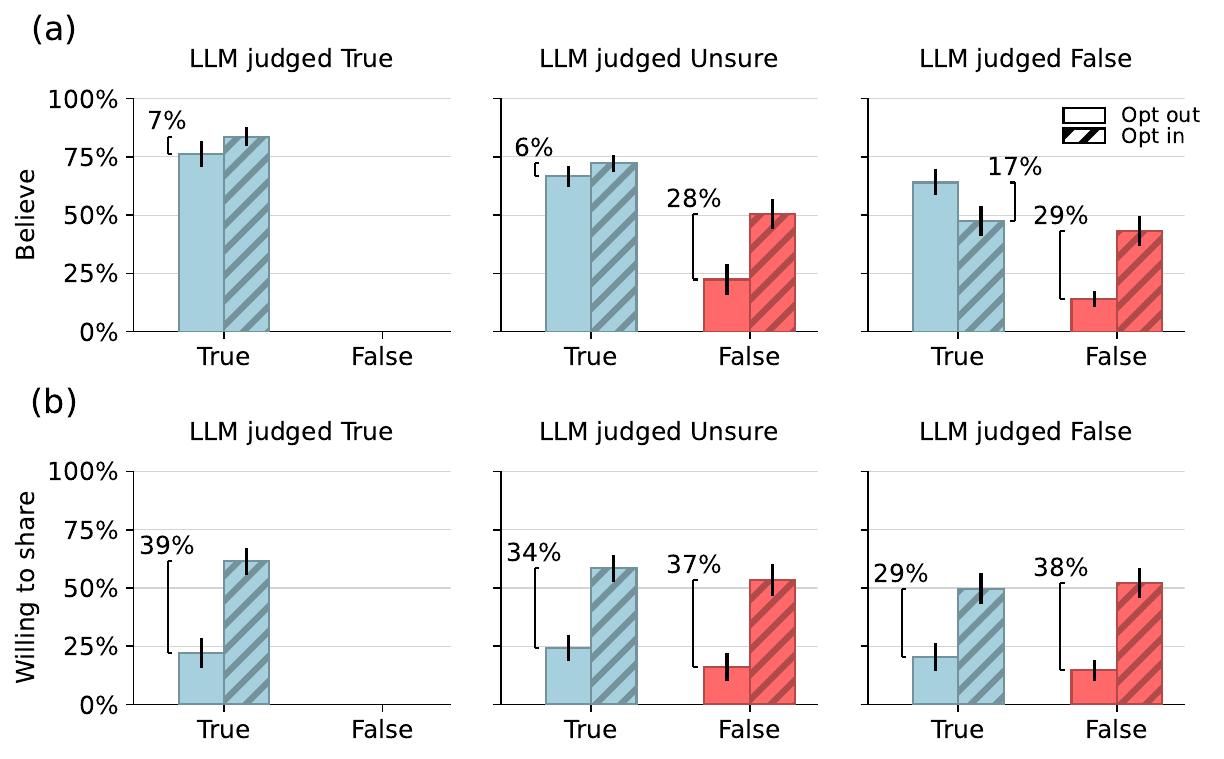}
    \caption{
    Proportions of headlines that participants in the optional condition indicated they (a) believed or (b) were willing to share on social media.
    These proportions are based on the headline's veracity, whether participants chose to see LLM fact-checking information (opt in) or not (opt out), and how the LLM judged the headlines (True, Unsure, False).
    No false headlines were judged as true.
    Error bars represent 95\% confidence intervals, calculated using a weighted bootstrapping method with 5,000 resamples.
    The mean difference between opt-in and opt-out groups (rounded to whole percentage) is annotated for each condition.
    }
    \label{fig:in_vs_out}
\end{figure*}

We note that this particular within-group analysis does not allow us to identify causal effects because participants are not randomly assigned to the treatment (opt in) or comparison (opt out) group for each headline.
Nonetheless, when participants viewed LLM-generated fact-checking information, they were more likely to share both true and false news.
Additionally, those who viewed this information were less likely to believe true news misjudged as false and more 
likely to believe false news, even when accurately identified as such by the model.

\subsection*{Attitudes toward AI and partisan congruence}
\pdfbookmark[1]{Attitudes toward AI and partisan congruence}{sec:res:interactions}
\label{sec:res:interactions}

Our preregistered analyses also examined the potential roles of individual attitudes toward AI (ATAI) and the partisan congruence of headlines. 
While we find minimal evidence that these variables significantly impacted the results of our first two analyses, we observed specific relationships when individuals had the option to view LLM fact-checking information.

In particular, we found clear evidence that participants with positive attitudes toward AI who chose to view LLM-generated fact checks were significantly more likely to share those headlines across all fact-checking scenarios.
However, the relationship between ATAI and participant belief was less clear.
Nonetheless, the tendency for participants to share and believe true news that the LLM was unsure about was more pronounced among those with positive attitudes towards AI when viewing AI fact checks.

When participants encountered politically incongruent true headlines that the LLM was unsure about, their likelihood of believing or sharing them diminished significantly. 
This relationship persisted irrespective of whether participants opted to access the fact-checking information.
We observed a similar negative relationship in only one other scenario: for incongruent false headlines when participants did not view LLM fact checks. 
For more details on these analyses, please refer to the Supplementary Information.

\section*{Discussion}
\pdfbookmark[0]{Discussion}{sec:discussion}
\label{sec:discussion}

% Limitations
While our experimental design allows us to assess the causal effects of LLM fact-checking information on the discernment of true and false headlines, it is important to exercise caution when generalizing these results to different contexts.
First, we use a specific version of ChatGPT to generate fact-checking information with a single prompt; these results may not apply to other AI models or prompting approaches.
Although our prompt aimed to reflect naturalistic usage, its realism is uncertain due to the lack of prior research on how people use LLMs for fact-checking in real-world settings.
Second, design choices intended to emulate a realistic social media environment---such as including headline sources and lede text---may contribute to people's assessments, although these effects should be equal for all experimental conditions.
Third, the survey setting of our experiment may not fully capture the complexities of real-world information consumption and sharing behaviors.
However, previous research has shown a correlation between self-reported willingness to share news in online surveys and actual sharing behavior on social media platforms~\cite{Mosleh2020Feb}.
Finally, while our study presents real headlines that replicate a common social media design, the results may not generalize beyond our relatively small selection of political news.
Nevertheless, the pretest conducted on these headlines ensured they are balanced with respect to dimensions known to be important to believing and sharing news (see \nameref{sec:methods}).

% Summarize findings
Despite these limitations, our study provides valuable insights into the complex interplay between humans and AI in the context of automated fact checking.
ChatGPT performs well at identifying false headlines while it mostly reports being unsure about true headlines, consistent with previous research~\cite{Quelle2023Oct, Hoes2023GPTrateClaims, Kuznetsova2023generative}.
Since we tested a limited number of headlines, it is difficult to determine why the model judges false headlines more accurately.
The model may be more likely to have seen false headline stimuli as their publication dates were less recent than those of true headlines.
This highlights a key limitation of large-scale automated fact-checking systems that we refer to as the ``breaking news problem'': developing news stories often discuss novel events the model has never been exposed to, making it difficult for AI to assess them accurately.
To this end, a promising future research direction is to augment LLMs with trusted data---e.g., via real-time search~\cite{Zhou2024Muse}---to improve their performance on new and evolving information~\cite{Lewis2020RAG}.

While the average belief and sharing discernment of participants was positively affected by viewing human fact checks, this was not the case for viewing LLM fact-checking information, whether or not such information was optional.
These results are surprising, considering previous research suggests that LLMs can persuade humans on controversial topics~\cite{bai_artificial_2023}.
However, we found that AI-generated fact checks can affect belief in and intent to share news headlines, contingent upon the accuracy of the AI's responses relative to the veracity of the headlines.
Consistent with literature showing that AI may be perceived as objective~\cite{sundar2008main, sundar_machine_2019}, participants tended to believe true headlines less when the LLM incorrectly labeled them as false. 
Furthermore, participants demonstrated an increased willingness to share true headlines that were correctly identified by the LLM.
The latter outcome is encouraging, as it supports efforts to enhance the acceptance of reliable information~\cite{Acerbi2022Jan}.
Since trusted content is far more abundant than misinformation, future research should investigate how the volume of different types of content interacts with model accuracy to impact overall information quality.

When the LLM expressed uncertainty about the veracity of false headlines, participants were more inclined to believe and share them.
This contradicts research suggesting that uncertain fact checks can be perceived as false~\cite{Park2021Jan}, and that expressions of uncertainty from an LLM can increase task accuracy~\cite{kim2024m}.
While expressing uncertainty has been considered a desirable quality in automated fact-checking systems~\cite{kotonya-toni-2020-explainable}, our results illustrate that unsure fact checks can lead to adverse outcomes. 
Given the impact of the format of fact checks~\cite{Li2022Bias, Bode2024Correction}, this conflicting evidence highlights an important question for future research: which formats and styles of AI-generated fact checks are most effective, and which prompting techniques can reliably create them?

The behavior of participants in the optional condition revealed a strong selection bias. 
When participants were given the choice to view LLM fact-checking information, those who chose to do so were significantly more likely to share both true and false news.
Furthermore, those who viewed this information were less likely to believe true news misjudged as false and more likely to believe false news.
These results suggest that individuals may have already formed their opinion about a headline before accessing the fact-checking information.
For example, they might wish to confirm what they believe to be true or see if the AI is wrong.
Of course, many factors may influence how one seeks and processes fact-checking information, including how well-informed~\cite{Li2020IndivFact} and confident they are~\cite{Pasek2015Aug}. 
Regardless, some participants subsequently disregard these fact checks.
This pattern is particularly evident with respect to false headlines, for which ChatGPT provides highly accurate information.
Despite being presented with helpful information indicating that these headlines were false, participants were still much more likely to report believing or being willing to share that content.
Further interaction analyses suggest that individual attitudes towards AI, as well as partisan congruence with headlines, are related to this behavior.
Although our study design cannot reveal the exact mechanism behind the outcomes of the optional condition, the findings suggest that this misinformation intervention design is unlikely to be helpful.

Future work could explore the effect of telling people whether a fact check comes from a human or AI. 
Similar questions have been investigated in the context of generic conversations~\cite{yin2024heard}, health prevention~\cite{lim2024effect}, advertising~\cite{Zhang_Gosline_2023}, and written content~\cite{irene2024effect}.
In these scenarios, disclosing the AI-generated source tends to lead to a negative perception of the content and a preference for human-generated content.
A dedicated investigation on the effect of fact-checking source disclosure will be required. 

% Conclusion
We present these results in the context of concerns raised by experts about the potential for AI to contribute to the digital misinformation problem~\cite{Spitale2023Jun, goldstein_generative_2023, Brewster2023chatgpt, Menczer2023NatMI}.
These concerns are well-founded; malicious AI-powered bots are virtually undetectable on social media~\cite{Yang2023Anatomy-AI-botnet} and even the developers of ChatGPT report that their technology is likely to be weaponized by malicious actors~\cite{Solaiman2019release, goldstein_generative_2023}.
To make matters worse, recent research indicates that state-of-the-art LLMs can persuade individuals on polarized topics~\cite{bai_artificial_2023, Karinshak2023Apr} and create persuasive propaganda~\cite{Goldstein2024Feb}, providing an incentive for their use in political information campaigns~\cite{goldstein_generative_2023}.

While the use of LLM-powered fact-checking to combat these concerns is enticing, our results reveal that the dynamics of human-AI interaction make this application potentially harmful, despite its accuracy.
This should not discourage us from exploring the potential of this technology to help us mitigate challenging problems.
Instead, as artificial intelligence becomes more deeply integrated into our information environment, it is crucial to fully understand both the risks and opportunities it presents.

\section*{Methods}
\pdfbookmark[0]{Methods}{sec:methods}
\label{sec:methods}

\paragraph{Participant sampling.}
\pdfbookmark[1]{Participant sampling}{sec:methods:sampling}
\label{sec:methods:sampling}

We utilized Qualtric's quota-matching system to ensure that our sample matched the United States population with respect to gender, age, race, education, and partisanship. 
We utilized 2020 U.S. Census~\cite{census2020educationalattainment} and Pew Research~\cite{pewresearch2020electorate} data as references for our quota criteria, which Qualtrics guaranteed with a $\pm$5\% accuracy.
We conducted $\chi^2$ tests to compare the distributions across the above dimensions for each experimental condition (control vs. LLM-optional vs. LLM-forced vs. human fact check), for both belief and sharing groups. 
We find one significant difference: in the belief group only, participants in the human fact check condition were more educated (i.e., held degrees) than those in the LLM-optional condition. 
Our analyses do not make comparisons between these two groups and our main results are confirmed by regression analyses that account for this and other factors.
Further details can be found in the Supplementary Information.
% Group   Condition
% Belief  Control      241
%         Optional     261
%         Forced       247
%         Human-FC     300
% Share   Control      267
%         Optional     263
%         Forced       269
%         Human-FC     311
After sampling, data for 2,159 participants were collected. 
In the belief group, the control, LLM-optional, LLM-forced, and human fact check conditions had 241, 261, 247, and 300 participants, respectively.
In the sharing group, the control, LLM-optional, LLM-forced, and human fact check conditions had 267, 263, 269, and 311 participants, respectively. 
The drop-out rate was low (between 1\%--6\%) across experimental conditions (see Supplementary Information).
All subjects confirmed their consent to participate in this study, which was approved by Indiana University's IRB (protocol 1307012383).

The data for the control, LLM-optional, and LLM-forced conditions were collected in March 2023.
At the request of reviewers, data for the human fact check conditions were gathered later, from March to June 2024.
All data were collected following the same protocols.
Participants were randomly assigned to one of the conditions at their respective times of collection.

\paragraph{News stories.}
\pdfbookmark[1]{News stories}{sec:methods:news_stories}
\label{sec:methods:news_stories}

We utilize 40 real news headlines that are related to US politics, balanced in terms of partisanship, believability, and the likelihood of being shared. 
These headlines were generated for another study~\cite{MercuryProject}.
Half were true and half false.
Each story included a headline, a lede sentence (if present), and an image.
All headline stimuli are included in our preregistration~\cite{DeVerna2022cgptOSF}. 
Further details can be found in the Supplementary Information.

\paragraph{LLM fact checks.}
\pdfbookmark[1]{LLM fact checks}{sec:methods:fact_checks}
\label{sec:methods:fact_checks}

Fact-checking information was generated by submitting to ChatGPT the prompt ``I saw something today that claimed $<$HEADLINE TEXT$>$. Do you think that this is likely to be true?'' 
This prompt was designed to capture a realistic scenario in which someone uses an AI chatbot to fact-check a headline to which they were exposed. 
All fact checks are included in our preregistration~\cite{DeVerna2022cgptOSF}.
To quantify and account for ChatGPT's fact-checking accuracy, the first three authors independently labeled the fact-checking information as either ``True,'' ``Unsure,'' or ``False.''
Final annotations were based on the majority labels (Krippendorff's $\alpha = $ 0.79).
Further details can be found in the Supplementary Information.

\paragraph{Human fact checks}
\pdfbookmark[1]{Human fact checks}{sec:methods:human_fcs}
\label{sec:methods:human_fcs}

Each human fact check begins with a clear statement about the truthfulness of the claim, such as ``The above claim is True'' or ``The above claim is False.''
Following this, the fact check addresses the publisher's reputation: if the headline is true, it mentions that the publisher is trustworthy; if false, it highlights the publisher's unreliability.
Brief supporting details are also provided to justify these assessments.
Further details can be found in the Supplementary Information.

\paragraph{Participant flow.}
\pdfbookmark[1]{Participant flow}{sec:methods:flow}
\label{sec:methods:flow}

All participants began by completing a brief survey, followed by exposure to their respective experimental conditions, followed by another brief survey and debriefing.
Regardless of the condition, all participants saw the same headlines in random order.
These stimuli were presented simultaneously with fact-checking information or questions about viewing fact checks (depending on experimental condition) along with questions regarding individual belief and sharing intention.
Participants who failed an attention check were excluded from the study.
Further details can be found in the Supplementary Information.

\paragraph{Preregistration}
\pdfbookmark[1]{Preregistration}{sec:methods:prereg}
\label{sec:methods:prereg}

Our preregistration~\cite{DeVerna2022cgptOSF} included the analysis plan and predicted outcomes related to results presented in the Ineffectiveness of LLM intervention section, excluding the human fact checks condition.
Data for this condition was collected later at the request of reviewers.
The preregistration also included various exploratory analyses without specific outcome predictions.
For all details, we refer the reader to the original preregistration document.

\section*{Code and data availability}
\pdfbookmark[0]{Code and data availability}{sec:methods:code_and_data}
\label{sec:methods:code_and_data}

All analysis code and data is available at: \href{https://github.com/osome-iu/AI_fact_checking}{github.com/osome-iu/AI\_fact\_checking}.

\section*{Acknowledgements}
\pdfbookmark[0]{Acknowledgements}{sec:methods:ack}
\label{sec:methods:ack}

The authors would like to acknowledge Lisa Fazio, David Rand, Stephan Lewandowsky, Adam Berinsky, Mark Susmann, and Gordon Pennycook, who kindly shared materials, as well as Byungkyu Lee, for helpful discussions during the early stages of this project.  
This research was supported in part by Knight Foundation and Volkswagen Foundation.

\phantomsection
\section*{Author contributions}
\pdfbookmark[0]{Author contributions}{sec:methods:contributions}
\label{sec:methods:contributions}

M.R.D. conceived of the research and developed the study design with help from H.Y.Y. and K.Y.
M.R.D. manually generated the fact checks.
M.R.D., H.Y.Y., and K.Y. developed the survey.
M.R.D. and H.Y.Y. conducted data analysis, with input from K.Y. and F.M.
M.R.D. wrote the paper, with input from all authors.

\phantomsection
\section*{Competing interests}
\pdfbookmark[0]{Competing interests}{sec:methods:contributions}
\label{sec:methods:competing_insterests}

The authors declare no competing interests.

\bibliography{ref.bib}

\begin{thebibliography}{85}
\providecommand{\natexlab}[1]{#1}
\providecommand{\url}[1]{\texttt{#1}}
\expandafter\ifx\csname urlstyle\endcsname\relax
  \providecommand{\doi}[1]{doi: #1}\else
  \providecommand{\doi}{doi: \begingroup \urlstyle{rm}\Url}\fi

\bibitem[Lewandowsky et~al.(2017)Lewandowsky, Ecker, and
  Cook]{lewandowsky2017beyond}
Stephan Lewandowsky, Ullrich~KH Ecker, and John Cook.
\newblock Beyond misinformation: Understanding and coping with the
  ``post-truth'' era.
\newblock \emph{Journal of Applied Research in Memory and Cognition},
  6\penalty0 (4):\penalty0 353--369, 2017.
\newblock URL \url{https://doi.org/10.1016/j.jarmac.2017.07.008}.

\bibitem[Lazer et~al.(2018)Lazer, Baum, Benkler, Berinsky, Greenhill, Menczer,
  Metzger, Nyhan, Pennycook, Rothschild, Schudson, Sloman, Sunstein, Thorson,
  Watts, and Zittrain]{Lazer2018Mar}
David M.~J. Lazer, Matthew~A. Baum, Yochai Benkler, Adam~J. Berinsky, Kelly~M.
  Greenhill, Filippo Menczer, Miriam~J. Metzger, Brendan Nyhan, Gordon
  Pennycook, David Rothschild, Michael Schudson, Steven~A. Sloman, Cass~R.
  Sunstein, Emily~A. Thorson, Duncan~J. Watts, and Jonathan~L. Zittrain.
\newblock {The science of fake news}.
\newblock \emph{Science}, 359\penalty0 (6380):\penalty0 1094--1096, March 2018.
\newblock \doi{10.1126/science.aao2998}.
\newblock URL \url{https://doi.org/10.1126/science.aao2998}.

\bibitem[Van~der Linden et~al.(2017)Van~der Linden, Leiserowitz, Rosenthal, and
  Maibach]{van2017inoculating}
Sander Van~der Linden, Anthony Leiserowitz, Seth Rosenthal, and Edward Maibach.
\newblock Inoculating the public against misinformation about climate change.
\newblock \emph{Global Challenges}, 1\penalty0 (2):\penalty0 1600008, 2017.
\newblock URL \url{https://doi.org/10.1002/gch2.201600008}.

\bibitem[Biddlestone et~al.(2022)Biddlestone, Azevedo, and {van der
  Linden}]{biddlestone2022climate}
Mikey Biddlestone, Flavio Azevedo, and Sander {van der Linden}.
\newblock Climate of conspiracy: A meta-analysis of the consequences of belief
  in conspiracy theories about climate change.
\newblock \emph{Current Opinion in Psychology}, 46:\penalty0 101390, 2022.
\newblock ISSN 2352-250X.
\newblock \doi{https://doi.org/10.1016/j.copsyc.2022.101390}.
\newblock URL
  \url{https://www.sciencedirect.com/science/article/pii/S2352250X22001099}.

\bibitem[Pierri et~al.(2022)Pierri, Perry, DeVerna, Yang, Flammini, Menczer,
  and Bryden]{pierri2022online}
Francesco Pierri, Brea~L Perry, Matthew~R DeVerna, Kai-Cheng Yang, Alessandro
  Flammini, Filippo Menczer, and John Bryden.
\newblock Online misinformation is linked to early {COVID}-19 vaccination
  hesitancy and refusal.
\newblock \emph{Scientific Reports}, 12\penalty0 (1):\penalty0 5966, 2022.
\newblock URL \url{https://doi.org/10.1038/s41598-022-10070-w}.

\bibitem[Rathje et~al.(2022)Rathje, He, Roozenbeek, Van~Bavel, and van~der
  Linden]{rathje2022social}
Steve Rathje, James~K He, Jon Roozenbeek, Jay~J Van~Bavel, and Sander van~der
  Linden.
\newblock Social media behavior is associated with vaccine hesitancy.
\newblock \emph{PNAS Nexus}, 1\penalty0 (4), 2022.
\newblock URL \url{https://doi.org/10.1093/pnasnexus/pgac207}.

\bibitem[Loomba et~al.(2021)Loomba, de~Figueiredo, Piatek, de~Graaf, and
  Larson]{loomba2021measuring}
Sahil Loomba, Alexandre de~Figueiredo, Simon~J Piatek, Kristen de~Graaf, and
  Heidi~J Larson.
\newblock Measuring the impact of {COVID}-19 vaccine misinformation on
  vaccination intent in the uk and usa.
\newblock \emph{Nature Human Behaviour}, 5\penalty0 (3):\penalty0 337--348,
  2021.
\newblock URL \url{https://doi.org/10.1038/s41562-021-01056-1}.

\bibitem[Tucker et~al.(2018)Tucker, Guess, Barberá, Vaccari, Siegel, Sanovich,
  Stukal, and Nyhan]{tucker2018social}
Joshua~A Tucker, Andrew Guess, Pablo Barberá, Cristian Vaccari, Alexandra
  Siegel, Sergey Sanovich, Denis Stukal, and Brendan Nyhan.
\newblock Social media, political polarization, and political disinformation: A
  review of the scientific literature.
\newblock SSRN[Preprint], 2018.
\newblock https://doi.org/10.2139/ssrn.3144139.

\bibitem[Van~Bavel et~al.(2021)Van~Bavel, Harris, Pärnamets, Rathje, Doell,
  and Tucker]{van2021political}
Jay~J Van~Bavel, Elizabeth~A Harris, Philip Pärnamets, Steve Rathje,
  Kimberly~C Doell, and Joshua~A Tucker.
\newblock Political psychology in the digital (mis)information age: A model of
  news belief and sharing.
\newblock \emph{Social Issues and Policy Review}, 15\penalty0 (1):\penalty0
  84--113, 2021.
\newblock URL \url{https://doi.org/10.1111/sipr.12077}.

\bibitem[Walter et~al.(2020)Walter, Cohen, Holbert, and Morag]{walter2020fact}
Nathan Walter, Jonathan Cohen, R~Lance Holbert, and Yasmin Morag.
\newblock Fact-checking: A meta-analysis of what works and for whom.
\newblock \emph{Political Communication}, 37\penalty0 (3):\penalty0 350--375,
  2020.
\newblock URL \url{https://doi.org/10.1080/10584609.2019.1668894}.

\bibitem[Brashier et~al.(2021)Brashier, Pennycook, Berinsky, and
  Rand]{brashier2021timing}
Nadia~M Brashier, Gordon Pennycook, Adam~J Berinsky, and David~G Rand.
\newblock Timing matters when correcting fake news.
\newblock \emph{Proceedings of the National Academy of Sciences}, 118\penalty0
  (5):\penalty0 e2020043118, 2021.
\newblock URL \url{https://doi.org/10.1073/pnas.2020043118}.

\bibitem[Nyhan et~al.(2020)Nyhan, Porter, Reifler, and Wood]{nyhan2020taking}
Brendan Nyhan, Ethan Porter, Jason Reifler, and Thomas~J Wood.
\newblock {Taking fact-checks literally but not seriously? The effects of
  journalistic fact-checking on factual beliefs and candidate favorability}.
\newblock \emph{Political Behavior}, 42:\penalty0 939--960, 2020.
\newblock URL \url{https://doi.org/10.1007/s11109-019-09528-x}.

\bibitem[Yaqub et~al.(2020)Yaqub, Kakhidze, Brockman, Memon, and
  Patil]{yaqub2020effects}
Waheeb Yaqub, Otari Kakhidze, Morgan~L. Brockman, Nasir Memon, and Sameer
  Patil.
\newblock Effects of credibility indicators on social media news sharing
  intent.
\newblock In \emph{Proceedings of the 2020 Conference on Human Factors in
  Computing Systems}, page 1–14, 2020.
\newblock ISBN 9781450367080.
\newblock \doi{10.1145/3313831.3376213}.
\newblock URL \url{https://doi.org/10.1145/3313831.3376213}.

\bibitem[Porter and Wood(2021)]{Porter2021Sep}
Ethan Porter and Thomas~J. Wood.
\newblock The global effectiveness of fact-checking: Evidence from simultaneous
  experiments in argentina, nigeria, south africa, and the united kingdom.
\newblock \emph{Proceedings of the National Academy of Sciences}, 118\penalty0
  (37):\penalty0 e2104235118, September 2021.
\newblock URL \url{https://doi.org/10.1073/pnas.2104235118}.

\bibitem[Pennycook and Rand(2021)]{pennycook2021psychology}
Gordon Pennycook and David~G Rand.
\newblock The psychology of fake news.
\newblock \emph{Trends in Cognitive Sciences}, 25\penalty0 (5):\penalty0
  388--402, 2021.
\newblock URL \url{https://doi.org/10.1016/j.tics.2021.02.007}.

\bibitem[Lee et~al.(2020)Lee, Li, Wang, Yih, Ma, and
  Khabsa]{lee_etal_2020_language}
Nayeon Lee, Belinda~Z. Li, Sinong Wang, Wen-tau Yih, Hao Ma, and Madian Khabsa.
\newblock Language models as fact checkers?
\newblock In \emph{Proceedings of the Third Workshop on Fact Extraction and
  VERification}, pages 36--41, 2020.
\newblock URL \url{https://aclanthology.org/2020.fever-1.5}.

\bibitem[Shu et~al.(2017)Shu, Sliva, Wang, Tang, and Liu]{shu2017fake}
Kai Shu, Amy Sliva, Suhang Wang, Jiliang Tang, and Huan Liu.
\newblock Fake news detection on social media: A data mining perspective.
\newblock \emph{SIGKDD Explor. Newsl.}, 19\penalty0 (1):\penalty0 22–36, sep
  2017.
\newblock ISSN 1931-0145.
\newblock \doi{10.1145/3137597.3137600}.
\newblock URL \url{https://doi.org/10.1145/3137597.3137600}.

\bibitem[Zhou and Zafarani(2020)]{zhou2020survey}
Xinyi Zhou and Reza Zafarani.
\newblock A survey of fake news: Fundamental theories, detection methods, and
  opportunities.
\newblock \emph{ACM Comput. Surv.}, 53\penalty0 (5), sep 2020.
\newblock ISSN 0360-0300.
\newblock \doi{10.1145/3395046}.
\newblock URL \url{https://doi.org/10.1145/3395046}.

\bibitem[Nakov et~al.(2021)Nakov, Corney, Hasanain, Alam, Elsayed,
  Barrón-Cedeño, Papotti, Shaar, and Da~San~Martino]{ijcai2021p619}
Preslav Nakov, David Corney, Maram Hasanain, Firoj Alam, Tamer Elsayed, Alberto
  Barrón-Cedeño, Paolo Papotti, Shaden Shaar, and Giovanni Da~San~Martino.
\newblock Automated fact-checking for assisting human fact-checkers.
\newblock In \emph{Proc. 30th Intl. Joint Conf. on Artificial Intelligence},
  pages 4551--4558, 8 2021.
\newblock URL \url{https://doi.org/10.24963/ijcai.2021/619}.

\bibitem[Yang et~al.(2021)Yang, Vega-Oliveros, Seibt, and Rocha]{Yang2021Dec}
Jing Yang, Didier Vega-Oliveros, Tais Seibt, and Anderson Rocha.
\newblock Scalable fact-checking with human-in-the-loop.
\newblock In \emph{2021 IEEE International Workshop on Information Forensics
  and Security (WIFS)}, pages 1--6, 2021.
\newblock \doi{10.1109/WIFS53200.2021.9648388}.
\newblock URL \url{https://doi.org/10.1109/WIFS53200.2021.9648388}.

\bibitem[Graves(2018)]{graves2018understanding}
D~Graves.
\newblock Understanding the promise and limits of automated fact-checking.
\newblock Technical report, Reuters Institute for the Study of Journalism,
  2018.
\newblock Publication Title: Reuters Institute for the Study of Journalism
  Series: Reuters Institute for the Study of Journalism Factsheets.

\bibitem[Hassan et~al.(2017)Hassan, Arslan, Li, and
  Tremayne]{2017NaeemulClaimBuster}
Naeemul Hassan, Fatma Arslan, Chengkai Li, and Mark Tremayne.
\newblock Toward automated fact-checking: Detecting check-worthy factual claims
  by claimbuster.
\newblock In \emph{Proceedings of the 23rd ACM SIGKDD International Conference
  on Knowledge Discovery and Data Mining}, page 1803–1812, 2017.
\newblock URL \url{https://doi.org/10.1145/3097983.3098131}.

\bibitem[Ciampaglia et~al.(2015)Ciampaglia, Shiralkar, Rocha, Bollen, Menczer,
  and Flammini]{Ciampaglia2015Jun}
Giovanni~Luca Ciampaglia, Prashant Shiralkar, Luis~M. Rocha, Johan Bollen,
  Filippo Menczer, and Alessandro Flammini.
\newblock Computational fact checking from knowledge networks.
\newblock \emph{PLoS One}, 10\penalty0 (6):\penalty0 e0128193, June 2015.
\newblock URL \url{https://doi.org/10.1371/journal.pone.0128193}.

\bibitem[Zeng et~al.(2021)Zeng, Abumansour, and Zubiaga]{zeng2021automated}
Xia Zeng, Amani~S Abumansour, and Arkaitz Zubiaga.
\newblock Automated fact-checking: A survey.
\newblock \emph{Language and Linguistics Compass}, 15\penalty0 (10):\penalty0
  e12438, 2021.
\newblock URL \url{https://doi.org/10.1111/lnc3.12438}.

\bibitem[Guo et~al.(2022)Guo, Schlichtkrull, and Vlachos]{guo2022survey}
Zhijiang Guo, Michael Schlichtkrull, and Andreas Vlachos.
\newblock {A Survey on Automated Fact-Checking}.
\newblock \emph{Transactions of the Association for Computational Linguistics},
  10:\penalty0 178--206, 2022.
\newblock URL \url{https://doi.org/10.1162/tacl_a_00454}.

\bibitem[Brown et~al.(2020)Brown, Mann, Ryder, Subbiah, Kaplan, Dhariwal,
  Neelakantan, Shyam, Sastry, Askell, Agarwal, Herbert-Voss, Krueger, Henighan,
  Child, Ramesh, Ziegler, Wu, Winter, Hesse, Chen, Sigler, Litwin, Gray, Chess,
  Clark, Berner, McCandlish, Radford, Sutskever, and Amodei]{brown2020language}
Tom Brown, Benjamin Mann, Nick Ryder, Melanie Subbiah, Jared~D Kaplan, Prafulla
  Dhariwal, Arvind Neelakantan, Pranav Shyam, Girish Sastry, Amanda Askell,
  Sandhini Agarwal, Ariel Herbert-Voss, Gretchen Krueger, Tom Henighan, Rewon
  Child, Aditya Ramesh, Daniel Ziegler, Jeffrey Wu, Clemens Winter, Chris
  Hesse, Mark Chen, Eric Sigler, Mateusz Litwin, Scott Gray, Benjamin Chess,
  Jack Clark, Christopher Berner, Sam McCandlish, Alec Radford, Ilya Sutskever,
  and Dario Amodei.
\newblock Language models are few-shot learners.
\newblock In \emph{Advances in Neural Information Processing Systems},
  volume~33, pages 1877--1901, 2020.
\newblock URL
  \url{https://proceedings.neurips.cc/paper_files/paper/2020/file/1457c0d6bfcb4967418bfb8ac142f64a-Paper.pdf}.

\bibitem[Ye et~al.(2023)Ye, Chen, Xu, Zu, Shao, Liu, Cui, Zhou, Gong, Shen,
  et~al.]{ye2023comprehensive}
Junjie Ye, Xuanting Chen, Nuo Xu, Can Zu, Zekai Shao, Shichun Liu, Yuhan Cui,
  Zeyang Zhou, Chao Gong, Yang Shen, et~al.
\newblock A comprehensive capability analysis of {GPT}-3 and {GPT}-3.5 series
  models.
\newblock arXiv[Preprint], 2023.
\newblock https://doi.org/10.48550/arXiv.2303.10420.

\bibitem[Qin et~al.(2023)Qin, Zhang, Zhang, Chen, Yasunaga, and
  Yang]{qin2023chatgpt}
Chengwei Qin, Aston Zhang, Zhuosheng Zhang, Jiaao Chen, Michihiro Yasunaga, and
  Diyi Yang.
\newblock Is {ChatGPT} a general-purpose natural language processing task
  solver?
\newblock arXiv[Preprint], 2023.
\newblock https://doi.org/10.48550/arXiv.2302.06476.

\bibitem[Katz et~al.(2023)Katz, Bommarito, Gao, and
  Arredondo]{Katz2023GPTpassbar}
Daniel~Martin Katz, Michael~James Bommarito, Shang Gao, and Pablo Arredondo.
\newblock {GPT}-4 passes the bar exam.
\newblock SSRN[Preprint], March 2023.
\newblock https://dx.doi.org/10.2139/ssrn.4389233.

\bibitem[{OpenAI}(2023)]{openai2023gpt4}
{OpenAI}.
\newblock {GPT}-4 technical report.
\newblock arXiv[Preprint], Mar 2023.
\newblock https://doi.org/10.48550/arXiv.2303.08774.

\bibitem[Yang and Menczer(2023)]{Yang2023LLMRating}
Kai-Cheng Yang and Filippo Menczer.
\newblock Large language models can rate news outlet credibility.
\newblock arXiv[Preprint], April 2023.
\newblock https://arxiv.org/abs/2304.00228.

\bibitem[Quelle and Bovet(2024)]{Quelle2023Oct}
Dorian Quelle and Alexandre Bovet.
\newblock The perils and promises of fact-checking with large language models.
\newblock \emph{Frontiers in Artificial Intelligence}, 7:\penalty0 1341697,
  2024.

\bibitem[Hoes et~al.(2023)Hoes, Altay, and Bermeo]{Hoes2023GPTrateClaims}
Emma Hoes, Sacha Altay, and Juan Bermeo.
\newblock Leveraging {ChatGPT} for efficient fact-checking.
\newblock PsyArXiv[Preprint], April 2023.
\newblock https://doi.org/10.31234/osf.io/qnjkf.

\bibitem[Kuznetsova et~al.(2023)Kuznetsova, Makhortykh, Vziatysheva, Stolze,
  Baghumyan, and Urman]{Kuznetsova2023generative}
Elizaveta Kuznetsova, Mykola Makhortykh, Victoria Vziatysheva, Martha Stolze,
  Ani Baghumyan, and Aleksandra Urman.
\newblock In generative {AI} we trust: Can chatbots effectively verify
  political information?
\newblock arXiv[Preprint], December 2023.
\newblock https://doi.org/10.48550/arXiv.2312.13096.

\bibitem[Zhou et~al.(2024)Zhou, Sharma, Zhang, and Althoff]{Zhou2024Muse}
Xinyi Zhou, Ashish Sharma, Amy~X. Zhang, and Tim Althoff.
\newblock {Correcting misinformation on social media with a large language
  model}.
\newblock arXiv[Preprint], 2024.
\newblock https://doi.org/10.48550/arXiv.2403.11169.

\bibitem[He et~al.(2023)He, Ahamad, and Kumar]{He2023Mar}
Bing He, Mustaque Ahamad, and Srijan Kumar.
\newblock Reinforcement learning-based counter-misinformation response
  generation: A case study of {COVID}-19 vaccine misinformation.
\newblock In \emph{{Proceedings of the ACM Web Conference 2023}}, pages
  2698--2709, 2023.
\newblock \doi{10.1145/3543507.3583388}.
\newblock URL \url{https://doi.org/10.1145/3543507.3583388}.

\bibitem[Taori et~al.(2023)Taori, Gulrajani, Zhang, Dubois, Li, Guestrin,
  Liang, and Hashimoto]{standfordAlpaca2023}
Rohan Taori, Ishaan Gulrajani, Tianyi Zhang, Yann Dubois, Xuechen Li, Carlos
  Guestrin, Percy Liang, and Tatsunori~B. Hashimoto.
\newblock Alpaca: A strong, replicable instruction-following model.
\newblock Stanford Center for Research on Foundation Models Blog, April 2023.
\newblock https://crfm.stanford.edu/2023/03/13/alpaca.html (accessed 7 April
  2023).

\bibitem[Conover et~al.(2023)Conover, Hayes, Mathur, Meng, Xie, Wan, Ghodsi,
  Wendell, and Zaharia]{databricks2023Dolly}
Mike Conover, Matt Hayes, Ankit Mathur, Xiangrui Meng, Jianwei Xie, Jun Wan,
  Ali Ghodsi, Patrick Wendell, and Matei Zaharia.
\newblock {Hello Dolly: Democratizing the magic of ChatGPT with open models}.
\newblock Databricks Blog, March 2023.
\newblock
  https://www.databricks.com/blog/2023/03/24/hello-dolly-democratizing-magic-chatgpt-open-models.html
  (accessed 7 April 2023).

\bibitem[Hancock et~al.(2020)Hancock, Naaman, and Levy]{Hancock2020Mar}
Jeffrey~T. Hancock, Mor Naaman, and Karen Levy.
\newblock {AI}-mediated communication: Definition, research agenda, and ethical
  considerations.
\newblock \emph{J Comput Mediat Commun}, 25\penalty0 (1):\penalty0 89--100,
  March 2020.
\newblock URL \url{https://doi.org/10.1093/jcmc/zmz022}.

\bibitem[Sundar(2008)]{sundar2008main}
S.~Shyam Sundar.
\newblock The {MAIN} model: A heuristic approach to understanding technology
  effects on credibility.
\newblock In M~Metzger and Andrew Flanagin, editors, \emph{Digital Media,
  Youth, and Credibility}, volume 2008, pages 73--100. MIT Press, 2008.

\bibitem[Sundar and Kim(2019)]{sundar_machine_2019}
S.~Shyam Sundar and Jinyoung Kim.
\newblock Machine heuristic: When we trust computers more than humans with our
  personal information.
\newblock In \emph{Proceedings of ACM Conference on Human Factors in Computing
  Systems}, pages 1--9, 2019.
\newblock \doi{10.1145/3290605.3300768}.
\newblock URL \url{https://doi.org/10.1145/3290605.3300768}.

\bibitem[Sundar(2020)]{sundar2020rise}
S~Shyam Sundar.
\newblock Rise of machine agency: A framework for studying the psychology of
  human–{AI} interaction ({HAII}).
\newblock \emph{Journal of Computer-Mediated Communication}, 25\penalty0
  (1):\penalty0 74--88, 2020.
\newblock URL \url{https://doi.org/10.1093/jcmc/zmz026}.

\bibitem[DeAndrea(2014)]{DeAndrea2014May}
David~C. DeAndrea.
\newblock Advancing warranting theory.
\newblock \emph{Communication Theory}, 24\penalty0 (2):\penalty0 186--204, May
  2014.
\newblock URL \url{https://doi.org/10.1111/comt.12033}.

\bibitem[Luger and Sellen(2016)]{Luger2016May}
Ewa Luger and Abigail Sellen.
\newblock ``like having a really bad {PA}'': The gulf between user expectation
  and experience of conversational agents.
\newblock In \emph{Proceedings of the 2016 CHI Conference on Human Factors in
  Computing Systems}, page 5286–5297, 2016.
\newblock \doi{10.1145/2858036.2858288}.
\newblock URL \url{https://doi.org/10.1145/2858036.2858288}.

\bibitem[Meurisch et~al.(2020)Meurisch, Mihale-Wilson, Hawlitschek, Giger,
  M\"{u}ller, Hinz, and M\"{u}hlh\"{a}user]{Meurisch2020Dec}
Christian Meurisch, Cristina~A. Mihale-Wilson, Adrian Hawlitschek, Florian
  Giger, Florian M\"{u}ller, Oliver Hinz, and Max M\"{u}hlh\"{a}user.
\newblock Exploring user expectations of proactive {AI} systems.
\newblock \emph{Proc. ACM Interact. Mob. Wearable Ubiquitous Technol.},
  4\penalty0 (4), dec 2020.
\newblock \doi{10.1145/3432193}.
\newblock URL \url{https://doi.org/10.1145/3432193}.

\bibitem[Shi et~al.(2022)Shi, Bhattacharya, Das, Lease, and
  Gwizdka]{Shi2022Mar}
Li~Shi, Nilavra Bhattacharya, Anubrata Das, Matt Lease, and Jacek Gwizdka.
\newblock The effects of interactive ai design on user behavior: An
  eye-tracking study of fact-checking {COVID}-19 claims.
\newblock In \emph{Proceedings of Conference on Human Information Interaction
  and Retrieval}, pages 315--320, 2022.
\newblock URL \url{https://doi.org/10.1145/3498366.3505786}.

\bibitem[Smith-Renner et~al.(2020)Smith-Renner, Fan, Birchfield, Wu,
  Boyd-Graber, Weld, and Findlater]{smith-renner_no_2020}
Alison Smith-Renner, Ron Fan, Melissa Birchfield, Tongshuang Wu, Jordan
  Boyd-Graber, Daniel~S. Weld, and Leah Findlater.
\newblock No explainability without accountability: An empirical study of
  explanations and feedback in interactive {ML}.
\newblock In \emph{Proceedings of the Conference on Human Factors in Computing
  Systems}, pages 1--13, 2020.
\newblock URL \url{https://doi.org/10.1145/3313831.3376624}.

\bibitem[Zhang et~al.(2020)Zhang, Liao, and Bellamy]{Zhang2020Jan}
Yunfeng Zhang, Q.~Vera Liao, and Rachel K.~E. Bellamy.
\newblock Effect of confidence and explanation on accuracy and trust
  calibration in {AI}-assisted decision making.
\newblock In \emph{Proceedings of Conference on Fairness, Accountability, and
  Transparency}, pages 295--305, 2020.
\newblock URL \url{https://doi.org/10.1145/3351095.3372852}.

\bibitem[Bansal et~al.(2021)Bansal, Wu, Zhou, Fok, Nushi, Kamar, Ribeiro, and
  Weld]{bansal_does_2021}
Gagan Bansal, Tongshuang Wu, Joyce Zhou, Raymond Fok, Besmira Nushi, Ece Kamar,
  Marco~Tulio Ribeiro, and Daniel Weld.
\newblock Does the whole exceed its parts? the effect of {AI} explanations on
  complementary team performance.
\newblock In \emph{Proceedings of Conference on Human Factors in Computing
  Systems}, pages 1--16, 2021.
\newblock URL \url{https://doi.org/10.1145/3411764.3445717}.

\bibitem[DeVerna et~al.(2023)DeVerna, Yan, Yang, and
  Menczer]{DeVerna2022cgptOSF}
Matthew~R. DeVerna, Harry~Yaojun Yan, Kai-Cheng Yang, and Filippo Menczer.
\newblock {ChatGPT} fact-checking as a misinformation intervention.
\newblock OSF Preregistration, March 2023.
\newblock https://osf.io/58rmu.

\bibitem[Pennycook et~al.(2021{\natexlab{a}})Pennycook, Epstein, Mosleh,
  Arechar, Eckles, and Rand]{Pennycook2021Shifting}
Gordon Pennycook, Ziv Epstein, Mohsen Mosleh, Antonio~A. Arechar, Dean Eckles,
  and David~G. Rand.
\newblock Shifting attention to accuracy can reduce misinformation online.
\newblock \emph{Nature}, 592:\penalty0 590--595, 2021{\natexlab{a}}.
\newblock URL \url{https://doi.org/10.1038/s41586-021-03344-2}.

\bibitem[Guay et~al.(2023)Guay, Berinsky, Pennycook, and Rand]{Guay2023Mar}
Brian Guay, Adam Berinsky, Gordon Pennycook, and David Rand.
\newblock How to think about whether misinformation interventions work.
\newblock \emph{Nature Human Behaviour}, 7, 2023.
\newblock \doi{10.1038/s41562-023-01667-w}.
\newblock URL \url{https://doi.org/10.1038/s41562-023-01667-w}.

\bibitem[DeVerna et~al.(2024)DeVerna, Guess, Berinsky, Tucker, and
  Jost]{DeVerna2022Rumors}
Matthew~R. DeVerna, Andrew~M. Guess, Adam~J. Berinsky, Joshua~A. Tucker, and
  John~T. Jost.
\newblock Rumors in retweet: Ideological asymmetry in the failure to correct
  misinformation.
\newblock \emph{Personality and Social Psychology Bulletin}, 50\penalty0
  (1):\penalty0 3--17, 2024.
\newblock \doi{10.1177/01461672221114222}.

\bibitem[Mosleh et~al.(2020)Mosleh, Pennycook, and Rand]{Mosleh2020Feb}
Mohsen Mosleh, Gordon Pennycook, and David~G. Rand.
\newblock Self-reported willingness to share political news articles in online
  surveys correlates with actual sharing on twitter.
\newblock \emph{PLoS One}, 15\penalty0 (2):\penalty0 e0228882, February 2020.
\newblock URL \url{https://doi.org/10.1371/journal.pone.0228882}.

\bibitem[Lewis et~al.(2020)Lewis, Perez, Piktus, Petroni, Karpukhin, Goyal,
  Küttler, Lewis, Yih, Rocktäschel, Riedel, and Kiela]{Lewis2020RAG}
Patrick Lewis, Ethan Perez, Aleksandra Piktus, Fabio Petroni, Vladimir
  Karpukhin, Naman Goyal, Heinrich Küttler, Mike Lewis, Wen-tau Yih, Tim
  Rocktäschel, Sebastian Riedel, and Douwe Kiela.
\newblock Retrieval-augmented generation for knowledge-intensive {NLP} tasks.
\newblock \emph{Advances in Neural Information Processing Systems},
  33:\penalty0 9459--9474, 2020.
\newblock URL
  \url{https://proceedings.neurips.cc/paper/2020/hash/6b493230205f780e1bc26945df7481e5-Abstract.html}.

\bibitem[Bai et~al.(2023)Bai, Voelkel, Eichstaedt, and
  Willer]{bai_artificial_2023}
(Max)~Hui Bai, Jan~G. Voelkel, Johannes~C. Eichstaedt, and Robb Willer.
\newblock Artificial intelligence can persuade humans on political issues.
\newblock OSF[Preprint], February 2023.
\newblock https://osf.io/stakv.

\bibitem[Acerbi et~al.(2022)Acerbi, Altay, and Mercier]{Acerbi2022Jan}
Alberto Acerbi, Sacha Altay, and Hugo Mercier.
\newblock Research note: Fighting misinformation or fighting for information?
\newblock \emph{Harvard Kennedy School Misinformation Review}, 3\penalty0 (1),
  January 2022.
\newblock URL \url{https//doi.org/10.37016/mr-2020-87}.

\bibitem[Park et~al.(2021)Park, Park, Kang, and Cha]{Park2021Jan}
Sungkyu Park, Jaimie~Yejean Park, Jeong-han Kang, and Meeyoung Cha.
\newblock {The presence of unexpected biases in online fact-checking}.
\newblock \emph{Harvard Kennedy School Misinformation Review}, 2, January 2021.
\newblock \doi{10.37016/mr-2020-53}.
\newblock URL \url{https://doi.org/10.37016/mr-2020-53}.

\bibitem[Kim et~al.(2024)Kim, Liao, Vorvoreanu, Ballard, and Vaughan]{kim2024m}
Sunnie~SY Kim, Q~Vera Liao, Mihaela Vorvoreanu, Stephanie Ballard, and
  Jennifer~Wortman Vaughan.
\newblock ``{I}'m not sure, but...'': Examining the impact of large language
  models' uncertainty expression on user reliance and trust.
\newblock arXiv[Preprint], 2024.
\newblock https://doi.org/10.48550/arXiv.2405.00623.

\bibitem[Kotonya and Toni(2020)]{kotonya-toni-2020-explainable}
Neema Kotonya and Francesca Toni.
\newblock Explainable automated fact-checking: A survey.
\newblock In \emph{Proceedings of 28th International Conference on
  Computational Linguistics}, pages 5430--5443, 2020.
\newblock \doi{10.18653/v1/2020.coling-main.474}.
\newblock URL \url{https://aclanthology.org/2020.coling-main.474}.

\bibitem[Li et~al.(2021)Li, Foley, Dumdum, and Wagner]{Li2022Bias}
Jianing Li, Jordan~M. Foley, Omar Dumdum, and Michael~W. Wagner.
\newblock {The Power of a Genre: Political News Presented as Fact-Checking
  Increases Accurate Belief Updating and Hostile Media Perceptions}.
\newblock \emph{Mass Communication and Society}, 25\penalty0 (2):\penalty0
  282--307, March 2021.
\newblock \doi{10.1080/15205436.2021.1924382}.
\newblock URL \url{https://doi.org/10.1080/15205436.2021.1924382}.

\bibitem[Bode et~al.(2024)Bode, Vraga, and Tang]{Bode2024Correction}
Leticia Bode, Emily~K. Vraga, and Rongwei Tang.
\newblock {User correction}.
\newblock \emph{Current Opinion in Psychology}, 56:\penalty0 101786, April
  2024.
\newblock \doi{10.1016/j.copsyc.2023.101786}.
\newblock URL \url{https://doi.org/10.1016/j.copsyc.2023.101786}.

\bibitem[Li and Wagner(2020)]{Li2020IndivFact}
Jianing Li and Michael~W. Wagner.
\newblock The value of not knowing: Partisan cue-taking and belief updating of
  the uninformed, the ambiguous, and the misinformed.
\newblock \emph{J Commun}, 70\penalty0 (5):\penalty0 646--669, October 2020.
\newblock \doi{10.1093/joc/jqaa022}.
\newblock URL \url{https://doi.org/10.1093/joc/jqaa022}.

\bibitem[Pasek et~al.(2015)Pasek, Sood, and Krosnick]{Pasek2015Aug}
Josh Pasek, Gaurav Sood, and Jon~A. Krosnick.
\newblock Misinformed about the affordable care act? leveraging certainty to
  assess the prevalence of misperceptions.
\newblock \emph{J Commun}, 65\penalty0 (4):\penalty0 660--673, August 2015.
\newblock \doi{10.1111/jcom.12165}.
\newblock URL \url{https://doi.org/10.1111/jcom.12165}.

\bibitem[Yin et~al.(2024)Yin, Jia, and Wakslak]{yin2024heard}
Yidan Yin, Nan Jia, and Cheryl~J. Wakslak.
\newblock {AI} can help people feel heard, but an {AI} label diminishes this
  impact.
\newblock \emph{Proceedings of the National Academy of Sciences}, 121\penalty0
  (14):\penalty0 e2319112121, 2024.
\newblock \doi{10.1073/pnas.2319112121}.
\newblock URL \url{https://www.pnas.org/doi/abs/10.1073/pnas.2319112121}.

\bibitem[Lim and Schmälzle(2024)]{lim2024effect}
Sue Lim and Ralf Schmälzle.
\newblock The effect of source disclosure on evaluation of {AI}-generated
  messages.
\newblock \emph{Computers in Human Behavior: Artificial Humans}, 2\penalty0
  (1):\penalty0 100058, 2024.
\newblock ISSN 2949-8821.
\newblock \doi{https://doi.org/10.1016/j.chbah.2024.100058}.
\newblock URL
  \url{https://www.sciencedirect.com/science/article/pii/S2949882124000185}.

\bibitem[Zhang and Gosline(2023)]{Zhang_Gosline_2023}
Yunhao Zhang and Renée Gosline.
\newblock Human favoritism, not {AI} aversion: People’s perceptions (and
  bias) toward generative {AI}, human experts, and human–{GAI} collaboration
  in persuasive content generation.
\newblock \emph{Judgment and Decision Making}, 18:\penalty0 e41, 2023.
\newblock \doi{10.1017/jdm.2023.37}.

\bibitem[Rae(2024)]{irene2024effect}
Irene Rae.
\newblock The effects of perceived {AI} use on content perceptions.
\newblock In \emph{Proceedings of the CHI Conference on Human Factors in
  Computing Systems}, CHI '24, New York, NY, USA, 2024. Association for
  Computing Machinery.
\newblock ISBN 9798400703300.
\newblock \doi{10.1145/3613904.3642076}.
\newblock URL \url{https://doi.org/10.1145/3613904.3642076}.

\bibitem[Spitale et~al.(2023)Spitale, Biller-Andorno, and
  Germani]{Spitale2023Jun}
Giovanni Spitale, Nikola Biller-Andorno, and Federico Germani.
\newblock Ai model {GPT}-3 (dis)informs us better than humans.
\newblock \emph{Science Advances}, 9\penalty0 (26):\penalty0 eadh1850, June
  2023.
\newblock URL \url{https://doi.org/10.1126/sciadv.adh1850}.

\bibitem[Goldstein et~al.(2023)Goldstein, Sastry, Musser, DiResta, Gentzel, and
  Sedova]{goldstein_generative_2023}
Josh~A. Goldstein, Girish Sastry, Micah Musser, Renee DiResta, Matthew Gentzel,
  and Katerina Sedova.
\newblock Generative language models and automated influence operations:
  Emerging threats and potential mitigations.
\newblock arXiv[Preprint], January 2023.
\newblock http://arxiv.org/abs/2301.04246.

\bibitem[Brewster et~al.(2023)Brewster, Arvanitis, and
  Sadeghi]{Brewster2023chatgpt}
Jack Brewster, Lorenzo Arvanitis, and McKenzie Sadeghi.
\newblock Could {ChatGPT} become a monster misinformation superspreader?
\newblock NewsGuard blog, March 2023.
\newblock https://www.newsguardtech.com/misinformation-monitor/jan-2023
  (accessed 28 March 2023).

\bibitem[Menczer et~al.(2023)Menczer, Crandall, Ahn, and
  Kapadia]{Menczer2023NatMI}
Filippo Menczer, David Crandall, Yong-Yeol Ahn, and Apu Kapadia.
\newblock Addressing the harms of {AI}-generated inauthentic content.
\newblock \emph{Nature Machine Intelligence}, 5:\penalty0 679--680, July 2023.
\newblock URL \url{https://doi.org/10.1038/s42256-023-00690-w}.

\bibitem[Yang and Menczer(2024)]{Yang2023Anatomy-AI-botnet}
Kai-Cheng Yang and Filippo Menczer.
\newblock Anatomy of an {AI}-powered malicious social botnet.
\newblock \emph{Journal of Quantitative Description: Digital Media}, 4, May
  2024.
\newblock \doi{10.51685/jqd.2024.icwsm.7}.
\newblock URL \url{https://journalqd.org/article/view/5848}.

\bibitem[Solaiman et~al.(2019)Solaiman, Brundage, Clark, Askell, Herbert-Voss,
  Wu, Radford, et~al.]{Solaiman2019release}
Irene Solaiman, Miles Brundage, Jack Clark, Amanda Askell, Ariel Herbert-Voss,
  Jeff Wu, Alec Radford, et~al.
\newblock Release strategies and the social impacts of language models.
\newblock arXiv[Preprint], November 2019.
\newblock https://doi.org/10.48550/arXiv.1908.09203.

\bibitem[Karinshak et~al.(2023)Karinshak, Liu, Park, and
  Hancock]{Karinshak2023Apr}
Elise Karinshak, Sunny~Xun Liu, Joon~Sung Park, and Jeffrey~T. Hancock.
\newblock Working with {AI} to persuade: Examining a large language model's
  ability to generate pro-vaccination messages.
\newblock \emph{{Proceedings of the ACM on Human-Computer Interaction}},
  7\penalty0 (CSCW1):\penalty0 1--29, 2023.
\newblock URL \url{https://doi.org/10.1145/3579592}.

\bibitem[Goldstein et~al.(2024)Goldstein, Chao, Grossman, Stamos, and
  Tomz]{Goldstein2024Feb}
Josh~A. Goldstein, Jason Chao, Shelby Grossman, Alex Stamos, and Michael Tomz.
\newblock {How persuasive is AI-generated propaganda?}
\newblock \emph{PNAS Nexus}, 3\penalty0 (2):\penalty0 pgae034, 2024.
\newblock \doi{10.1093/pnasnexus/pgae034}.
\newblock URL \url{https://doi.org/10.1093/pnasnexus/pgae034}.

\bibitem[{U.S. Census Bureau}(2020)]{census2020educationalattainment}
{U.S. Census Bureau}.
\newblock Educational attainment in the united states: 2020, 2020.
\newblock
  https://www.census.gov/data/tables/2020/demo/educational-attainment/cps-detailed-tables.html
  (accessed 7 April 2023).

\bibitem[{Pew Research Center}(2020)]{pewresearch2020electorate}
{Pew Research Center}.
\newblock What the 2020 electorate looks like by party, race and ethnicity,
  age, education and religion, 2020.
\newblock
  https://www.pewresearch.org/short-reads/2020/10/26/what-the-2020-electorate-looks-like-by-party-race-and-ethnicity-age-education-and-religion
  (accessed 7 April 2023).

\bibitem[Fazio et~al.(2024)Fazio, Rand, Lewandowsky, Susmann, Berinsky, Guess,
  Kendeou, Lyons, Miller, Newman, Pennycook, and
  Swire-Thompson]{MercuryProject}
Lisa Fazio, David Rand, Stephan Lewandowsky, Mark Susmann, Adam~J. Berinsky,
  Andrew Guess, Panayiota Kendeou, Benjamin Lyons, Joanne Miller, Eryn Newman,
  Gordon Pennycook, and Briony Swire-Thompson.
\newblock {Combating misinformation: A megastudy of nine interventions designed
  to reduce the sharing of and belief in false and misleading headlines}.
\newblock OSF Preprint, 2024.
\newblock URL \url{https://doi.org/10.31234/osf.io/uyjha}.

\bibitem[Faul et~al.(2007)Faul, Erdfelder, Lang, and Buchner]{faul2007gpower}
Franz Faul, Egon Erdfelder, Achim-Gerd Lang, and Axel Buchner.
\newblock G*power 3: A flexible statistical power analysis program for the
  social, behavioral, and biomedical sciences.
\newblock \emph{{Behavior Research Methods}}, 39\penalty0 (2):\penalty0
  175--191, 2007.
\newblock \doi{10.3758/BF03193146}.

\bibitem[Pennycook et~al.(2021{\natexlab{b}})Pennycook, Binnendyk, Newton, and
  Rand]{Pennycook2021Jan}
Gordon Pennycook, Jabin Binnendyk, Christie Newton, and David~G. Rand.
\newblock {A Practical Guide to Doing Behavioral Research on Fake News and
  Misinformation}.
\newblock \emph{Collabra: Psychology}, 7\penalty0 (1), January
  2021{\natexlab{b}}.
\newblock URL \url{https://doi.org/10.1525/collabra.25293}.

\bibitem[Pennycook and Rand(2019)]{pennycook2019lazy}
Gordon Pennycook and David~G Rand.
\newblock Lazy, not biased: Susceptibility to partisan fake news is better
  explained by lack of reasoning than by motivated reasoning.
\newblock \emph{Cognition}, 188:\penalty0 39--50, 2019.
\newblock URL \url{https://doi.org/10.1016/j.cognition.2018.06.011}.

\bibitem[Pennycook et~al.(2020)Pennycook, Bear, Collins, and
  Rand]{pennycook2020implied}
Gordon Pennycook, Adam Bear, Evan~T Collins, and David~G Rand.
\newblock The implied truth effect: Attaching warnings to a subset of fake news
  headlines increases perceived accuracy of headlines without warnings.
\newblock \emph{Management science}, 66\penalty0 (11):\penalty0 4944--4957,
  2020.
\newblock URL \url{https://doi.org/10.1287/mnsc.2019.3478}.

\bibitem[OpenAI(2023)]{ChatGPT_ReleaseNotes}
OpenAI.
\newblock {ChatGPT --- Release Notes}, May 2023.
\newblock URL
  \url{https://help.openai.com/en/articles/6825453-chatgpt-release-notes}.
\newblock [Online; accessed 16. May 2023].

\bibitem[Sindermann et~al.(2021)Sindermann, Sha, Zhou, Wernicke, Schmitt, Li,
  Sariyska, Stavrou, Becker, and Montag]{sindermann_assessing_2021}
Cornelia Sindermann, Peng Sha, Min Zhou, Jennifer Wernicke, Helena~S. Schmitt,
  Mei Li, Rayna Sariyska, Maria Stavrou, Benjamin Becker, and Christian Montag.
\newblock {Assessing the Attitude Towards Artificial Intelligence: Introduction
  of a Short Measure in German, Chinese, and English Language}.
\newblock \emph{Kunstliche Intelligenz}, 35\penalty0 (1):\penalty0 109--118,
  March 2021.
\newblock URL \url{https://doi.org/10.1007/s13218-020-00689-0}.

\end{thebibliography}

\clearpage

\setcounter{page}{1}
\appendix

% Everything in this block helps to renew the pdf-rendering link for SI Figures and Tables, while also prefixing them with "S" and restarting the counter at zero.
% -------
\counterwithin{figure}{section}
\counterwithin{table}{section}

% This prevents figure and table numbering from restarting in each section of the appendix
\counterwithout{figure}{section}
\counterwithout{table}{section}

% These prefix the rendered Figure and Table numbers with an "S"
\renewcommand{\thefigure}{S\arabic{figure}}
\renewcommand{\thetable}{S\arabic{table}}

% This resets the counter at 0
\setcounter{figure}{0}
% (not needed for table because we do not have tables in the main text)

% Numbering style for sections
\renewcommand{\thesection}{\arabic{section}}

% Numbering style for subsections
\renewcommand{\thesubsection}{\thesection.\arabic{subsection}}
% -------

\section*{Supplementary information}
\pdfbookmark[0]{Supplementary information}{sec:si}
\label{sec:si}

% Add TOC
\tableofcontents

% Start on next page
\clearpage

\section{Supplementary methods}
% \pdfbookmark[0]{Supplementary methods}{sec:si:supp-methods}
\label{sec:si:supp-methods}

\subsection{Sampling details}
% \pdfbookmark[0]{Sampling details}{sec:si:sampling}
\label{sec:si:sampling}

In our final sample, females comprised 53.40\% of the sample, males 46.46\%, and other genders 0.14\%.
Age segments were 65+ (20.66\%), 55-64 (16.78\%), 45-54 (17.74\%), 35-44 (16.91\%), 25-34 (18.25\%), and 18-24 (9.68\%). 
Race percentages were: White (60.17\%), Hispanic or Latino/a (17.46\%), Black or African American (13.43\%), Asian (5.51\%), and Other (3.43\%).
Slightly more than half of the sample (51.92\%) had less than a college education, while 48.08\% had a college degree.
With respect to party identification, 50.72\% identified as Democrat or Democrat-leaning, 43.68\% as Republican or Republican-leaning, and 5.60\% as Independent.

The sampling plan for the control, LLM-optional, and LLM-forced conditions, for both the belief and sharing groups, was preregistered\cite{DeVerna2022cgptOSF} with the goal of obtaining .95 power to detect a small effect size of .1 at the standard .05 error probability with two-by-three-level between-subject manipulations ([Belief vs. Sharing groups] $\times$ [Control, LLM-forced, LLM-optional conditions]).
Power analysis by the G*Power\cite{faul2007gpower} software suggested a minimum number of 44 subjects per condition ($N = 264$), but we aimed for a larger target sample size of $N = 1{,}500$ (250 participants per condition) to increase the precision of our measurements.
As noted in the Materials and Methods, data for the human fact check conditions were gathered later at reviewers' request.
Gathering this data required larger sample sizes to meet the minimum spending threshold of our survey partner (Qualtrics) for both the belief and sharing conditions (300 participants per condition).

\subsection{Stimuli curation}
\label{sec:stimuli_curation}

The news headlines used as stimuli were selected from a project aimed at comparing misinformation interventions\cite{MercuryProject}. 
Specifically, 40 headlines were selected from a set of 317 political news stories using a pretest approach\cite{Pennycook2021Jan,pennycook2019lazy,pennycook2020implied} to balance the selected headlines in terms of perceived partisanship, impact, familiarity, sensationalism, and the likelihood of being shared and believed.

The 20 false headlines were originally selected from a third-party fact-checking website (\href{https://www.snopes.com/}{snopes.com}), ensuring their falsehood.
The 20 true headlines were all accurate and selected from mainstream news outlets (e.g., \textit{New York Times}, \textit{Washington Post}, \textit{Fox News}, and \textit{Wall Street Journal}) to be roughly contemporary with the false news headlines.

The claims were presented in a digital format resembling popular social media platforms, commonly known as the ``Facebook format''\cite{Pennycook2021Shifting}, which includes an image, the article headline, and a lede sentence (if present).
See the \nameref{sec:headlines-fcs} section for all stimuli text.

\subsection{LLM fact check generation}

A new ChatGPT session was created on the publicly available OpenAI website (\href{https://chat.openai.com/}{chat.openai.com}), where the headline text was inserted into a prompt asking, ``I saw something today that claimed $<$HEADLINE TEXT$>$. Do you think that this is likely to be true?''
The source of an article (e.g., ``nytimes.com'') was excluded.
If an article's lede sentence was shown in the stimulus image, it was also included in the prompt, separated by a colon. 
The prompt for each headline was provided to ChatGPT only once, and the response was saved as a screenshot.
All headlines were generated on January 25, 2023, between 12:30--8:00pm Eastern Standard Time.
According to the release notes\cite{ChatGPT_ReleaseNotes}, the language model utilized by ChatGPT at that time was a version of \texttt{GPT-3.5} that has since been updated and is no longer available.
See the \nameref{sec:headlines-fcs} section for the text of all fact checks as well as the \nameref{sec:prompt_eng} section for further analysis of model accuracy.

\subsection{Human fact check generation}

Human fact checks were generated to create a uniform structure with clear judgments, as outlined in the Materials and Methods.
Fact checks for false headlines were selected from the same misinformation intervention study\cite{MercuryProject} from which headline stimuli were selected.
Since that study did not create fact checks for true headlines, one of the authors manually created these by reading each article to identify accurate and relevant information to support the veracity label.
See the \nameref{sec:headlines-fcs} section for the text of all fact checks.

\subsection{Attrition}

Drop out rates varied between 1\%--6\% across experimental conditions, as reported in Table~\ref{tab:attrition}.

\begin{table}
    \centering
    \begin{tabular}{lcc}
    \hline
    Condition     &  Drop-out & Attention-check failure \\
    \hline
    Belief control           & 1.46\%  & 63.25\% \\
    Belief LLM-forced        & 3.98\%  & 55.06\% \\
    Belief LLM-optional      & 5.27\%  & 54.26\%\\
    Belief human fact check  & 3.33\%  & 67.32\% \\
    \hline
    Sharing control          & 1.80\% & 50.27\% \\
    Sharing LLM-forced       & 6.06\% & n/a \\
    Sharing LLM-optional     & 3.62\% & 54.96\%\\
    Sharing human fact check & 3.19\% & 60.05\% \\
    \hline
    \end{tabular}
    \caption{Drop out and attention-check failure rates for each experimental condition.}
    \label{tab:attrition}
\end{table}

We incorporated an attention-check question that involved a headline created by the researchers stating that the color of the sky is yellow.
Prior to viewing any headlines, participants were informed about this specific headline and instructed to later answer ``Yes'' when asked if they believed the headline or were willing to share it, depending on their respective experimental conditions.
To minimize the distinction between the attention check and the regular experimental stimuli, this question was formatted in the same manner as all other headlines.
This attention check headline was then presented randomly within the 40 stimuli headlines.
Participants who answered this question incorrectly were not included in analyses. 
Table~\ref{tab:attrition} reports on the attention-check failure rates in the different groups.
In one group this rate is not available due to a Qualtrics data collection error.

Using $\chi^2$ tests, we compared the attrition rates between the control and experimental conditions (LLM-forced, LLM-optional, human fact check).
The LLM-forced condition within the sharing group was excluded from this analysis due to the data collection issues mentioned above.
This analysis revealed significant differences in attrition between the control and human fact check conditions in the sharing group (Bonferroni adjusted $P < 0.001$).
Despite matching experimental groups on key demographic characteristics and maintaining identical experimental protocols, these attrition differences may have resulted from different participant recruitment procedures employed by Qualtrics at different times.
No other evidence of differential attrition was found.

Table \ref{tab:screenout} lists the number of participants who were screened out for other reasons prior to being assigned to an experimental group.

\begin{table}
    \centering
    \begin{tabular}{lc}
        Screen-out type & Num. of participants \\
        \hline
        Did not consent & 551 \\
        Age ($<$ 18 y/o) & 58 \\
        Non-US resident & 37 \\
        Would not agree to give their best answers & 86 \\
        \hline
    \end{tabular}
    \caption{Screen-out attrition. These participants were never assigned to an experimental group.}
    \label{tab:screenout}
\end{table}

\section{Covariates}
\label{sec:si:covariates}

\subsection{Education}
\label{sec:si:edu}

% Operationalization
The participants' level of education was assessed by asking the following question: ``What is the highest level of education you have completed?''
The provided options, numbered by their corresponding recoded values for our regression analyses (see Section \nameref{sec:si:regress_confirm} for details), are listed below:
\begin{enumerate}
    \item Less than high school degree
    \item High school graduate (high school diploma or equivalent including GED)
    \item Some college but no degree
    \item Associate degree in college (2-year)
    \item Bachelor's degree in college (4-year)
    \item Master's degree
    \item Doctoral degree
    \item Professional degree (JD, MD)
\end{enumerate}

\subsection{Attitude towards AI}
\label{sec:si:atai}

% Operationalization
Participants' attitudes towards artificial intelligence (ATAI) were estimated with a four-item battery that is a slightly altered version of one developed by \citeauthor{sindermann_assessing_2021}\cite{sindermann_assessing_2021}.
Specifically, it included the following four items:
\begin{enumerate}
    \item I fear artificial intelligence
    \item I trust artificial intelligence
    \item Artificial intelligence will destroy humankind
    \item Artificial intelligence will benefit humankind
\end{enumerate}
Questions were answered with a seven-point Likert scale ranging from ``strongly disagree'' to ``strongly agree.''
Items 1 and 3 were reverse coded such that higher values on all items indicated greater trust in artificial intelligence.
For our regression analyses (see the \nameref{sec:si:regress_confirm} section for details), each participant's ATAI is calculated as the mean value of their responses to this battery.

\subsection{Headline congruence}
\label{sec:si:congruence}

A headline is considered ``congruent'' with a participant's partisan perspective if it is typically considered to be favorable towards the political party that they are affiliated with.
Headlines are either pro-Democrat or pro-Republican, based on the pretest described in the main text.
Thus, a congruent headline for a Democrat (Republican) would be one that is pro-Democrat (pro-Republican).
Conversely, an incongruent headline for a Democrat (Republican) would be one that is pro-Republican (pro-Democrat).
In regression analyses, we recode congruent headlines as 0 and incongruent headlines as 1.

We estimated partisanship by asking participants the following question: ``Generally speaking do you think of yourself as a Republican, a Democrat, an Independent, or what?''
Possible answers were ``Democrat,'' ``Republican,'' ``Independent,'' ``No Preference,'' ``Don't know,'' and ``Other'' (with a text box to fill if this option is selected).
If ``Democrat'' or ``Republican'' was not selected as their answer to this question they were then asked, ``Do you think of yourself as closer to the Republican or Democratic Party?''
Possible answers were ``Republican Party,'' ``Democratic Party,'' ``Don't know,'' and ``Neither.''
We consider participants who answered ``Democrat'' for the first question or ``Democratic Party'' for the second question as Democrats.
We consider as Republicans those who answered ``Republican'' for the first question or ``Republican Party'' for the second question.
In other words, those who lean towards Democrats (Republicans) were recoded as Democrats (Republicans) in our analysis.
All others are considered Independents.

\section{Regression analyses}
\label{sec:si:regress_confirm}

In this section, we aim to reproduce the results presented in the main text via regression analysis. 

In our preregistered research design, we proposed an exploratory analysis employing logistic cross-classified multilevel modeling (MLM) to predict item-level response accuracy.
This approach categorizes responses into two distinct groups: those considered desirable (i.e., believing or sharing true news, and not believing or sharing false news) and those deemed undesirable (believing or sharing false news, and not believing or sharing true news). 
However, we later noticed two problems with this approach that drove us to pursue a different exploratory analysis.
First, the MLM experienced issues converging properly, raising doubts about its reliability. 
Second, we recognized that this methodology does not allow us to separately analyze responses to true and false news, crucial to assessing discernment. 
Consequently, to align with the analysis in the main text, we opted to employ linear regression with clustered standard errors, focusing on participant responses as the dependent variable.
This deviation from our preregistered exploratory design brings our methodology in line with precedents set in the literature\cite{Pennycook2021Shifting, Pennycook2021Jan}.
The dependent variable in all models is the participant's response indicating belief or willingness to share a specific headline, coded as 1 for ``Yes'' and 0 for ``No.'' 
Age and Education level (as described in \nameref{sec:si:edu}) are included as covariates in all analyses.
Participants' age is scaled by a factor of 10 to facilitate the interpretation of the coefficients, allowing for a more straightforward understanding of the effects associated with each decade of age rather than each individual year.
Finally, for the sake of brevity, we sometimes use ``Optional'' and ``Forced'' interchangeably with ``LLM-optional'' and ``LLM-forced'' to describe these experimental conditions.

% Figure 1: Main effects
\subsection{Ineffectiveness of LLM fact checks}
\label{sec:si:ineffectiveness}

To examine the robustness of our findings related to the effects of different treatments on participants' average discernment, our model incorporates dummy variables for the experimental conditions and headline veracity, as well as a term accounting for their interactions\cite{Guay2023Mar, Pennycook2021Shifting}. 

Tables \ref{tab:ineffectiveness_B} and \ref{tab:ineffectiveness_S} display the results obtained from fitting our data to this model for the belief and share groups, respectively.
Of particular relevance to our primary findings, the interaction terms of interest, namely ``Condition(Forced):Veracity(True)'' and ``Condition(Optional):Veracity(True),'' are not significant predictors in either model.
On the other hand, the ``Condition(HumanFC):Veracity(True)'' interaction term is significant in both models.
As shown previously\cite{Guay2023Mar, Pennycook2021Shifting}, the coefficients of such interaction terms directly quantify the average change in discernment driven by each respective experimental treatment.
Therefore, this analysis reinforces our finding that exposure to LLM fact-checking information did not significantly affect average discernment, whereas human fact checks led to an increase in average discernment.

\begin{table}
\centering
\caption{Ineffectiveness of LLM Fact Checks Coefficients (Belief Group; $F= 1454.23$, $R^2=0.24$, $P < 0.001$)}
\begin{tabular}{lccccc}
\hline
Variable & Estimate & Std. Error & $t$ value & $P$ & Sig. \\
\hline
(Intercept)                     & 0.545 & 0.048 & 11.266 & $< 0.001$ & *** \\
Condition(Forced)                 & 0.035 & 0.030 & 1.160  & 0.246 &  \\
Condition(Optional)               & 0.005 & 0.030 & 0.183  & 0.855 &  \\
Condition(HumanFC)               & -0.011 & 0.029 & -0.379  & 0.705 &  \\
Veracity(True)                    & 0.393 & 0.027 & 14.776 & $< 0.001$ & *** \\
Age                             & -0.006 & 0.001 & -7.250 & $< 0.001$ & *** \\
Education                    & 0.009 & 0.005 & 1.909  & 0.056 & $\cdot$ \\
Condition(Forced):Veracity(True)    & -0.045 & 0.036 & -1.259 & 0.208 &  \\
Condition(Optional):Veracity(True)  & -0.003 & 0.033 & -0.083 & 0.934 &  \\
Condition(HumanFC):Veracity(True)  & 0.181 & 0.032 & 5.696 & $< 0.001$ & *** \\
\hline
\multicolumn{5}{l}{Significance codes: *** $P< 0.001$, ** $P< 0.01$, * $P< 0.05$, $\cdot$ $P< 0.1$} \\
\hline
\end{tabular}
\label{tab:ineffectiveness_B}
\end{table}

\begin{table}
\centering
\caption{Ineffectiveness of LLM Fact Checks Coefficients (Share Group; $F= 599.84$, $R^2=0.11$, $P < 0.001$)}
\begin{tabular}{lccccc}
\hline
Variable & Estimate & Std. Error & $t$ value & $P$ & Sig. \\
\hline
(Intercept)                     & 0.826 & 0.041 & 20.392 & $< 0.001$  & *** \\
Condition(Forced)                 & 0.049 & 0.029 & 1.698  & 0.090  & $\cdot$ \\
Condition(Optional)               & -0.008 & 0.031 & -0.262  & 0.794 & \\
Condition(HumanFC)               & 0.020 & 0.029 & 0.682  & 0.495 & \\
Veracity(True)                    & 0.085 & 0.018 & 4.710 & $< 0.001$  & *** \\
Age                             & -0.008 & 0.001 & -13.980 & $< 0.001$  & *** \\
Education                    & -0.016 & 0.007 & -2.389  & 0.017  & * \\
Condition(Forced):Veracity(True)    & -0.004 & 0.017 & -0.262 & 0.793 & \\
Condition(Optional):Veracity(True)  & -0.007 & 0.019 & -0.351 & 0.725 & \\
Condition(HumanFC):Veracity(True)  & 0.090 & 0.021 & 4.357 & $< 0.001$ & *** \\
\hline
\multicolumn{5}{l}{Significance codes: *** $P<$ 0.001, ** $P<$ 0.01, * $P<$ 0.05, $\cdot$ $P<$ 0.1} \\
\hline
\end{tabular}
\label{tab:ineffectiveness_S}
\end{table}

% Figure 2: Accounting for ChatGPT Accuracy
\subsection{Accounting for LLM accuracy}
\label{sec:si:accnt_for_ai}

To incorporate the accuracy of the LLM fact checks, we include an interaction between experimental condition and fact-checking (FC) scenario (True/False $\times$ Correct/Incorrect/Unsure).
These variables capture the five scenarios found in our data.
We remind the reader that no false headlines were judged to be true in our data.
To match the analysis from the main text and highlight the potential effects of LLM fact-checking information, we focus on the forced and control conditions.

Tables \ref{tab:AI_judge_B} and \ref{tab:AI_judge_S} present the results of fitting our data to this model for the belief and share groups, respectively.\footnote{Note that a few standard errors cannot be computed leading to `NaN' values.
This occurs only for some terms related to the ``False $\times$ unsure'' scenario, likely due to the low number of headlines (two) in that scenario.} 
Some significant interaction terms are observed for specific FC scenarios.
This tells us that the Condition $\times$ FC Scenario relationship is significantly different in these scenarios  relative to the ``reference group'' FC Scenario (False $\times$ false)---not shown in the table. 
However, this is not the appropriate reference group: we wish to specifically test the significance of this interaction within each fact-checking scenario. 
To this end, we conduct post-hoc comparisons similar to those presented within the main text for each group.
Utilizing the fitted models, estimated marginal mean values for the Control and Forced groups are calculated and compared in each headline scenario, adjusting $P$ values with Bonferroni's method.
The results of these post-hoc comparisons for the belief and share groups are shown in Tables \ref{tab:AI_judge_posthoc_B} and \ref{tab:AI_judge_posthoc_S}, respectively.
We observe significant mean differences for fact-checking scenarios in both groups that are consistent with those presented in the main text.
However, we also observe significant mean differences suggesting that the LLM fact-checking information is harmful in additional fact checking scenarios within both the belief (False $\times$ false) and sharing (False $\times$ false, False $\times$ unsure, and True $\times$ unsure) groups.
To remain conservative in our analyses, we do not report these results in the main text as they are inconsistent with the corresponding analysis based on mean differences calculated from the raw data.

\begin{table}
\centering
\caption{Account for LLM Accuracy Coefficients (Belief Group; $F= 428.65$, $R^2=0.19$, $P < 0.001$)}
\small
\begin{tabular}{lccccc}
\hline
Variable & Estimate & Std. Error & $t$ value & $P$ & Sig. \\
\hline
(Intercept)                                       & 0.556 & 0.055 & 10.079 & $< 0.001$ & *** \\
Cond.(Forced)                                 & 0.029 & 0.030 & 0.960  & 0.337 &  \\
FC Scen.(False $\times$ unsure)                       & 0.025 & 0.009 & 2.782  & 0.005 & ** \\
FC Scen.(True $\times$ false)                         & 0.360 & 0.034 & 10.574 & $< 0.001$ & *** \\
FC Scen.(True $\times$ true)                          & 0.448 & 0.026 & 17.161 & $< 0.001$ & *** \\
FC Scen.(True $\times$ unsure)                        & 0.394 & 0.030 & 13.198 & $< 0.001$ & *** \\
Age                                              & -0.007& 0.001 & -7.334 & $< 0.001$ & *** \\
Education                                         & 0.017 & 0.008 & 2.269  & 0.023 & * \\
Cond.(Forced):FC Scen.(False $\times$ unsure)     & 0.070 & 0.024 & 2.864  & 0.004 & ** \\
Cond.(Forced):FC Scen.(True $\times$ false)       & -0.149& 0.040 & -3.749 & $< 0.001$ & *** \\
Cond.(Forced):FC Scen.(True $\times$ true)        & 0.020 & 0.034 & 0.572  & 0.567 &  \\
Cond.(Forced):FC Scen.(True $\times$ unsure)      & -0.017& 0.040 & -0.434 & 0.664 &  \\
\hline
\multicolumn{6}{l}{Significance codes: *** $P< 0.001$, ** $P< 0.01$, * $P< 0.05$, $\cdot$ $P< 0.1$} \\
\hline
\end{tabular}
\label{tab:AI_judge_B}
\end{table}

\begin{table}
\centering
\caption{Account for LLM Accuracy Coefficients (Share Group; $F= 259.48$, $R^2=0.12$, $P < 0.001$)}
\small
\begin{tabular}{lccccc}
\hline
Variable & Estimate & Std. Error & $t$ value & $P$ & Sig. \\
\hline
(Intercept)                                       & 0.858 & 0.050 & 17.262 & $< 0.001$ & *** \\
Cond.(Forced)                                 & 0.047 & 0.029 & 1.613  & 0.107 &  \\
FC Scen.(False $\times$ unsure)                       & -0.010& NaN   &     &    &  \\
FC Scen.(True $\times$ false)                         & 0.063 & 0.017 & 3.721  & $< 0.001$ & *** \\
FC Scen.(True $\times$ true)                          & 0.049 & 0.033 & 1.486  & 0.137 &  \\
FC Scen.(True $\times$ unsure)                        & 0.098 & 0.020 & 4.904  & $< 0.001$ & *** \\
Age                                              & -0.008& 0.001 & -8.226 & $< 0.001$ & *** \\
Education                                         & -0.037& 0.011 & -3.481 & $< 0.001$ & *** \\
Cond.(Forced):FC Scen.(False $\times$ unsure)     & 0.031 & NaN   &     &    &  \\
Cond.(Forced):FC Scen.(True $\times$ false)       & -0.032& 0.006 & -5.068 & $< 0.001$ & *** \\
Cond.(Forced):FC Scen.(True $\times$ true)        & 0.044 & 0.016 & 2.717  & 0.007 & ** \\
Cond.(Forced):FC Scen.(True $\times$ unsure)      & -0.002& 0.018 & -0.133 & 0.894 &  \\
\hline
\multicolumn{6}{l}{Significance codes: *** $P< 0.001$, ** $P< 0.01$, * $P< 0.05$, $\cdot$ $P< 0.1$} \\
\hline
\end{tabular}
\label{tab:AI_judge_S}
\end{table}

\begin{table}
\centering
\caption{Post-hoc analysis of mean belief in headlines, accounting for LLM accuracy}
\vspace{.1em}
\begin{tabular}{lcccccc}
\hline
Headline Scenario & Forced $-$ Control & Std. Err. & df & $t$ ratio & Adj. $P^{\dagger}$ & Sig. \\
\hline
True $\times$ False & -0.120 & 0.020 & 19508 & -5.915 & $<$ 0.001 & *** \\
True $\times$ Unsure & 0.012 & 0.011 & 19508 & 1.021 & 1.000 & \\
True $\times$ True & 0.048 & 0.023 & 19508 & 2.063 & 0.196 & \\
False $\times$ False & 0.029 & 0.010 & 19508 & 2.995 & 0.014 & * \\
False $\times$ Unsure & 0.098 & 0.029 & 19508 & 3.427 & 0.003 & ** \\
\hline
\multicolumn{7}{l}{Significance codes: *** $P< 0.001$, ** $P< 0.01$, * $P< 0.05$, $\cdot$ $P< 0.1$} \\
\multicolumn{7}{l}{$\dagger$ Bonferroni's method comparing a family of 5 estimates} \\
\hline
\end{tabular}
\label{tab:AI_judge_posthoc_B}
\end{table}

\begin{table}
\centering
\caption{Post-hoc analysis of mean intent to share headlines, accounting for LLM accuracy}
\vspace{.1em}
\begin{tabular}{lcccccc}
\hline
Headline Scenario & Forced $-$ Control & Std. Err. & df & $t$ ratio & Adj. $P^{\dagger}$ & Sig. \\
\hline
True $\times$ False & 0.015 & 0.020 & 21428 & 0.756 & 1.000 &  \\
True $\times$ Unsure & 0.045 & 0.011 & 21428 & 3.992 & $<$ 0.001 & *** \\
True $\times$ True & 0.091 & 0.023 & 21428 & 3.922 & $<$ 0.001 & *** \\
False $\times$ False & 0.047 & 0.010 & 21428 & 4.943 & $<$ 0.001 & *** \\
False $\times$ Unsure & 0.078 & 0.028 & 21428 & 2.739 & 0.0310 & *  \\
\hline
\multicolumn{7}{l}{Significance codes: *** $P< 0.001$, ** $P< 0.01$, * $P< 0.05$, $\cdot$ $P< 0.1$} \\
\multicolumn{7}{l}{$\dagger$ Bonferroni's method comparing a family of 5 estimates} \\
\hline
\end{tabular}
\label{tab:AI_judge_posthoc_S}
\end{table}

% Figure 3: Optional
\subsection{Opt in versus opt out}

To provide support for our analysis related to the LLM-optional condition, we now incorporate an interaction between whether a participant in this condition chose to see LLM fact-checking information (opt in) or not (opt out) and the fact checking scenario.

Tables \ref{tab:opt_in_reg_B} and \ref{tab:opt_in_reg_S} present the results of fitting our data for the belief and share groups, respectively.
To confirm the results presented in the main text, we utilize the models to perform the same comparisons of estimated marginal means.
These post-hoc comparisons further support our findings, and are shown for the belief and share groups in Tables \ref{tab:opt_in_posthoc_B} and \ref{tab:opt_in_posthoc_S}, respectively.

\begin{table}
\centering
\caption{Opt In versus Opt Out Coefficients (Belief Group; $F= 286.42$, $R^2=0.23$, $P < 0.001$)}
\begin{tabular}{lccccc}
\hline
Variable        & Estimate & Std. Error & $t$ value & $P$ & Sig. \\
\hline
(Intercept)                & 0.628 & 0.061 & 10.251 & $< 0.001$ & *** \\
Option(opt out)           & -0.243 & 0.039 & -6.294 & $< 0.001$ & *** \\
FC Scen.(False $\times$ unsure)            & 0.082   & 0.069   &  1.196  &  0.232  &  \\
FC Scen.(True $\times$ false)              & 0.052  & 0.019 & 2.704  & 0.007  & ** \\
FC Scen.(True $\times$ true)               & 0.409 & 0.031 & 13.082 & $< 0.001$ & *** \\
FC Scen.(True $\times$ unsure)             & 0.301 & 0.037 & 8.119  & $< 0.001$ & *** \\
Age                        & -0.004 & 0.001 & -3.825 & $< 0.001$   & *** \\
Education                  & -0.008 & 0.009 & -0.841 & 0.400 & \\
Option(opt out):FC Scen.(False $\times$ unsure) & $<$ 0.000  & 0.072 & 0.002 & 0.999  &  \\
Option(opt out):FC Scen.(True $\times$ false) & 0.442 & 0.043 & 10.380 & $< 0.001$  & *** \\
Option(opt out):FC Scen.(True $\times$ true) & 0.216 & 0.046 & 4.673 & $< 0.001$  & *** \\
Option(opt out):FC Scen.(True $\times$ unsure) & 0.223 & 0.053 & 4.212 & $< 0.001$  & *** \\
\hline
\multicolumn{6}{l}{Significance codes: *** $P< 0.001$, ** $P< 0.01$, * $P< 0.05$, $\cdot$ $P< 0.1$} \\
\hline
\end{tabular}
\label{tab:opt_in_reg_B}
\end{table}

\begin{table}
\centering
\caption{Opt In versus Opt Out Coefficients (Share Group; $F= 217.95$, $R^2=0.19$, $P < 0.001$)}
\begin{tabular}{lccccc}
\hline
Variable                   & Estimate & Std. Error & $t$ value & $P$                     & Sig. \\
\hline
(Intercept)                & 0.805 & 0.067 & 12.066 & $< 0.001$ & *** \\
Option(opt out)           & -0.305 & 0.039 & -7.873 & $< 0.001$ & *** \\
FC Scen.(False $\times$ unsure)            & 0.017 & NaN &  &  &  \\
FC Scen.(True $\times$ false)              & -0.016 & 0.005 & -2.905 & 0.004 & ** \\
FC Scen.(True $\times$ true)               & 0.104 & 0.021 & 5.022 & $< 0.001$ & *** \\
FC Scen.(True $\times$ unsure)             & 0.072 & 0.018 & 4.128  & $< 0.001$ & *** \\
Age                        & -0.007 & 0.001 & -5.258 & $< 0.001$  & *** \\
Education                  & 0.001 & 0.014 & 0.053 & 0.958 & \\
Option(opt out):FC Scen.(False $\times$ unsure) & -0.007 & 0.013 & -0.543 & 0.587 &  \\
Option(opt out):FC Scen.(True $\times$ false) & 0.065 & 0.021 & 3.104 & 0.002 & ** \\
Option(opt out):FC Scen.(True $\times$ true) & -0.039 & 0.029 & -1.348 & 0.178 & \\
Option(opt out):FC Scen.(True $\times$ unsure) & 0.016 & 0.026 & 0.598 & 0.550 &  \\
\hline
\multicolumn{6}{l}{Significance codes: *** $P< 0.001$, ** $P< 0.01$, * $P< 0.05$, $\cdot$ $P< 0.1$} \\
\hline
\end{tabular}
\label{tab:opt_in_reg_S}
\end{table}

\begin{table}
\centering
\caption{Post-hoc analysis of mean belief in headlines in the Optional condition}
\vspace{.1em}
\begin{tabular}{lcccccc}
\hline
Headline scenario & Opt in $-$ Opt out & Std. Error & df & $t$ ratio & Adj. $P^{\dagger}$ & Sig. \\
\hline
True $\times$ False & -0.200 & 0.027 & 10428 & -7.288 & $<$ 0.001 & *** \\
True $\times$ Unsure & 0.020 & 0.015 & 10428 & 1.277 & 1.000 &  \\
True $\times$ True & 0.027 & 0.032 & 10428 & 0.837 & 1.000 &  \\
False $\times$ False & 0.243 & 0.013 & 10428 & 18.414 & $<$ 0.001 & *** \\
False $\times$ Unsure & 0.243 & 0.039 & 10428 & 6.226 & $<$ 0.001 & ***  \\
\hline 
\multicolumn{7}{l}{Significance codes: *** $P< 0.001$, ** $P< 0.01$, * $P< 0.05$, $\cdot$ $P< 0.1$} \\
\multicolumn{7}{l}{$\dagger$ Bonferroni's method comparing a family of 5 estimates} \\
\hline
\end{tabular}
\label{tab:opt_in_posthoc_B}
\end{table}

\begin{table}
\centering
\caption{Post-hoc analysis of mean intent to share headlines in the Optional condition}
\vspace{.1em}
\begin{tabular}{lcccccc}
\hline
Headline scenario & Opt in $-$ Opt out & Std. Error & df & $t$ ratio & Adj. $P^{\dagger}$ & Sig.\\
\hline
True $\times$ False & 0.239 & 0.028 & 10508 & 8.501 & $<$ 0.001 & *** \\
True $\times$ Unsure & 0.289 & 0.016 & 10508 & 18.347 & $<$ 0.001 & *** \\
True $\times$ True & 0.344 & 0.032 & 10508 & 10.630 & $<$ 0.001 & *** \\
False $\times$ False & 0.305 & 0.013 & 10508 & 22.876 & $<$ 0.001 & *** \\
False $\times$ Unsure & 0.312 & 0.039 & 10508 & 7.917 & $<$ 0.001 & ***  \\
\hline
\multicolumn{7}{l}{Significance codes: *** $P< 0.001$, ** $P< 0.01$, * $P< 0.05$, $\cdot$ $P< 0.1$} \\
\multicolumn{7}{l}{$\dagger$ Bonferroni's method comparing a family of 5 estimates} \\
\hline
\end{tabular}
\label{tab:opt_in_posthoc_S}
\end{table}

\section{Interaction analyses}
\label{sec:si:interactions}

In this section, we explore the potential moderation effects of two factors on our main results: attitude towards AI (ATAI) and headline congruence (see the \nameref{sec:si:covariates} section for details).
We employ linear regression with robust standard errors clustered on participant and headline for each key finding discussed in the main text.
Each analysis covered in the \nameref{sec:si:regress_confirm} section is revisited to incorporate these variables and create three-way interactions.
Covariates that were included in the earlier analyses (Age and Education level) are included again.
The belief and sharing group data are modeled separately.

\subsection{Attitude towards AI}
\label{sec:si:atai_interaction}

\begin{figure}
    \centering
    \includegraphics[width=\linewidth]{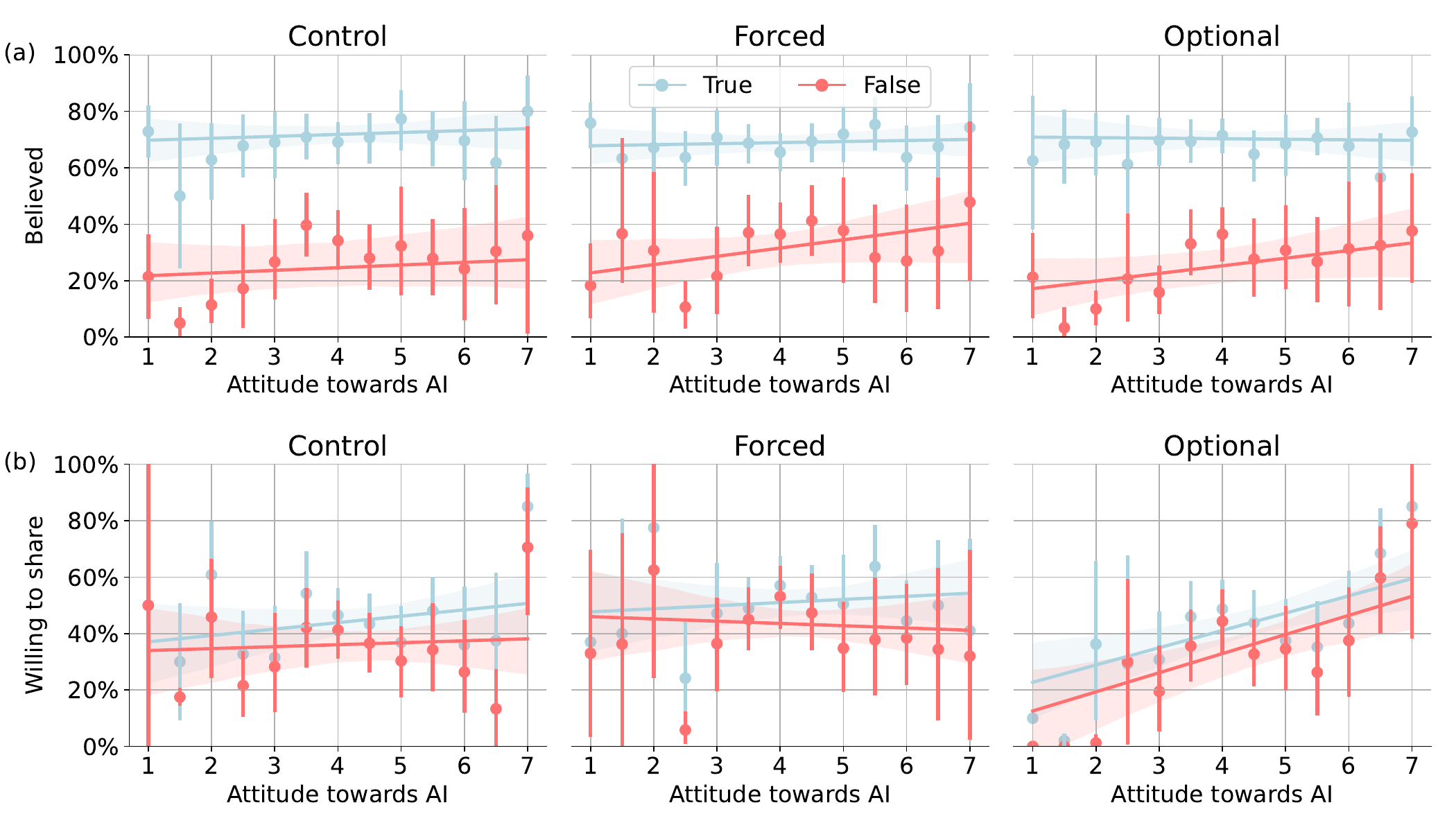}
    \caption{
    Relationship between participants' ATAI and their (a) belief in and (b) intent to share headlines for all conditions.
    Responses are binned with a size of .5 and centers at $[1, 1.5, 2, \ldots, 7]$, which does not affect the regression fit.
    Headline veracity is indicated by the color of the data.
    }
    \label{fig:si_main_effects_atai}
\end{figure}

% Figure 1: Main effects
% % % % % % % % % % % % % % % % % % 
We begin by examining whether LLM fact-checking information remains ineffective amongst individuals with varying levels of ATAI.
Therefore, we test the three-way interaction between Condition, Veracity, and ATAI (Condition $\times$ Veracity $\times$ ATAI).
The human fact checking group is excluded from this analysis, as there is no reason to believe that participants' interactions with human-generated fact checks would vary based on their attitudes toward artificial intelligence.
Figure~\ref{fig:si_main_effects_atai} presents the relationship between participants' ATAI and their belief in (panel a) and intent to share (panel b) true versus false headlines across all conditions.
The results of our modeling analysis indicate that there is no significant three-way interaction between ATAI and either belief in (Table \ref{tab:ineffective_atai_B}) or intent to share (Table \ref{tab:ineffective_atai_S}) headlines for all conditions.

\begin{table}
\centering
\caption{Ineffectiveness of LLM Fact Checks Coefficients (ATAI interaction; Belief Group; $F= 526.74$, $R^2=0.19$, $P < 0.001$)}
\small
\begin{tabular}{lcccccc}
\hline
Variable & Estimate & Std. Error & $t$ value & $P$ & Sig. \\
\hline
(Intercept) & 0.559 & 0.082 & 6.793 & $< 0.001$ & *** \\
Condition(Forced) & -0.012 & 0.090 & -0.136 & 0.892 &  \\
Condition(Optional) & -0.052 & 0.087 & -0.596 & 0.551 &  \\
Veracity(True) & 0.424 & 0.067 & 6.303 & $< 0.001$ & *** \\
ATAI & 0.001 & 0.015 & 0.089 & 0.929 &  \\
Age & -0.006 & 0.001 & -7.294 & $< 0.001$ & *** \\
Education & 0.008 & 0.006 & 1.363 & 0.173 &  \\
Condition(Forced):Veracity(True) & 0.027 & 0.085 & 0.315 & 0.753 &  \\
Condition(Optional):Veracity(True) & 0.116 & 0.078 & 1.481 & 0.139 &  \\
Condition(Forced):ATAI & 0.011 & 0.022 & 0.516 & 0.606 &  \\
Condition(Optional):ATAI & 0.013 & 0.021 & 0.633 & 0.527 &  \\
Veracity(True):ATAI & -0.008 & 0.015 & -0.499 & 0.618 &  \\
Condition(Forced):Veracity(True):ATAI & -0.017 & 0.020 & -0.815 & 0.415 &  \\
Condition(Optional):Veracity(True):ATAI & -0.026 & 0.019 & -1.408 & 0.159 &  \\
\hline
\multicolumn{6}{l}{Significance codes: *** $P< 0.001$, ** $P< 0.01$, * $P< 0.05$, $\cdot$ $P< 0.1$} \\
\hline
\end{tabular}
\label{tab:ineffective_atai_B}
\end{table}

\begin{table}
\centering
\caption{Ineffectiveness of LLM Fact Checks Coefficients (ATAI interaction; Share Group; $F= 318.67$, $R^2=0.11$, $P < 0.001$)}
\small
\begin{tabular}{lcccccc}
\hline
Variable & Estimate & Std. Error & $t$ value & $P$ & Sig. \\
\hline
(Intercept)         & 0.853 & 0.093 & 9.180 & $< 0.001$ & *** \\
Condition(Forced)   & 0.089 & 0.119 & 0.752 & 0.452 &  \\
Condition(Optional) & -0.268 & 0.118 & -2.265 & 0.023 & * \\
Veracity(True)      & 0.015 & 0.055 & 0.276 & 0.782 &  \\
ATAI                & -0.005 & 0.018 & -0.279 & 0.781 &  \\
Age                 & -0.008 & 0.001 & -10.190 & $< 0.001$ & *** \\
Education           & -0.024 & 0.009 & -2.712 & 0.007 & ** \\
Condition(Forced):Veracity(True)        & -0.017 & 0.066 & -0.261 & 0.794 &  \\
Condition(Optional):Veracity(True)      & 0.087 & 0.066 & 1.322 & 0.186 &  \\
Condition(Forced):ATAI                  & -0.009 & 0.026 & -0.349 & 0.727 &  \\
Condition(Optional):ATAI                & 0.059 & 0.026 & 2.243 & 0.025 & * \\
Veracity(True):ATAI                     & 0.016 & 0.012 & 1.330 & 0.183 &  \\
Condition(Forced):Veracity(True):ATAI   & 0.003 & 0.016 & 0.211 & 0.833 &  \\
Condition(Optional):Veracity(True):ATAI & -0.021 & 0.014 & -1.478 & 0.140 &  \\
\hline
\multicolumn{6}{l}{Significance codes: *** $P< 0.001$, ** $P< 0.01$, * $P< 0.05$, $\cdot$ $P< 0.1$} \\
\hline
\end{tabular}
\label{tab:ineffective_atai_S}
\end{table}

\begin{figure}
    \centering
    \includegraphics[width=\linewidth]{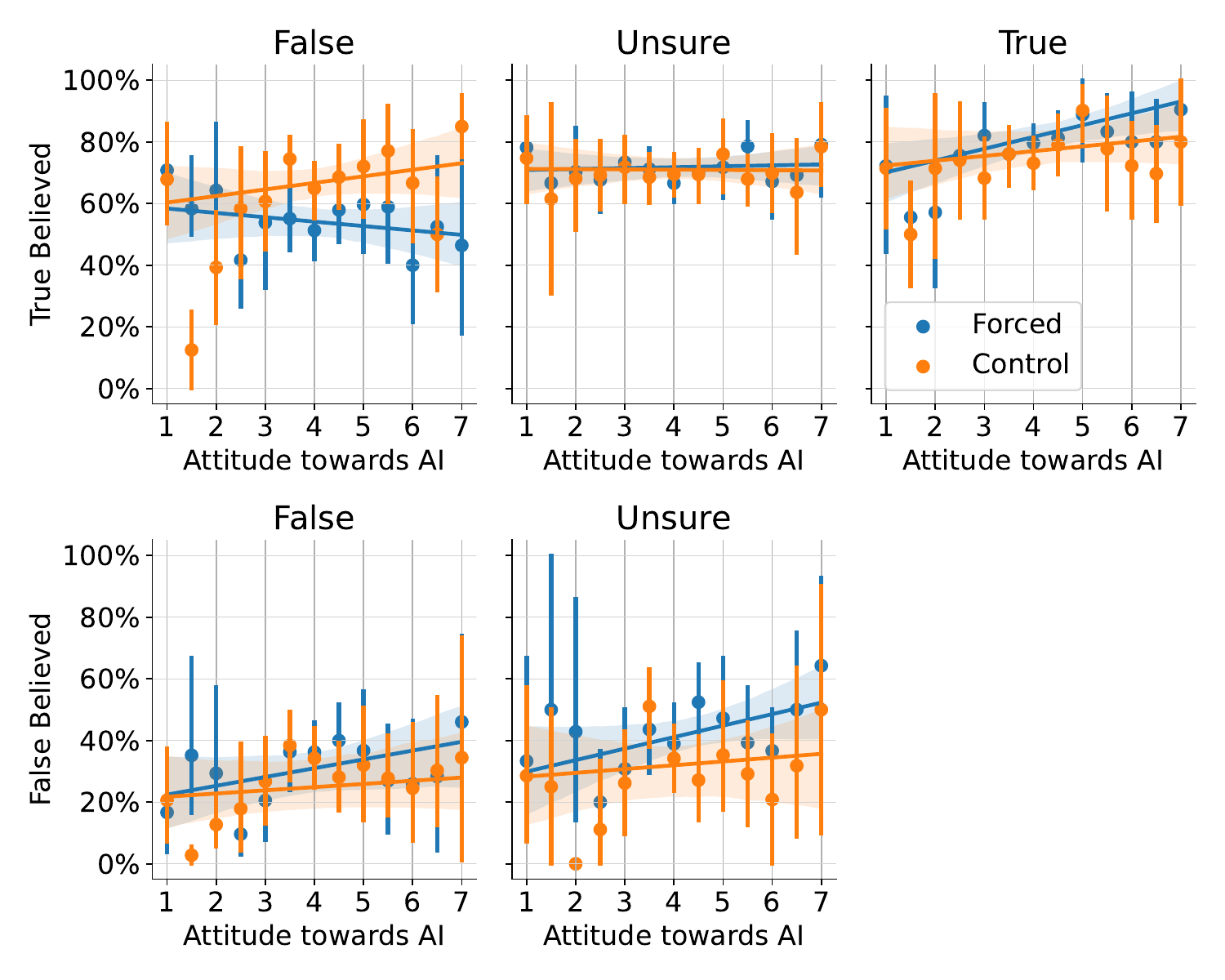}
    \caption{
    Relationship between belief in headlines and ATAI for the control and forced conditions.
    Panels are representative of participants' responses to different types of headlines.
    The top and bottom panel rows represent true and false headlines, respectively.
    The left, center, and right panel columns represent ChatGPT's judgment of those headlines as false, unsure, and true, respectively.
    The bottom right panel is excluded as this type of headline (false headlines judged by ChatGPT to be true) does not exist in our data.
    Responses are binned with a size of .5 and centers at $[1, 1.5, 2, \ldots, 7]$, which does not affect the regression fit.
    }
    \label{fig:si_5way_atai_belief}
\end{figure}

\begin{figure}
    \centering
    \includegraphics[width=\linewidth]{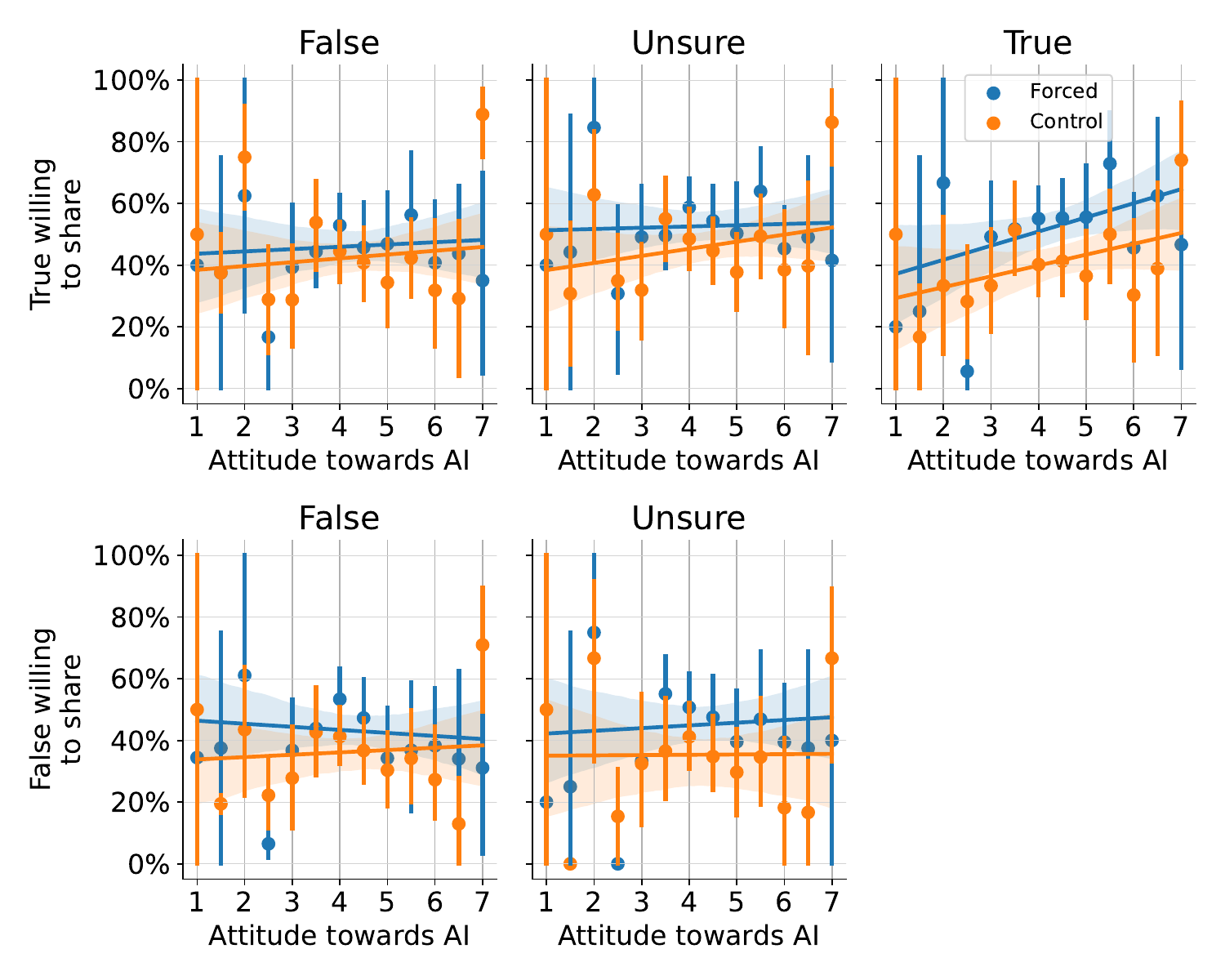}
    \caption{
    Relationship between headline sharing intent and ATAI for the control and forced conditions.
    Panels are representative of participants' responses to different types of headlines.
    The top and bottom panel rows represent true and false headlines, respectively.
    The left, center, and right panel columns represent ChatGPT's judgment of those headlines as false, unsure, and true, respectively.
    The bottom right panel is excluded as this type of headline (false headline judged by ChatGPT to be true) does not exist in our data.
    Responses are binned with a size of .5 and centers at $[1, 1.5, 2, \ldots, 7]$, which does not affect the regression fit.
    }
    \label{fig:si_5way_atai_share}
\end{figure}

% Figure 2: Accounting for ChatGPT accuracy
% % % % % % % % % % % % % % % % % % % % % % % % % % 
Next, we examine whether the effects of fact-checking scenarios stay consistent among people with different ATAI (the three-way interaction Condition $\times$ FC Scenario $\times$ ATAI).
Again, we focus on the forced and control conditions and exclude data for the optional participants when fitting each model.
Figure~\ref{fig:si_5way_atai_belief} illustrates the relationship between belief in headlines and ATAI for the control and forced conditions in each fact-checking scenario.
The same relationship is presented with respect to sharing intent in Figure~\ref{fig:si_5way_atai_share}.
The result of fitting the belief and share group models are found in Tables \ref{tab:fiveway_atai_B} and \ref{tab:fiveway_atai_S}, respectively.
These models are then utilized for post-hoc comparisons similar to those presented within the main text for each group.
However, to test for an ATAI interaction, this analysis compares the slopes of the Control and Forced groups predicted response line, given different values of ATAI.
These results, shown in Tables \ref{tab:fiveway_atai_B_posthoc} and \ref{tab:fiveway_atai_S_posthoc} for the belief and share groups, respectively, validate our results by illustrating that participants did not respond differently depending on ATAI.

\begin{table}
\centering
\caption{Account for LLM Accuracy Coefficients (ATAI interaction, Belief Group; $F= 225.85$, $R^2=0.20$, $P < 0.001$)}
\small
\begin{tabular}{lcccccc}
\hline
Variable & Estimate & Std. Error & $t$ value & $P$ & Sig. \\
\hline
(Intercept) & 0.556 & 0.087 & 6.423 & $< 0.001$ & *** \\
Cond.(Forced) & -0.012 & 0.091 & -0.135 & 0.893 &  \\
FC Scen.(False $\times$ unsure) & 0.041 & NaN &  &  &  \\
FC Scen.(True $\times$ false) & 0.335 & 0.095 & 3.516 & $< 0.001$ & *** \\
FC Scen.(True $\times$ true) & 0.436 & 0.102 & 4.298 & $< 0.001$ & *** \\
FC Scen.(True $\times$ unsure) & 0.455 & 0.069 & 6.568 & $< 0.001$ & *** \\
ATAI & $<$ 0.001 & 0.015 & 0.013 & 0.990 &  \\
Age & -0.007 & 0.001 & -7.277 & $< 0.001$ & *** \\
Education & 0.017 & 0.008 & 2.278 & 0.023 & * \\
Cond.(Forced):FC Scen.(False $\times$ unsure) & 0.012 & 0.115 & 0.101 & 0.920 &  \\
Cond.(Forced):FC Scen.(True $\times$ false) & 0.054 & 0.102 & 0.535 & 0.593 &  \\
Cond.(Forced):FC Scen.(True $\times$ true) & 0.007 & 0.097 & 0.072 & 0.943 &  \\
Cond.(Forced):FC Scen.(True $\times$ unsure) & 0.024 & 0.091 & 0.268 & 0.788 &  \\
Cond.(Forced):ATAI & 0.010 & 0.022 & 0.443 & 0.658 &  \\
FC Scen.(False $\times$ unsure):ATAI & -0.004 & NaN &  &  &  \\
FC Scen.(True $\times$ false):ATAI & 0.006 & 0.018 & 0.345 & 0.730 &  \\
FC Scen.(True $\times$ true):ATAI & 0.003 & 0.023 & 0.119 & 0.905 &  \\
FC Scen.(True $\times$ unsure):ATAI & -0.015 & 0.016 & -0.942 & 0.346 &  \\
Cond.(Forced):FC Scen.(False $\times$ unsure):ATAI & 0.014 & 0.033 & 0.418 & 0.676 &  \\
Cond.(Forced):FC Scen.(True $\times$ false):ATAI & -0.048 & 0.022 & -2.210 & 0.027 & * \\
Cond.(Forced):FC Scen.(True $\times$ true):ATAI & 0.003 & 0.023 & 0.130 & 0.897 &  \\
Cond.(Forced):FC Scen.(True $\times$ unsure):ATAI & -0.009 & 0.022 & -0.435 & 0.664 &  \\
\hline
\multicolumn{6}{l}{Significance codes: *** $P< 0.001$, ** $P< 0.01$, * $P< 0.05$, $\cdot$ $P< 0.1$} \\
\hline
\end{tabular}
\label{tab:fiveway_atai_B}
\end{table}

\begin{table}
\centering
\caption{Account for LLM Accuracy Coefficients (ATAI interaction, Share Group; $F= 137.04$, $R^2=0.12$, $P < 0.001$)}
\begin{tabular}{lccccc}
\hline
Variables & Estimate & Std. Error & $t$ value & $P$ & Sig. \\
\hline
(Intercept) & 0.862 & 0.096 & 8.998 & $< 0.001$ & *** \\
Cond.(Forced) & 0.103 & 0.118 & 0.875 & 0.382 & \\
FC Scen.(False $\times$ unsure) & 0.019 & NaN &  &  & \\
FC Scen.(True $\times$ false) & 0.042 & 0.049 & 0.847 & 0.397 & \\
FC Scen.(True $\times$ true) & -0.073 & 0.047 & -1.558 & 0.119 & \\
FC Scen.(True $\times$ unsure) & 0.030 & 0.053 & 0.567 & 0.571 & \\
ATAI & -0.001 & 0.018 & -0.047 & 0.962 & \\
Age & -0.007 & 0.001 & -8.023 & $< 0.001$ & *** \\
Education & -0.037 & 0.011 & -3.496 & $< 0.001$ & *** \\
Cond.(Forced):FC Scen.(False $\times$ unsure) & -0.079 & 0.052 & -1.526 & 0.127 & \\
Cond.(Forced):FC Scen.(True $\times$ false) & -0.086 & 0.037 & -2.316 & 0.021 & * \\
Cond.(Forced):FC Scen.(True $\times$ true) & -0.076 & 0.088 & -0.865 & 0.387 & \\
Cond.(Forced):FC Scen.(True $\times$ unsure) & 0.005 & 0.067 & 0.074 & 0.941 & \\
Cond.(Forced):ATAI & -0.013 & 0.026 & -0.496 & 0.620 & \\
FC Scen.(False $\times$ unsure):ATAI & -0.007 & NaN &  &  & \\
FC Scen.(True $\times$ false):ATAI & 0.005 & 0.008 & 0.576 & 0.565 & \\
FC Scen.(True $\times$ true):ATAI & 0.028 & 0.013 & 2.069 & 0.039 & * \\
FC Scen.(True $\times$ unsure):ATAI & 0.015 & 0.012 & 1.331 & 0.183 & \\
Cond.(Forced):FC Scen.(False $\times$ unsure):ATAI & 0.025 & 0.017 & 1.482 & 0.138 & \\
Cond.(Forced):FC Scen.(True $\times$ false):ATAI & 0.013 & 0.006 & 2.007 & 0.045 & * \\
Cond.(Forced):FC Scen.(True $\times$ true):ATAI & 0.028 & 0.020 & 1.385 & 0.166 & \\
Cond.(Forced):FC Scen.(True $\times$ unsure):ATAI & -0.001 & 0.016 & -0.088 & 0.930 & \\
\hline
\multicolumn{6}{l}{Significance codes: *** $P< 0.001$, ** $P< 0.01$, * $P< 0.05$, $\cdot$ $P< 0.1$} \\
\hline
\end{tabular}
\label{tab:fiveway_atai_S}
\end{table}

\begin{table}
\centering
\caption{Post-hoc comparison of belief slopes fit to different condition and ATAI values, accounting for LLM accuracy}
\vspace{.1em}
\begin{tabular}{lcccccc}
\hline
Headline Scenario & Forced $-$ Control & Std. Err. & df & $t$ ratio & Adj. $P^{\dagger}$ & Sig. \\
\hline
False $\times$ false & 0.010 & 0.008 & 19498 & 1.234 & 1.000& \\
False $\times$ unsure & 0.023 & 0.023 & 19498 & 1.000 & 1.000& \\
True $\times$ false & -0.038 & 0.016 & 19498 & -2.313 & 0.104 & \\
True $\times$ unsure & $<$ 0.001 & 0.009 & 19498 & 0.013 & 1.000 & \\
True $\times$ true & 0.013 & 0.019 & 19498 & 0.658 & 1.000& \\
\hline
\multicolumn{7}{l}{Significance codes: *** $P< 0.001$, ** $P< 0.01$, * $P< 0.05$, $\cdot$ $P< 0.1$} \\
\multicolumn{7}{l}{$\dagger$ Bonferroni's method comparing a family of 5 estimates} \\
\hline
\end{tabular}
\label{tab:fiveway_atai_B_posthoc}
\end{table}

\begin{table}
\centering
\caption{Post-hoc comparison of sharing intent slopes fit to different condition and ATAI values, accounting for LLM accuracy}
\vspace{.1em}
\begin{tabular}{lcccccc}
\hline
Headline Scenario & Forced $-$ Control & Std. Err. & df & $t$ ratio & Adj. $P^{\dagger}$ & Sig. \\
\hline
False $\times$ false  & -0.013 & 0.008 & 21418 & -1.537 & 0.622 & \\
False $\times$ unsure  & 0.012 & 0.025 & 21418 & 0.495 & 1.000 & \\
True $\times$ false & $<$ 0.001 & 0.018 & 21418 & -0.017 & 1.000 & \\
True $\times$ unsure & -0.014 & 0.010 & 21418 & -1.449 & 0.736 & \\
True $\times$ true & 0.015 & 0.021 & 21418 & 0.746 & 1.000 & \\
\hline
\multicolumn{7}{l}{Significance codes: *** $P< 0.001$, ** $P< 0.01$, * $P< 0.05$, $\cdot$ $P< 0.1$} \\
\multicolumn{7}{l}{$\dagger$ Bonferroni's method comparing a family of 5 estimates} \\
\hline
\end{tabular}
\label{tab:fiveway_atai_S_posthoc}
\end{table}

% Figure 3: Optional Analysis
% % % % % % % % % % % % % % % % % % % %
Next, we examine whether behavior in the optional condition depends on ATAI by introducing a three-way interaction term involving whether a participant chose to view LLM fact checks (opt in vs. opt out), fact checking scenario, and individual attitude towards AI (Opt-Condition $\times$ FC Scenario $\times$ ATAI).
The results of fitting these models for the belief and share groups are presented in Tables \ref{tab:opt_atai_B} and \ref{tab:opt_atai_S}, respectively.
We conduct a post-hoc analysis that compares the slopes of the opt-in and opt-out conditions across varying levels of ATAI for the belief (Table~\ref{tab:opt_atai_B_slopes}) and sharing (Table~\ref{tab:opt_atai_S_slopes}) groups, respectively.
Results of the post-hoc comparisons can be found in Tables \ref{tab:opt_atai_B_post} and \ref{tab:opt_atai_S_post} for the belief and sharing groups, respectively.

We observe clear evidence suggesting that participants with more favorable ATAI are significantly more inclined to share news headlines (mean $b=0.044$) when viewing LLM fact-checking information, irrespective of the fact checking scenario.
However, this relationship does not extend to belief.
Instead, we find that ATAI has a significant and negative influence on belief in True headlines that are not identified as such for participants who opt out (True $\times$ false: $b = -.040$, $P = .014$; True $\times$ unsure: $b = -.033$, $P < .001$).
In other words, when participants decide to not view LLM fact-checking information, they are less likely to believe incorrectly labeled True headlines if their attitudes towards AI are more positive.
It would be interesting for future research to further explore the underlying psychological mechanisms that drive this complex relationship between attitudes towards AI, belief in True headlines, and the decision to engage with LLM fact-checking information.

Finally, we observe some evidence of a significant positive interaction between ATAI within the True $\times$ unsure fact checking scenario in both the belief and sharing groups.
Specifically, when the LLM provided unsure fact-checking information about true headlines, participants with higher levels of ATAI tended to believe and be willing to share those headlines more often (belief: $b = .032$, sharing: $b = .044$).

\begin{table}
\centering
\caption{Opt In versus Opt Out Coefficients (ATAI interaction, Belief Group; $F= 151.22.53$, $R^2=0.23$, $P < 0.001$)}
\begin{tabular}{lccccc}
\toprule
Variables & Estimate & Std. Error & $t$ value & $P$ & Sig. \\
\midrule
(Intercept)             &  0.593  & 0.125 & 4.735 & $< 0.001$ & *** \\
Option(opt out)         & -0.190  & 0.123 & -1.550 & 0.121 & \\
FC Scen.(False $\times$ unsure)     & 0.081 & NaN &  &  & \\
FC Scen.(True $\times$ false)       & 0.142 & 0.053 & 2.647 & 0.008 & ** \\
FC Scen.(True $\times$ true)        & 0.425 & 0.130 & 3.271 & 0.001 & *** \\
FC Scen.(True $\times$ unsure)      & 0.345 & 0.084 & 4.126 & $< 0.001$ & *** \\
ATAI                    &  0.008  & 0.023 & 0.347 & 0.729 & \\
Age                     & -0.004  & 0.001 & -3.907 & $< 0.001$ & *** \\
Education               & -0.007  & 0.009 & -0.714 & 0.475 & \\
Option(opt out):FC Scen.(False $\times$ unsure)     & -0.060  & 0.021 & -2.906 & 0.004 & ** \\
Option(opt out):FC Scen.(True $\times$ false)   & 0.508  & 0.101 & 5.046 & $< 0.001$ & *** \\
Option(opt out):FC Scen.(True $\times$ true)    & 0.182  & 0.221 & 0.822 & 0.411 & \\
Option(opt out):FC Scen.(True $\times$ unsure)  & 0.303  & 0.109 & 2.788 & 0.005 & ** \\
Option(opt out):ATAI            &  -0.011  & 0.026 & -0.424 & 0.671 &  \\
FC Scen.(False $\times$ unsure):ATAI    & $<$ 0.001 & NaN &  &  & \\
FC Scen.(True $\times$ false):ATAI      & -0.019 & 0.007 & -2.712 & 0.007 & ** \\
FC Scen.(True $\times$ true):ATAI       & -0.003 & 0.027 & -0.122 & 0.903 & \\
FC Scen.(True $\times$ unsure):ATAI     & -0.009 & 0.016 & -0.593 & 0.553 & \\
Option(opt out):FC Scen.(False $\times$ unsure):ATAI    & 0.015 & NaN &  &  & \\
Option(opt out):FC Scen.(True $\times$ false):ATAI      & -0.018 & 0.021 & -0.836 & 0.403 & \\
Option(opt out):FC Scen.(True $\times$ true):ATAI       & 0.008 & 0.044 & 0.173 & 0.863 &  \\
Option(opt out):FC Scen.(True $\times$ unsure):ATAI     & -0.020 & 0.022 & -0.932 & 0.351 & \\
\hline
\multicolumn{6}{l}{Significance codes: *** $P< 0.001$, ** $P< 0.01$, * $P< 0.05$, $\cdot P< 0.1$} \\
\hline
\end{tabular}
\label{tab:opt_atai_B}
\end{table}

\begin{table}
\centering
\caption{Opt In versus Opt Out Coefficients (ATAI interaction, Share Group; $F= 273.28$, $R^2=0.19$, $P < 0.001$)}
\begin{tabular}{lccccc}
\toprule
Variables & Estimate & Std. Error & $t$ value & $P$ & Sig. \\
\midrule
(Intercept)             & 0.622  & 0.144 & 4.309 & $< 0.001$ & *** \\
Option(opt out)         & -0.165 & 0.143 & -1.154 & 0.248 & \\
FC Scen.(False $\times$ unsure)     & -0.080 & NaN &  &  & \\
FC Scen.(True $\times$ false)       & -0.017 & NaN &  &  &  \\
FC Scen.(True $\times$ true)        & 0.082  & 0.082 & 0.991 & 0.322 &  \\
FC Scen.(True $\times$ unsure)      & 0.045  & 0.065 & 0.686 & 0.493 &  \\
ATAI                    & 0.038  & 0.027 & 1.418 & 0.156 & \\
Age                     & -0.007 & 0.001 & -5.057 & $< 0.001$ & *** \\
Education               & $< 0.001$  & 0.014 & -0.021 & 0.983 & \\
Option(opt out):FC Scen.(False $\times$ unsure)     & 0.131  & NaN &  &  &  \\
Option(opt out):FC Scen.(True $\times$ false)   & 0.067  & 0.047 & 1.436 & 0.151 & \\
Option(opt out):FC Scen.(True $\times$ true)    & 0.008  & 0.082 & 0.098 & 0.922 & \\
Option(opt out):FC Scen.(True $\times$ unsure)  & 0.080  & 0.093 & 0.859 & 0.391 & \\
Option(opt out):ATAI            & -0.029 & 0.030 & -0.943 & 0.346 &  \\
FC Scen.(False $\times$ unsure):ATAI    & 0.021 & NaN &  &  &  \\
FC Scen.(True $\times$ false):ATAI      & $< 0.001$ & NaN &  &  &  \\
FC Scen.(True $\times$ true):ATAI       & 0.005  & 0.017 & 0.321 & 0.748 & \\
FC Scen.(True $\times$ unsure):ATAI     & 0.006  & 0.014 & 0.464 & 0.642 & \\
Option(opt out):FC Scen.(False $\times$ unsure):ATAI    & -0.031 & NaN &  &  & \\
Option(opt out):FC Scen.(True $\times$ false):ATAI      & -0.001 & 0.013 & -0.053 & 0.958 & \\
Option(opt out):FC Scen.(True $\times$ true):ATAI       & -0.012 & 0.019 & -0.618 & 0.537 &  \\
Option(opt out):FC Scen.(True $\times$ unsure):ATAI     & -0.015 & 0.020 & -0.772 & 0.440 & \\
\hline
\multicolumn{6}{l}{Significance codes: *** $P< 0.001$, ** $P< 0.01$, * $P< 0.05$, $\cdot P< 0.1$} \\
\hline
\end{tabular}
\label{tab:opt_atai_S}
\end{table}

\begin{table}
\centering
\caption{Opt In versus Opt Out ATAI interaction slopes (Belief Group)}
\begin{tabular}{lccccccc}
  \hline
    Option & Headline Scenario & $b$ & Std. Err. & df & $t$-ratio & $P$ & Sig. \\ 
    \hline
    Opt in & False $\times$ false & 0.008 & 0.007 & 10418 & 1.200  & 0.230 & \\ 
    Opt out & False $\times$ false & -0.003 & 0.007 & 10418 & -0.437  & 0.662 & \\ 
    Opt in & False $\times$ unsure & 0.008 & 0.019 & 10418 & 0.440  & 0.660 & \\ 
    Opt out & False $\times$ unsure & 0.012 & 0.023 & 10418 & 0.512 & 0.609 & \\ 
    Opt in & True $\times$ false & -0.011 & 0.013 & 10418 & -0.817 & 0.414 & \\ 
    Opt out & True $\times$ false & -0.040 & 0.016 & 10418 & -2.464 & 0.014 & ** \\ 
    Opt in & True $\times$ true & 0.005 & 0.016 & 10418 & 0.307 & 0.759 & \\ 
    Opt out & True $\times$ true & 0.001 & 0.018 & 10418 & 0.068 & 0.946 & \\ 
    Opt in & True $\times$ unsure & -0.001 & 0.007 & 10418 & -0.166 & 0.868 & \\ 
    Opt out & True $\times$ unsure & -0.033 & 0.009 & 10418 & -3.552 & $<$ 0.001 & *** \\ 
    \hline
    \multicolumn{8}{l}{Significance codes: *** $P< 0.001$, ** $P< 0.01$, * $P< 0.05$, $\cdot P< 0.1$} \\
    \hline
\end{tabular}
\label{tab:opt_atai_B_slopes}
\end{table}

\begin{table}
\centering
\caption{Opt In versus Opt Out ATAI interaction slopes (Share Group)}
\begin{tabular}{lccccccc}
    \hline
    Option & Headline Scenario & $b$ & Std. Err. & df & $t$-ratio & $P$ & Sig. \\ 
    \hline
    Opt in & False $\times$ false & 0.038 & 0.007 & 10498 & 5.269 & $<$ .001 & *** \\ 
    Opt out & False $\times$ false & 0.009 & 0.008 & 10498 & 1.183 & 0.237 & \\ 
    Opt in & False $\times$ unsure & 0.059 & 0.021 & 10498 & 2.789 & 0.005 & ** \\ 
    Opt out & False $\times$ unsure & -4.71$\times$ 10$^{-5}$ & 0.026 & 10498 & -0.002 & 0.999 & \\ 
    Opt in & True $\times$ false & 0.038 & 0.015 & 10498 & 2.665 & 0.008 & ** \\ 
    Opt out & True $\times$ false & 0.009 & 0.019 & 10498 & 0.508 & 0.612 & \\ 
    Opt in & True $\times$ true & 0.043 & 0.017 & 10498 & 2.560 & 0.011 & * \\ 
    Opt out & True $\times$ true & 0.003 & 0.021 & 10498 & 0.181 & 0.856 & \\ 
    Opt in & True $\times$ unsure & 0.044 & 0.008 & 10498 & 5.497 & $<$ .001 & *** \\ 
    Opt out & True $\times$ unsure & 0.001 & 0.010 & 10498 & 0.108 & 0.914 &  \\
    \hline
    \multicolumn{8}{l}{Significance codes: *** $P< 0.001$, ** $P< 0.01$, * $P< 0.05$, $\cdot P< 0.1$} \\
    \hline
\end{tabular}
\label{tab:opt_atai_S_slopes}
\end{table}

\begin{table}
\centering
\caption{Post-hoc comparison of belief slopes fit to different ATAI values in the Optional condition}
\vspace{.1em}
\begin{tabular}{lcccccc}
\hline
Headline Scenario & Opt in $-$ Opt out & Std. Err. & df & $t$ ratio & Adj. $P^{\dagger}$ & Sig. \\
\hline
False $\times$ false  & 0.011 & 0.010 & 10418 & 1.142 & 1.000 & \\
False $\times$ unsure  & -0.003 & 0.030 & 10418 & -0.110 & 1.000 & \\
True $\times$ false & 0.029 & 0.021 & 10418 & 1.369 & 0.8550 & \\
True $\times$ true & 0.004 & 0.024 & 10418 & 0.147 & 1.000 & \\
True $\times$ unsure & 0.032 & 0.012 & 10418 & 2.678 & 0.037 & * \\
\hline 
\multicolumn{7}{l}{Significance codes: *** $P< 0.001$, ** $P< 0.01$, * $P< 0.05$, $\cdot P< 0.1$} \\
\multicolumn{7}{l}{$\dagger$ Bonferroni's method comparing a family of 5 estimates} \\
\hline
\end{tabular}
\label{tab:opt_atai_B_post}
\end{table}

\begin{table}
\centering
\caption{Post-hoc comparison of sharing intent slopes fit to different ATAI values in the Optional condition}
\vspace{.1em}
\begin{tabular}{lcccccc}
\hline
Headline Scenario & Opt in $-$ Opt out & Std. Err. & df & $t$ ratio & Adj. $P^{\dagger}$ & Sig. \\
\hline
False $\times$ false  & 0.029 & 0.011 & 10498 & 2.571 & 0.051 & $\cdot$ \\
False $\times$ unsure  & 0.059 & 0.033 & 10498 & 1.766 & 0.387 & \\
True $\times$ false & 0.029 & 0.024 & 10498 & 1.217 & 1.000 & \\
True $\times$ true & 0.040 & 0.027 & 10498 & 1.481 & 0.693 & \\
True $\times$ unsure & 0.044 & 0.013 & 10498 & 3.315 & 0.005 & ** \\
\hline 
\multicolumn{7}{l}{Significance codes: *** $P< 0.001$, ** $P< 0.01$, * $P< 0.05$, $\cdot P< 0.1$} \\
\multicolumn{7}{l}{$\dagger$ Bonferroni's method comparing a family of 5 estimates} \\
\hline
\end{tabular}
\label{tab:opt_atai_S_post}
\end{table}

\subsection{Headline congruence}
\label{sec:si:congruence_interaction}

\begin{figure}
    \centering
    \includegraphics[width=\linewidth]{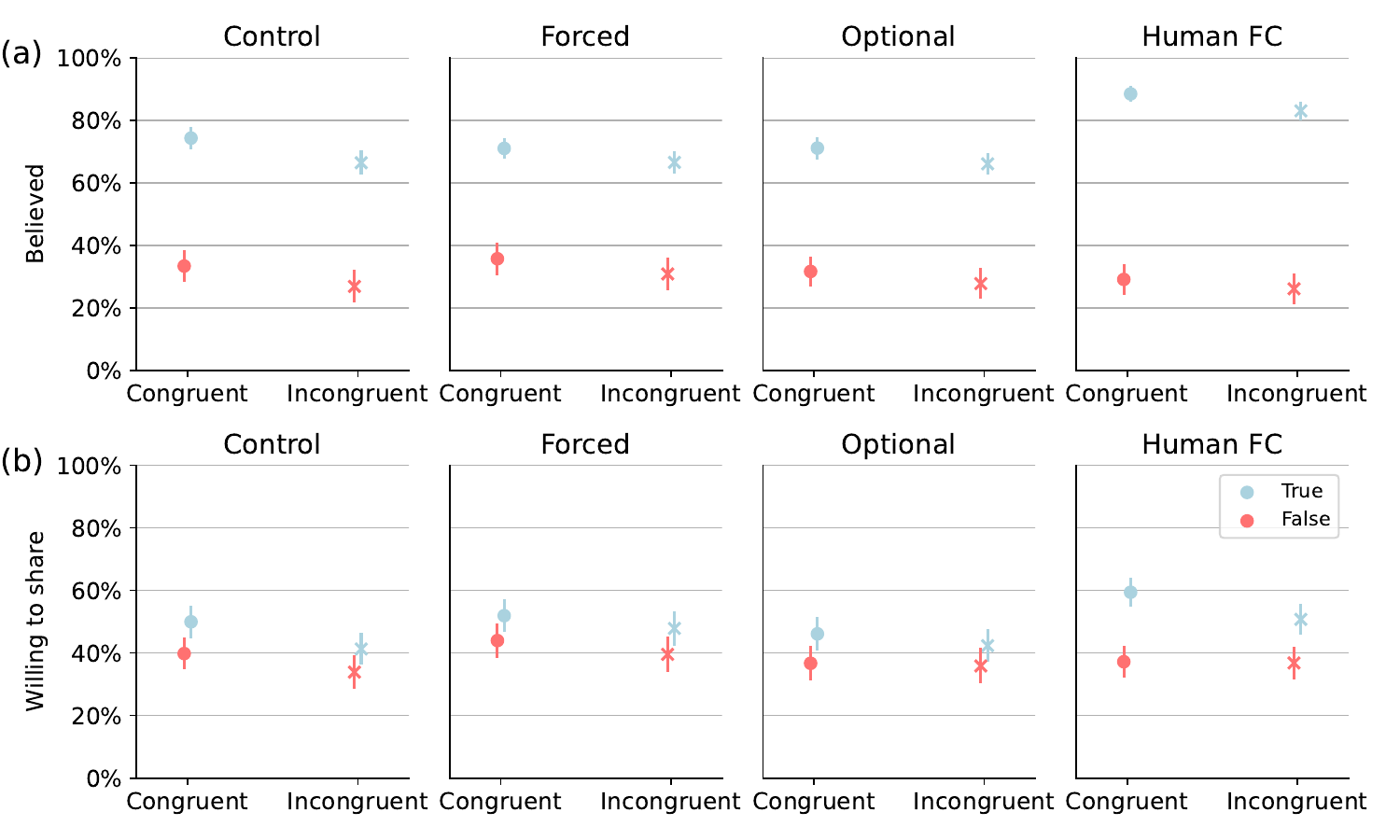}
    \caption{
    Relationship between (a) belief in and (b) intent to share headlines and their congruency across all conditions.
    Headline congruency is shown along the x-axis.
    }
    \label{fig:si_main_effects_congruency}
\end{figure}

% Figure 1: Main effects (Congruence)
% % % % % % % % % % % % % % % % % %
We now examine the potential moderating effects of headline congruence on participants' belief in and intention to share them, shown in Figure~\ref{fig:si_main_effects_congruency}.
We model this relationship by including a three-way interaction between Condition, Veracity, and headline Congruence (Condition $\times$ Veracity $\times$ Congruence).
The results related to belief and sharing intent can be found in Tables \ref{tab:ineffective_congru_B} and \ref{tab:ineffective_congru_S}, respectively.
We find no evidence of a significant three-way interaction between headline congruence in either group, suggesting that average discernment is not altered by the effects of headline congruence.

\begin{table}
\centering
\caption{Ineffectiveness of LLM Fact Checks Coefficients (Congruence interaction; Belief Group; $F= 762.09$, $R^2=0.25$, $P < 0.001$)}
\begin{tabular}{lccccc}
\hline
Variable & Estimate & Std. Error & $t$ value & $P$ & Sig. \\
\hline
(Intercept) & 0.582 & 0.049 & 11.857 & $< 0.001$ & *** \\
Cond.(Forced) & 0.029 & 0.031 & 0.930 & 0.352 & \\
Cond.(Optional) & -0.001 & 0.031 & -0.029 & 0.977 & \\
Cond.(HumanFC) & -0.030 & 0.030 & -1.006 & 0.314 & \\
Veracity(True) & 0.409 & 0.031 & 13.311 & $< 0.001$ & *** \\
Congr.(Inc.) & -0.065 & 0.014 & -4.646 & $< 0.001$ & *** \\
Age & -0.006 & 0.001 & -7.513 & $< 0.001$ & *** \\
Education & 0.009 & 0.005 & 1.874 & 0.061 & $\cdot$ \\
Cond.(Forced):Veracity(True) & -0.056 & 0.042 & -1.341 & 0.180 & \\
Cond.(Optional):Veracity(True) & -0.015 & 0.038 & -0.389 & 0.697 & \\
Cond.(HumanFC):Veracity(True) & 0.184 & 0.034 & 5.389 & $< 0.001$ & *** \\
Cond.(Forced):Congr.(Inc.) & 0.017 & 0.011 & 1.560 & 0.119 & \\
Cond.(Optional):Congr.(Inc.) & 0.026 & 0.006 & 4.058 & $< 0.001$ & *** \\
Cond.(HumanFC):Congr.(Inc.) & 0.035 & 0.007 & 4.718 & $< 0.001$ & *** \\
Veracity(True):Congr.(Inc.) & -0.014 & 0.024 & -0.573 & 0.567 & \\
Cond.(Forced):Veracity(True):Congr.(Inc.) & 0.018 & 0.023 & 0.759 & 0.448 & \\
Cond.(Optional):Veracity(True):Congr.(Inc.) & 0.002 & 0.027 & 0.082 & 0.935 & \\
Cond.(HumanFC):Veracity(True):Congr.(Inc.) & -0.010 & 0.017 & -0.620 & 0.535 & \\
\hline
\multicolumn{6}{l}{Significance codes: *** $P< 0.001$, ** $P< 0.01$, * $P< 0.05$, $\cdot P< 0.1$} \\
\hline
\end{tabular}
\label{tab:ineffective_congru_B}
\end{table}

\begin{table}
\centering
\caption{Ineffectiveness of LLM Fact Checks Coefficients (Congruence interaction; Share Group; $F= 313.41$, $R^2=0.11$, $P < 0.001$)}
\begin{tabular}{lccccc}
\hline
Variable & Estimate & Std. Error & $t$ value & $P$ & Sig. \\
\hline
(Intercept) & 0.879 & 0.044 & 20.031 & $< 0.001$ & *** \\
Cond.(Forced) & 0.023 & 0.032 & 0.722 & 0.470 & \\
Cond.(Optional) & -0.041 & 0.034 & -1.208 & 0.227 & \\
Cond.(HumanFC) & -0.017 & 0.036 & -0.460 & 0.646 & \\
Veracity(True) & 0.102 & 0.026 & 3.913 & $< 0.001$ & *** \\
Congr.(Inc.) & -0.059 & 0.013 & -4.748 & $< 0.001$ & *** \\
Age & -0.008 & 0.001 & -13.845 & $< 0.001$ & *** \\
Education & -0.018 & 0.007 & -2.542 & 0.011 & * \\
Cond.(Forced):Veracity(True) & -0.022 & 0.024 & -0.928 & 0.354 & \\
Cond.(Optional):Veracity(True) & -0.008 & 0.024 & -0.309 & 0.757 & \\
Cond.(HumanFC):Veracity(True) & 0.121 & 0.034 & 3.554 & $< 0.001$ & *** \\
Cond.(Forced):Congr.(Inc.) & 0.015 & 0.016 & 0.988 & 0.323 & \\
Cond.(Optional):Congr.(Inc.) & 0.051 & 0.013 & 3.982 & $< 0.001$ & *** \\
Cond.(HumanFC):Congr.(Inc.) & 0.056 & 0.028 & 1.970 &  0.049 & * \\
Veracity(True):Congr.(Inc.) & -0.027 & 0.023 & -1.186 & 0.236 & \\
Cond.(Forced):Veracity(True):Congr.(Inc.) & 0.031 & 0.022 & 1.368 & 0.171 & \\
Cond.(Optional):Veracity(True):Congr.(Inc.) & -0.002 & 0.012 & -0.186 & 0.852 & \\
Cond.(HumanFC):Veracity(True):Congr.(Inc.) & -0.056 & 0.037 & -1.504 & 0.133 & \\
\hline
\multicolumn{6}{l}{Significance codes: *** $P< 0.001$, ** $P< 0.01$, * $P< 0.05$, $\cdot P< 0.1$} \\
\hline
\end{tabular}
\label{tab:ineffective_congru_S}
\end{table}

% Figure 2: Accounting for LLM (Congruence)
% % % % % % % % % % % % % % % % % % % % % % % % % %

\begin{figure}
    \centering
    \includegraphics[width=\linewidth]{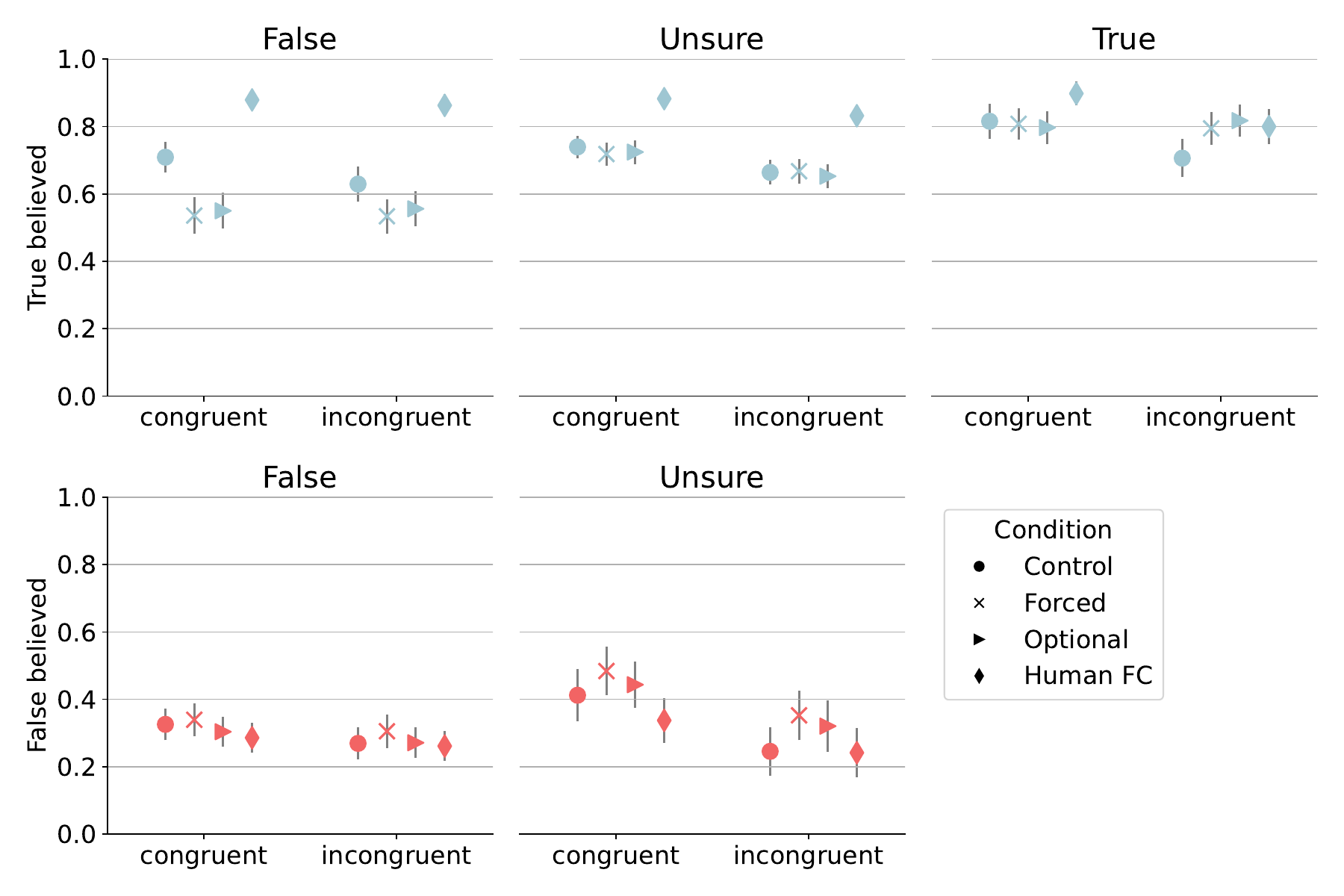}
    \caption{
    Relationship between belief in headlines and their congruency across all fact-checking scenarios.
    Experimental conditions are grouped along the x-axis based on headline congruency.
    The top and bottom panel rows represent true and false headlines, respectively.
    The left, center, and right panel columns represent ChatGPT's judgment of those headlines as false, unsure, and true, respectively.
    The bottom right panel is excluded as this type of headline (false headline judged by ChatGPT to be true) does not exist in our data.
    }
    \label{fig:si_5way_congruence_belief}
\end{figure}

\begin{figure}
    \centering
    \includegraphics[width=\linewidth]{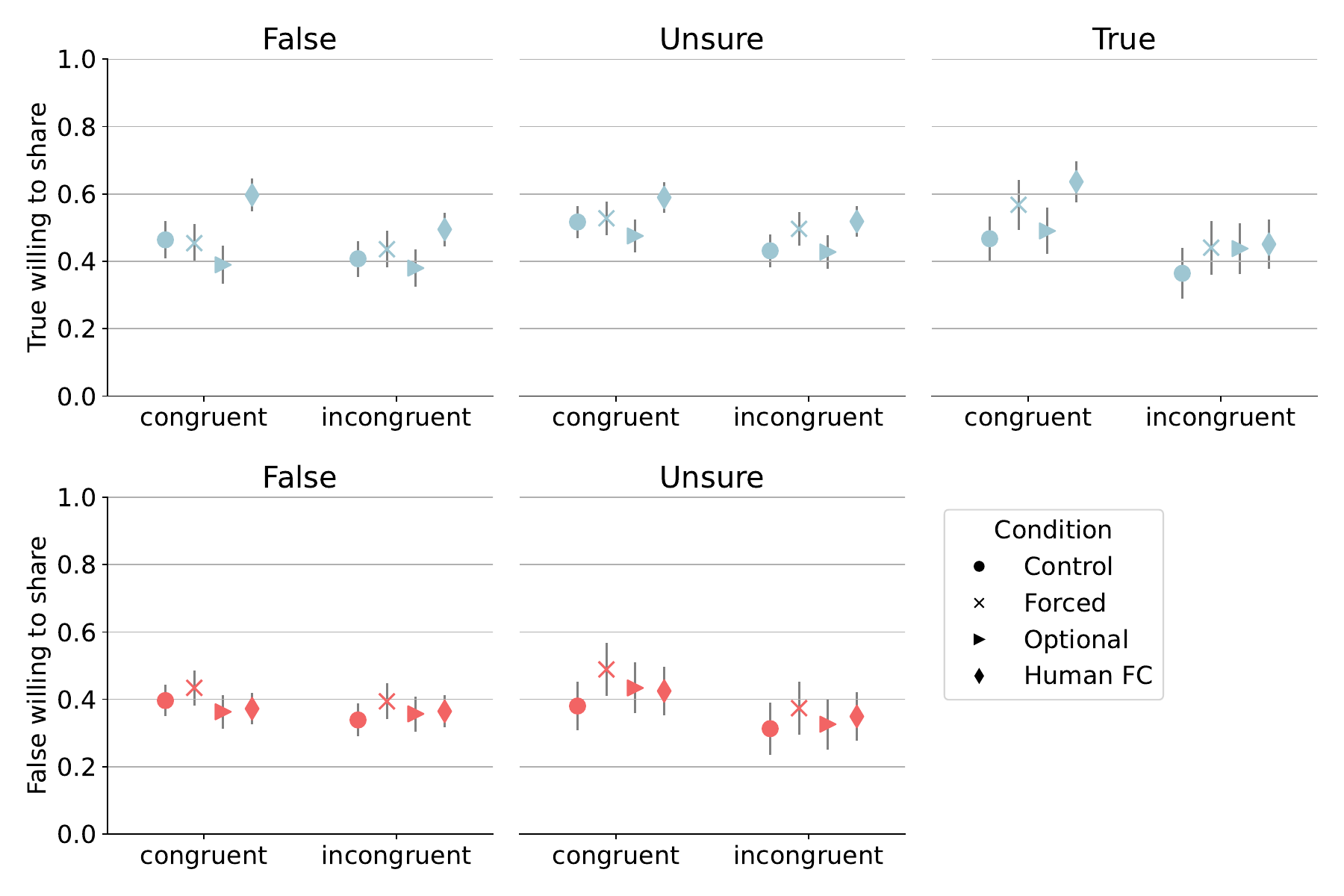}
    \caption{
    Relationship between intent to share headlines and their congruency across all conditions.
    Experimental conditions are grouped along the x-axis based on headline congruency.
    The top and bottom panel rows represent true and false headlines, respectively.
    The left, center, and right panel columns represent ChatGPT's judgment of those headlines as false, unsure, and true, respectively.
    The bottom right panel is excluded as this type of headline (false headline judged by ChatGPT to be true) does not exist in our data.
    }
    \label{fig:si_5way_congruence_share}
\end{figure}

Figure~\ref{fig:si_5way_congruence_belief} illustrates the relationship between belief in headlines and their congruency across all fact-checking scenarios and experimental conditions.
The same relationship is presented with respect to sharing intent in Figure~\ref{fig:si_5way_congruence_share}.
We model this relationship using a three-way interaction between condition, fact-checking scenario, and headline congruence (Condition $\times$ FC Scenario $\times$ Congruence).
Again, we focus on the forced and control conditions and exclude data for the optional participants when fitting each model.
The results of fitting the belief and share group models are found in Tables \ref{tab:fiveway_congru_B} and \ref{tab:fiveway_congru_S}.
However, we again must utilize these models for post-hoc comparisons similar to those presented in the main text for each group.
To do this, we compare headline congruence fitted slopes between the Control and Forced groups.
These results are shown in Tables \ref{tab:fiveway_congru_B_post} and \ref{tab:fiveway_congru_S_post} for the belief and share group, respectively.
We found no evidence of significant interactions within the belief group.
However, in the sharing group, some significant interactions were observed for a specific fact-checking scenario.
Participants who were forced to view unsure LLM fact checks about politically incongruent true headlines (True $\times$ unsure) were more likely to report a willingness to share these headlines compared to participants in the control group who viewed similar headlines.
This was true despite the fact that, within each group, the tendency was to report a willingness to share incongruent headlines less than congruent headlines (Control: $b=-0.10$; LLM-forced: $b=-0.03$).

\begin{table}
\centering
\caption{Account for LLM Accuracy Coefficients (Congruence interaction, Belief Group; $F= 233.48$, $R^2=0.21$, $P < 0.001$)}
\small
\begin{tabular}{lccccc}
\hline
Variable & Estimate & Std. Error & $t$ value & $P$ value & Sig. \\
\hline
(Intercept) & 0.589 & 0.057 & 10.377 & $< 0.001$ & *** \\
Cond.(Forced) & 0.022 & 0.031 & 0.719 & 0.472 &  \\
FC Scen.(False $\times$ unsure) & 0.081 & 0.029 & 2.792 & 0.005 & ** \\
FC Scen.(True $\times$ false) & 0.383 & 0.035 & 10.847 & $< 0.001$ & *** \\
FC Scen.(True $\times$ true) & 0.484 & 0.037 & 13.030 & $< 0.001$ & *** \\
FC Scen.(True $\times$ unsure) & 0.412 & 0.034 & 12.013 & $< 0.001$ & *** \\
Congr.(Inc.) & -0.055 & 0.014 & -3.945 & $< 0.001$ & *** \\
Age & -0.007 & 0.001 & -7.507 & $< 0.001$ & *** \\
Education & 0.016 & 0.008 & 2.161 & 0.031 & * \\
Cond.(Forced):FC Scen.(False $\times$ unsure) & 0.061 & 0.035 & 1.739 & 0.082 & $\cdot$ \\
Cond.(Forced):FC Scen.(True $\times$ false) & -0.190 & 0.051 & -3.683 & $< 0.001$ & *** \\
Cond.(Forced):FC Scen.(True $\times$ true) & -0.019 & 0.039 & -0.487 & 0.626 &  \\
Cond.(Forced):FC Scen.(True $\times$ unsure) & -0.014 & 0.049 & -0.294 & 0.769 &  \\
Cond.(Forced):Congr.(Inc.) & 0.017 & 0.013 & 1.302 & 0.193 &  \\
FC Scen.(False $\times$ unsure):Congr.(Inc.) & -0.102 & 0.040 & -2.536 & 0.011 & * \\
FC Scen.(True $\times$ false):Congr.(Inc.) & -0.025 & 0.073 & -0.340 & 0.734 &  \\
FC Scen.(True $\times$ true):Congr.(Inc.) & -0.045 & 0.034 & -1.313 & 0.189 &  \\
FC Scen.(True $\times$ unsure):Congr.(Inc.) & -0.019 & 0.027 & -0.724 & 0.469 &  \\
Cond.(Forced):FC Scen.(False $\times$ unsure):Congr.(Inc.) & 0.008 & 0.045 & 0.173 & 0.863 &  \\
Cond.(Forced):FC Scen.(True $\times$ false):Congr.(Inc.) & 0.061 & 0.054 & 1.126 & 0.260 &  \\
Cond.(Forced):FC Scen.(True $\times$ true):Congr.(Inc.) & 0.069 & 0.035 & 1.962 & 0.050 & * \\
Cond.(Forced):FC Scen.(True $\times$ unsure):Congr.(Inc.) & -0.005 & 0.031 & -0.171 & 0.864 &  \\
\hline
\multicolumn{6}{l}{Significance codes: *** $P< 0.001$, ** $P< 0.01$, * $P< 0.05$, $\cdot P< 0.1$} \\
\hline
\end{tabular}
\label{tab:fiveway_congru_B}
\end{table}

\begin{table}
\centering
\caption{Account for LLM Accuracy Coefficients (Congruence interaction, Share Group; $F= 133.51$, $R^2=0.12$, $P < 0.001$)}
\small
\begin{tabular}{lccccc}
\hline
Variable & Estimate & Std. Error & $t$ value & $P$ & Sig. \\
\hline
(Intercept) & 0.916 & 0.054 & 17.043 & $< 0.001$ & *** \\
Cond.(Forced) & 0.015 & 0.032 & 0.477 & 0.633 & \\
FC Scen.(False $\times$ unsure) & -0.036 & 0.021 & -1.693 & 0.090 & $\cdot$ \\
FC Scen.(True $\times$ false) & 0.062 & 0.040 & 1.530 & 0.126 & \\
FC Scen.(True $\times$ true) & 0.051 & 0.055 & 0.919 & 0.358 & \\
FC Scen.(True $\times$ unsure) & 0.122 & 0.027 & 4.468 & $< 0.001$ & *** \\
Congr.(Inc.) & -0.063 & 0.015 & -4.253 & $< 0.001$ & *** \\
Age & -0.008 & 0.001 & -7.974 & $< 0.001$ & *** \\
Education & -0.041 & 0.011 & -3.690 & $< 0.001$  & *** \\
Cond.(Forced):FC Scen.(False $\times$ unsure) & 0.082 & 0.018 & 4.674 & $< 0.001$ & *** \\
Cond.(Forced):FC Scen.(True $\times$ false) & -0.042 & 0.027 & -1.565 & 0.118 & \\
Cond.(Forced):FC Scen.(True $\times$ true) & 0.074 & 0.046 & 1.604 & 0.109 & \\
Cond.(Forced):FC Scen.(True $\times$ unsure) & -0.027 & 0.023 & -1.147 & 0.251 & \\
Cond.(Forced):Congr.(Inc.) & 0.025 & 0.018 & 1.369 & 0.171 & \\
FC Scen.(False $\times$ unsure):Congr.(Inc.) & 0.029 & 0.039 & 0.739 & 0.460 & \\
FC Scen.(True $\times$ false):Congr.(Inc.) & 0.007 & 0.048 & 0.140 & 0.889 & \\
FC Scen.(True $\times$ true):Congr.(Inc.) & -0.006 & 0.068 & -0.093 & 0.926 & \\
FC Scen.(True $\times$ unsure):Congr.(Inc.) & -0.041 & 0.027 & -1.481 & 0.139 & \\
Cond.(Forced):FC Scen.(False $\times$ unsure):Congr.(Inc.) & -0.088 & 0.056 & -1.571 & 0.116 & \\
Cond.(Forced):FC Scen.(True $\times$ false):Congr.(Inc.) & 0.013 & 0.038 & 0.351 & 0.726 & \\
Cond.(Forced):FC Scen.(True $\times$ true):Congr.(Inc.) & -0.064 & 0.075 & -0.855 & 0.393 & \\
Cond.(Forced):FC Scen.(True $\times$ unsure):Congr.(Inc.) & 0.046 & 0.026 & 1.783 & 0.075 & $\cdot$ \\
\hline
\multicolumn{6}{l}{Significance codes: *** $P< 0.001$, ** $P< 0.01$, * $P< 0.05$, $\cdot P< 0.1$} \\
\hline
\end{tabular}
\label{tab:fiveway_congru_S}
\end{table}

\begin{table}
\centering
\caption{Post-hoc comparison of belief slopes fit different FC scenarios and headline congruence}
\vspace{.1em}
\begin{tabular}{lcccccc}
\hline
Headline Scenario & Forced $-$ Control & Std. Error & df & $t$ ratio & Adj. $P^{\dagger}$ & Sig. \\
\hline
False $\times$ false & 0.017 & 0.019 & 18738 & 0.873 & 1.000 & \\
False $\times$ unsure & 0.025 & 0.058 & 18738 & 0.425 & 1.000 & \\
True $\times$ false & 0.078 & 0.041 & 18738 & 1.887 & 0.296 & \\
True $\times$ true & 0.086 & 0.048 & 18738 & 1.804 & 0.356 & \\
True $\times$ unsure & 0.012 & 0.023 & 18738 & 0.511 & 1.000 & \\
\hline
\multicolumn{7}{l}{Significance codes: *** $P< 0.001$, ** $P< 0.01$, * $P< 0.05$, $\cdot$ $P< 0.1$} \\
\multicolumn{7}{l}{$\dagger$ Bonferroni's method comparing a family of 5 estimates} \\
\hline
\end{tabular}
\label{tab:fiveway_congru_B_post}
\end{table}

\begin{table}
\centering
\caption{Post-hoc comparison of sharing slopes fit to different FC scenarios and headline congruence}
\vspace{.1em}
\begin{tabular}{lcccccc}
\hline
Headline Scenario & Forced $-$ Control & Std. Error & df & $t$ ratio & Adj. $P^{\dagger}$ & Sig. \\
\hline
False $\times$ false & 0.025 & 0.020 & 19938 & 1.278 & 1.000 & \\
False $\times$ unsure & -0.062 & 0.059 & 19938 & -1.055 & 1.000 & \\
True $\times$ false & 0.038 & 0.042 & 19938 & 0.919 & 1.000 & \\
True $\times$ true & -0.039 & 0.048 & 19938 & -0.815 & 1.000 & \\
True $\times$ unsure & 0.071 & 0.023 & 19938 & 3.085 & 0.010 & * \\
\hline
\multicolumn{7}{l}{Significance codes: *** $P< 0.001$, ** $P< 0.01$, * $P< 0.05$, $\cdot$ $P< 0.1$} \\
\multicolumn{7}{l}{$\dagger$ Bonferroni's method comparing a family of 5 estimates} \\
\hline
\end{tabular}
\label{tab:fiveway_congru_S_post}
\end{table}

% Figure 3: Optional (Congruence)
% % % % % % % % % % % % % % % % % % % % % %
Next, we examine whether behaviors in the optional condition differ based on the congruence of headlines by introducing a three-way interaction term involving whether a participant chose to view LLM fact checks (opt in vs. opt out), fact-checking scenario, and headline congruence (Opt-Condition $\times$ FC Scenario $\times$ Congruence).
Tables \ref{tab:opt_congru_B} and \ref{tab:opt_congru_S} show the results of fitting these models for the belief and intent to share groups, respectively.
We perform a post-hoc comparison of the belief (Table~\ref{tab:opt_congru_B_slopes}) and sharing (Table~\ref{tab:opt_congru_S_slopes}) group slopes, fit to the opt-in and opt-out conditions across different levels of congruence.
The results of these post-hoc comparisons are shown in Tables \ref{tab:opt_congru_B_post} and \ref{tab:opt_congru_S_post}, respectively.

We observe that partisan incongruency is significantly negatively related to participants' belief (Opt in $b = -.086$, $P < .001$; Opt out $b = -.097$, $P < 0.001$) and sharing intent (Opt in $b = -.038$, $P = .050$; Opt out $b = -.088$, $P < 0.001$) with respect to True headlines that the model was unsure about, regardless of whether participants chose to view the LLM fact-checking information.
Additionally, we find that participants who did not view the LLM fact-checking information for false headlines were significantly less likely to believe incongruent headlines (False $\times$ false: $b = -.057$, $P = .002$; False $\times$ unsure: $b = -.195$, $P = .001$).
In other words, when participants encountered politically incongruent true headlines that the LLM was unsure about, their likelihood of believing or being willing to share them diminished significantly. 
This relationship persisted irrespective of whether participants opted to access the fact-checking information. 
This relationship does not hold for accurately identified True headlines in either the belief or sharing groups. 
However, we do find evidence of a similar relationship for false headlines, but only when participants did not view LLM-generated fact checks.

\begin{table}
\centering
\caption{Opt In versus Opt Out Coefficients (Congruency interaction, Belief Group; $F= 146.91$, $R^2=0.24$, $P < 0.001$)}
\small
\begin{tabular}{lccccc}
\hline
Variable & Estimate & Std. Error & $t$ value & $P$ & Sig. \\
\midrule
(Intercept)         & 0.637  & 0.063 & 10.175 & $< 0.001$  & *** \\
Option(opt out)     & -0.217 & 0.046 & -4.765 & $< 0.001$ & *** \\
FC Scen.(False $\times$ unsure)     & 0.072 & 0.083 & 0.864 & 0.388      &  \\
FC Scen.(True $\times$ false)       & 0.034 & 0.018 & 1.856 & 0.063      & $\cdot$ \\
FC Scen.(True $\times$ true)        & 0.379 & 0.034 & 11.154 & $< 0.001$ & *** \\
FC Scen.(True $\times$ unsure)      & 0.344 & 0.047 & 7.365  & $< 0.001$ & *** \\
Congr.(Inc.)    & -0.003 & 0.023 & -0.108 & 0.914 & \\
Age             & -0.004 & 0.001 & -3.926 & $< 0.001$ & *** \\
Education       & -0.009 & 0.009 & -0.929 & 0.353 & \\
Option(opt out):FC Scen.(False $\times$ unsure)     & 0.068 & 0.096 & 0.710  & 0.478  &  \\
Option(opt out):FC Scen.(True $\times$ false)       & 0.457 & 0.044 & 10.325 & $< 0.001$ & *** \\
Option(opt out):FC Scen.(True $\times$ true)        & 0.193 & 0.049 & 3.907  & $< 0.001$ & *** \\
Option(opt out):FC Scen.(True $\times$ unsure)      & 0.191 & 0.055 & 3.444 & 0.001 & *** \\
Option(opt out):Congr.(Inc.)                & -0.054 & 0.043 & -1.271 & 0.204  & \\
FC Scen.(False $\times$ unsure):Congr.(Inc.)    & 0.027 & 0.053 & 0.503 & 0.615     &  \\
FC Scen.(True $\times$ false):Congr.(Inc.)      & 0.043 & 0.017 & 2.500 & 0.012     & * \\
FC Scen.(True $\times$ true):Congr.(Inc.)       & 0.072 & 0.038 & 1.871 & 0.061     & $\cdot$ \\
FC Scen.(True $\times$ unsure):Congr.(Inc.)     & -0.083 & 0.039 & -2.153 & 0.031   & * \\
Option(opt out):FC Scen.(False $\times$ unsure):Congr.(Inc.)    & -0.165 & 0.073 & -2.266 & 0.023  & * \\
Option(opt out):FC Scen.(True $\times$ false):Congr.(Inc.)      & -0.032 & 0.052 & -0.605 & 0.545  &  \\
Option(opt out):FC Scen.(True $\times$ true):Congr.(Inc.)       & 0.024 & 0.064 & 0.382 & 0.703 &  \\
Option(opt out):FC Scen.(True $\times$ unsure):Congr.(Inc.)     & 0.042 & 0.060 & 0.703 & 0.482 &  \\

\hline
\multicolumn{6}{l}{Significance codes: *** $P< 0.001$, ** $P< 0.01$, * $P< 0.05$, $\cdot P< 0.1$} \\
\hline
\end{tabular}
\label{tab:opt_congru_B}
\end{table}

\begin{table}
\centering
\caption{Opt In versus Opt Out Coefficients (Congruency interaction, Share Group; $F= 113.92$, $R^2=0.19$, $P < 0.001$)}
\small
\begin{tabular}{lcccccc}
\hline
Variable & Estimate & Std. Error & $t$ value & $P$ & Sig. \\
\midrule
(Intercept)         & 0.798 & 0.068 & 11.778 & $< 0.001$ & *** \\
Option(opt out)     & -0.304 & 0.039 & -7.822 & $< 0.001$ & *** \\
FC Scen.(False $\times$ unsure)     & 0.007 & 0.018 & 0.396 & 0.692     &  \\
FC Scen.(True $\times$ false)       & -0.005 & 0.010 & -0.462 & 0.644   &  \\
FC Scen.(True $\times$ true)        & 0.084 & 0.039 & 2.179 & 0.029     & * \\
FC Scen.(True $\times$ unsure)      & 0.098 & 0.023 & 4.253 & $< 0.001$ & *** \\
Congr.(Inc.)    & 0.013 & 0.010 & 1.256 & 0.209 & \\
Age             & -0.006 & 0.001 & -5.026 & $< 0.001$ & *** \\
Education       & -0.001 & 0.014 & -0.074 & 0.941 & \\
Option(opt out):FC Scen.(False $\times$ unsure)     & 0.043  & 0.029 & 1.472  & 0.141 &  \\
Option(opt out):FC Scen.(True $\times$ false)       & 0.058  & 0.036 & 1.596  & 0.111 &  \\
Option(opt out):FC Scen.(True $\times$ true)        & -0.008 & 0.068 & -0.111 & 0.911 &  \\
Option(opt out):FC Scen.(True $\times$ unsure)      & 0.030  & 0.034 & 0.897  & 0.370 &  \\
Option(opt out):Congr.(Inc.)                & -0.022 & 0.014 & -1.587 & 0.113  & \\
FC Scen.(False $\times$ unsure):Congr.(Inc.)    & 0.036  & 0.057 & 0.623  & 0.533       &  \\
FC Scen.(True $\times$ false):Congr.(Inc.)      & -0.023 & NaN   &        &             &  \\
FC Scen.(True $\times$ true):Congr.(Inc.)       & 0.060  & 0.059 & 1.030  & 0.303       &  \\
FC Scen.(True $\times$ unsure):Congr.(Inc.)     & -0.051 & 0.013 & -3.937 & $< 0.001$   & ***  \\
Option(opt out):FC Scen.(False $\times$ unsure):Congr.(Inc.)    & -0.105 & 0.070 & -1.494 & 0.135 & \\
Option(opt out):FC Scen.(True $\times$ false):Congr.(Inc.)      & 0.021 & 0.039 & 0.553 & 0.580 &  \\
Option(opt out):FC Scen.(True $\times$ true):Congr.(Inc.)       & -0.069 & 0.097 & -0.713 & 0.476 &  \\
Option(opt out):FC Scen.(True $\times$ unsure):Congr.(Inc.)     & -0.028 & 0.030 & -0.941 & 0.347 &  \\
\hline
\multicolumn{6}{l}{Significance codes: *** $P< 0.001$, ** $P< 0.01$, * $P< 0.05$, $\cdot P< 0.1$} \\
\hline
\end{tabular}
\label{tab:opt_congru_S}
\end{table}

\begin{table}
\centering
\caption{Opt In versus Opt Out congruency interaction slopes (Belief Group)}
\begin{tabular}{lccccccc}
    \hline
    Option & Headline Scenario & $b$ & Std. Err. & df & $t$-ratio & $P$ & Sig. \\ 
    \hline
    Opt in & False $\times$ false & -0.003 & 0.019 & 9978 & -0.136 & 0.891 & \\ 
    Opt out & False $\times$ false & -0.057 & 0.018 & 9978 & -3.077 & 0.002 & ** \\ 
    Opt in & False $\times$ unsure & 0.024 & 0.052 & 9978 & 0.465 & 0.642 & \\ 
    Opt out & False $\times$ unsure & -0.195 & 0.061 & 9978 & -3.208 & 0.001 & ** \\ 
    Opt in & True $\times$ false & 0.040 & 0.037 & 9978 & 1.081 & 0.280 &  \\ 
    Opt out & True $\times$ false & -0.046 & 0.042 & 9978 & -1.098 & 0.272 & \\ 
    Opt in & True $\times$ true & 0.069 & 0.043 & 9978 & 1.617 & 0.106 & \\ 
    Opt out & True $\times$ true & 0.039 & 0.049 & 9978 & 0.813 & 0.416 & \\ 
    Opt in & True $\times$ unsure & -0.086 & 0.020 & 9978 & -4.257 & $<$ .001 & *** \\ 
    Opt out & True $\times$ unsure & -0.097 & 0.024 & 9978 & -4.085 & $<$ .001 & *** \\
    \hline
    \multicolumn{8}{l}{Significance codes: *** $P< 0.001$, ** $P< 0.01$, * $P< 0.05$, $\cdot P< 0.1$} \\
    \hline
\end{tabular}
\label{tab:opt_congru_B_slopes}
\end{table}

\begin{table}
\centering
\caption{Opt In versus Opt Out congruency interaction slopes (Sharing Group)}
\begin{tabular}{lccccccc}
    \hline
    Option & Headline Scenario & $b$ & Std. Err. & df & $t$-ratio & $P$ & Sig. \\ 
    \hline
    Opt in & False $\times$ false & 0.013 & 0.017 & 10138 & 0.729 & 0.466 & \\ 
    Opt out & False $\times$ false & -0.010 & 0.020 & 10138 & -0.480 & 0.632 & \\ 
    Opt in & False $\times$ unsure & 0.048 & 0.052 & 10138 & 0.933 & 0.351 & \\ 
    Opt out & False $\times$ unsure & -0.079 & 0.063 & 10138 & -1.258 & 0.209 & \\ 
    Opt in & True $\times$ false & -0.010 & 0.035 & 10138 & -0.300 & 0.765 &  \\ 
    Opt out & True $\times$ false & -0.011 & 0.045 & 10138 & -0.251 & 0.802 & \\ 
    Opt in & True $\times$ true & 0.073 & 0.042 & 10138 & 1.758 & 0.079 & \\ 
    Opt out & True $\times$ true & -0.018 & 0.052 & 10138 & -0.352 & 0.725 & \\ 
    Opt in & True $\times$ unsure & -0.038 & 0.019 & 10138 & -1.961 & 0.050 & * \\ 
    Opt out & True $\times$ unsure & -0.088 & 0.025 & 10138 & -3.501 & $<$ 0.001 & *** \\ 
    \hline
    \multicolumn{8}{l}{Significance codes: *** $P< 0.001$, ** $P< 0.01$, * $P< 0.05$, $\cdot P< 0.1$} \\
    \hline
\end{tabular}
\label{tab:opt_congru_S_slopes}
\end{table}

\begin{table}
\centering
\caption{Post-hoc comparison of belief slopes for different headline congruence in the Optional condition}
\vspace{.1em}
\begin{tabular}{lcccccc}
\hline
Headline scenario & Opt in $-$ Opt out & Std. Error & df & $t$ ratio & Adj. $P^{\dagger}$ & Sig. \\
    \hline
    True $\times$ False     & 0.086 & 0.056 & 9978 & 1.540 & 0.619 &  \\
    True $\times$ Unsure    & 0.012 & 0.031 & 9978 & 0.371 & 1.000 &  \\
    True $\times$ True      & 0.030 & 0.065 & 9978 & 0.459 & 1.000 &  \\
    False $\times$ False    & 0.054 & 0.026 & 9978 & 2.071 & 0.192 &  \\
    False $\times$ Unsure   & 0.219 & 0.080 & 9978 & 2.745 & 0.030 & * \\
\hline 
\multicolumn{7}{l}{Significance codes: *** $P< 0.001$, ** $P< 0.01$, * $P< 0.05$, $\cdot$ $P< 0.1$} \\
\multicolumn{7}{l}{$\dagger$ Bonferroni's method comparing a family of 5 estimates} \\
\hline
\end{tabular}
\label{tab:opt_congru_B_post}
\end{table}

\begin{table}
\centering
\caption{Post-hoc comparison of sharing intent slopes for different headline congruence in the Optional condition}
\vspace{.1em}
\begin{tabular}{lcccccc}
\hline
Headline scenario & Opt in $-$ Opt out & Std. Error & df & $t$ ratio & Adj. $P^{\dagger}$ & Sig. \\
    \hline
    False $\times$ False    & 0.022 & 0.026 & 10138 & 0.838 & 1.000 &  \\
    False $\times$ Unsure   & 0.127 & 0.081 & 10138 & 1.564 & 0.589 &  \\
    True $\times$ False     & 0.001 & 0.057 & 10138 & 0.015 & 1.000 &  \\
    True $\times$ True      & 0.091 & 0.066 & 10138 & 1.375 & 0.845 &  \\
    True $\times$ Unsure    & 0.050 & 0.032 & 10138 & 1.577 & 0.574 &  \\
\hline 
\multicolumn{7}{l}{Significance codes: *** $P< 0.001$, ** $P< 0.01$, * $P< 0.05$, $\cdot$ $P< 0.1$} \\
\multicolumn{7}{l}{$\dagger$ Bonferroni's method comparing a family of 5 estimates} \\
\hline
\end{tabular}
\label{tab:opt_congru_S_post}
\end{table}

\section{Opt-in behavior}

\begin{figure}
    \centering
    \includegraphics[width=.75\linewidth]{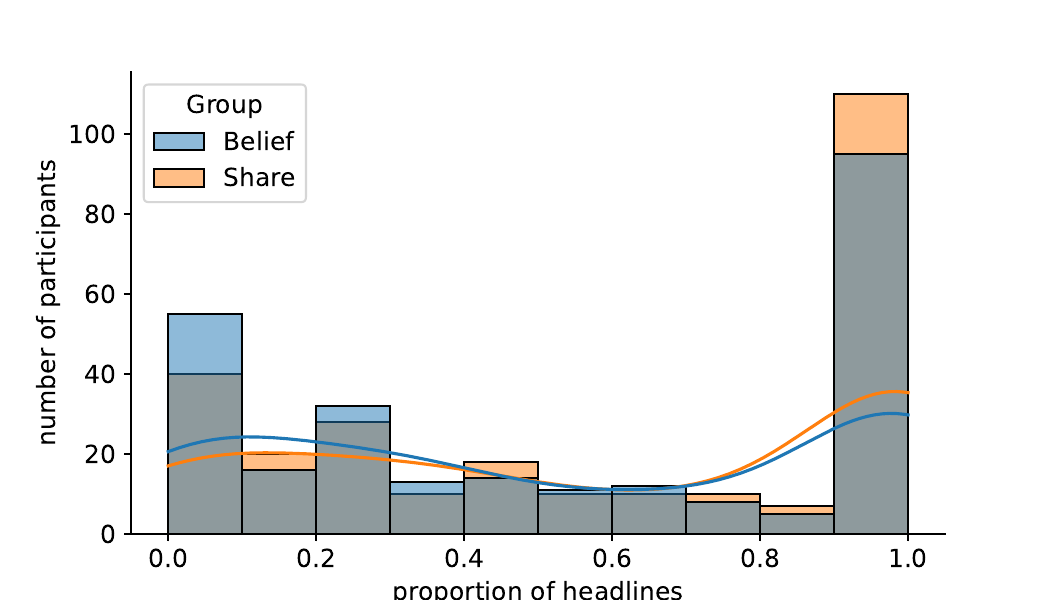}
    \caption{
    Distribution of the proportion of headlines for which participants chose to view LLM-generated fact checking information by experimental group.
    }
    \label{fig:si_opt_in_distribtions_by_group}
\end{figure}

\begin{figure}
    \centering
    \includegraphics[width=\linewidth]{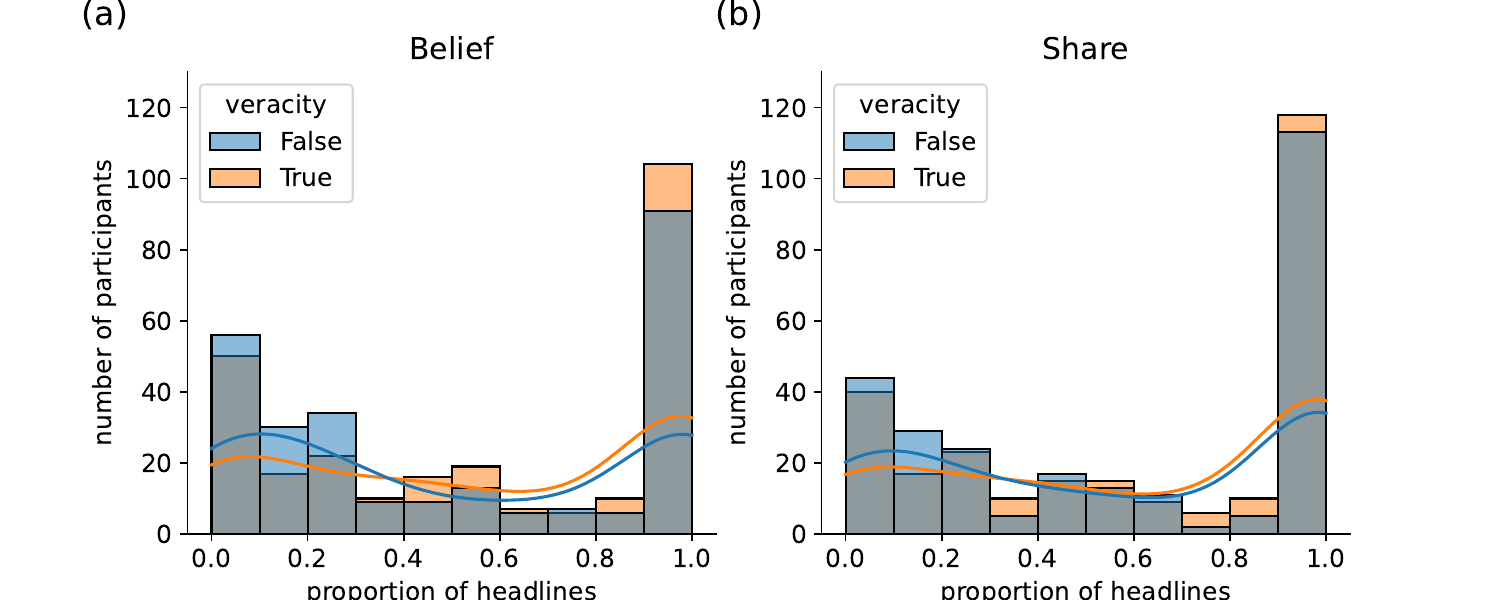}
    \caption{
    Distribution of the proportion of headlines for which participants chose to view LLM-generated fact checking information by veracity for the Belief (a) and Share (b) experimental groups.
    }
    \label{fig:si_opt_in_distribtions_by_veracity}
\end{figure}

Figure~\ref{fig:si_opt_in_distribtions_by_group} presents the distributions of headlines that participants chose to view when in the LLM-optional condition.
Figure~\ref{fig:si_opt_in_distribtions_by_veracity} presents the same information by headline veracity for each experimental group.
Mann-Whitney U tests show that there is no significant difference in the average number of headlines opted into by the belief and sharing groups ($P = 0.10$). 
Additionally, we observe no significant difference in the average number of true versus false headlines chosen by participants in either group (belief: $P = 0.13$; sharing: $P = 0.55$).
Table \ref{tab:opt-in-all} displays statistical results for all opt in versus opt out comparisons discussed in the main text.


\begin{table}[ht!]
\centering
\begin{tabular}{lllcccc}
\toprule
\textbf{Group} & \textbf{Veracity} & \textbf{Judged} & \textbf{Opt in $-$ Opt out} & \textbf{Adj. $P^\dagger$} & \textbf{Cohen's d} & \textbf{95\% CI (\%)} \\
\midrule
Belief & True  & True   & 7.46\%  & 0.0411    & 0.18  & [-1.50\%, 16.35\%] \\
Belief & True  & False  & -16.59\% & $<$ 0.001   & -0.35 & [-26.34\%, -7.24\%] \\
Belief & True  & Unsure & 5.51\%  & 1.0       & 0.12  & [-3.70\%, 14.47\%] \\
Belief & False & False  & 29.35\% & 0.0049    & 0.63  & [20.81\%, 37.93\%] \\
Belief & False & Unsure & 28.12\% & $<$ 0.001   & 0.64  & [18.43\%, 38.12\%] \\
\hline
Share  & True  & True   & 39.46\% & $<$ 0.001   & 0.85 & [30.00\%, 49.15\%] \\
Share  & True  & False  & 29.39\% & $<$ 0.001   & 0.62  & [19.76\%, 38.75\%] \\
Share  & True  & Unsure & 34.24\% & $<$ 0.001   & 0.71  & [25.18\%, 43.17\%] \\
Share  & False & False  & 37.63\% & $<$ 0.001   & 0.74  & [28.30\%, 46.83\%] \\
Share  & False & Unsure & 37.39\% & $<$ 0.001   & 0.81  & [26.80\%, 47.13\%] \\
\bottomrule
\multicolumn{7}{l}{$\dagger$ Bonferroni's method comparing a family of 10 estimates}
\end{tabular}
\caption{Comparisons of the weighted mean difference in the percentage of headlines believed or willing to be shared when participants chose to view versus not view LLM fact-checking information, split by group, headline veracity, and veracity judgment of the LLM.}
\label{tab:opt-in-all}
\end{table}

\clearpage
\section{Accuracy of different prompt methods}
\label{sec:prompt_eng}

To investigate the accuracy of different prompting methods, we conducted three additional experiments in 2024 to test ChatGPT-3.5's ability to correctly predict the veracity of our headline stimuli.
Below we briefly introduce their setups:

\begin{enumerate}[itemsep=0em]\addtocounter{enumi}{-1}
    \item \textbf{Original prompt via web in 2023}: This is the original, manual approach utilized to generate the fact-checking information used in our experiment.
    \item \textbf{Original prompt via API in 2024}: We reproduced the original prompt with the OpenAI application programming interface (API) available in 2024.
    \item \textbf{Forced binary via API in 2024}: The model is forced to report a judgment of either ``True'' or ``False'' and nothing else.
    \item \textbf{Forced binary $+$ rationale via API in 2024}: The model is forced to report a judgment of either ``True'' or ``False'' as well as include the rationale for its judgment.
\end{enumerate}

Approach \#1 evaluates differences between using the general public-facing website and the programmable API options.
When we performed the original experiment in 2023, ChatGPT was only available through the website.
The web version of the model has a system prompt that defines the chatbot's default behavior.
However, the system prompt is not publicly available.
The API, on the other hand, allows us to define the system prompt ourselves, giving us better control over the experiment setup.
Approach \#2 attempts to capture a binary design that has been proposed within the literature\cite{Hoes2023GPTrateClaims}, while Approach \#3 builds on Approach \#2 by investigating whether asking the model to include a rationale for its judgments leads to clearer thinking and more accurate responses.\footnote{Per OpenAI's official prompt engineering guide: \url{https://platform.openai.com/docs/guides/prompt-engineering}.}

\begin{table}[h!]
    \centering
    \caption{Counts of ChatGPT's judgments across different prompts.
    For each approach, from left to right, we report the prompt style, interface, time of the experiment, ground-truth veracity of the headlines, numbers of ``True,'' ``Unsure,'' and ``False'' judgments, percentage of ``Unsure'' responses, and the accuracy and F1 scores of ChatGPT (excluding ``Unsure'' responses).}
    \resizebox{\columnwidth}{!}{
    \begin{tabular}{llll|lrrr|rrr}
         \toprule
         \textbf{Approach}  & \textbf{Prompt style} & \textbf{Interface} & \textbf{Time} & \textbf{Veracity} & \textbf{True} & \textbf{Unsure} & \textbf{False} & \textbf{\% Unsure} & \textbf{Accuracy} & \textbf{F1}\\
         \midrule
         \multirow{2}{*}{\#0} & \multirow{2}{*}{Original} & \multirow{2}{*}{Web} & \multirow{2}{*}{2023} & True & 3 & 13 & 4  & \multirow{2}{*}{37.5\%} & \multirow{2}{*}{0.84} & \multirow{2}{*}{0.90}\\
         & &  &  & False & 0 & 2 & 18 & & & \\
         \hline
         \multirow{2}{*}{\#1} & \multirow{2}{*}{Original} & \multirow{2}{*}{API} & \multirow{2}{*}{2024} & True & 1 & 19 & 0 & \multirow{2}{*}{77.5\%} & \multirow{2}{*}{1.00} & \multirow{2}{*}{1.00} \\
         & &  &  & False & 0 & 12 & 8 & & &\\
         \hline
         \multirow{2}{*}{\#2} & \multirow{2}{*}{Binary} & \multirow{2}{*}{API} & \multirow{2}{*}{2024} & True & 7 & 0 & 13 & \multirow{2}{*}{0\%} & \multirow{2}{*}{0.63} & \multirow{2}{*}{0.71} \\
         & &  &  & False & 2 & 0 & 18 & & &\\
         \hline
         \multirow{2}{*}{\#3} & \multirow{2}{*}{Binary + rationale} & \multirow{2}{*}{API} & \multirow{2}{*}{2024} & True & 8 & 0 & 12 & \multirow{2}{*}{0\%} & \multirow{2}{*}{0.65} & \multirow{2}{*}{0.72} \\
         & &  &  & False & 2 & 0 & 18 & & &\\
         \bottomrule
    \end{tabular}
    }
    \label{tab:prompt_engineering_raw_counts}
\end{table}

In Table~\ref{tab:prompt_engineering_raw_counts}, we report the accuracy and F1 scores of ChatGPT's judgments across the four prompt approaches in terms of identifying false headlines.
To calculate these metrics for Approaches \#0 and \#1, we ignore the ``Unsure'' responses, as this label does not conform to standard accuracy measures.
Accuracy is defined as the portion of correct judgments among all cases and reflects the overall performance of ChatGPT in different setups.
The F1 score is the harmonic mean of precision and recall and serves as another metric to quantify the performance of ChatGPT in identifying false news headlines.

Excluding ``Unsure'' headline responses, we find that ChatGPT was more accurate with Approach \#1 as compared to Approach \#0.
However, Approach \#1 had a much higher number of ``Unsure'' responses (77.5\% of the headlines versus 37.5\% for Approach \#0).
Approaches \#2 forced ChatGPT to dichotomize the unsure cases,  yielding lower accuracy.
Asking ChatGPT to generate rationale together with the judgment (Approach \#3) improved the accuracy marginally. 

Caution is necessary when generalizing these findings to AI-based fact-checking accuracy at scale; a robust evaluation would require a much larger number of test cases\cite{Quelle2023Oct, Hoes2023GPTrateClaims}. 
Recent advancements employing retrieval-augmented generation approaches achieve better performance across a broader range of claim topics and modalities\cite{Zhou2024Muse}. 
While research continues rapidly in improving the accuracy of these models, AI model accuracy will still be constrained when encountering new information that was not included in training data. 
The main contribution of our study is not to benchmark the model's accuracy but to investigate how people interact with and respond to this information, contextualized by its accuracy. 

With these caveats, our results suggest that forcing conventional fact-checking responses (by reducing uncertainty) leads to more erroneous assessments. Therefore the potential risks of AI-based fact checks  highlighted in our experiment may not be easily addressed by prompt engineering efforts. 

\clearpage
\section{Survey questions and participant flow}

Here we include all survey questions in the order they are asked, as well as their associated response options and additional information about participant flow.

Participants begin by reading a consent form and are then asked the following questions.

\begin{description}
    \item[Q1] \textbf{Question:} After reading the information sheet, do you agree to participate in this study? \\
    \textbf{Response Options:} ``Yes'' OR ``No'' \\
    \textbf{Comments:} Participants who answered ``No'' were screened out.

    \item[Q2] \textbf{Question:} We care about the quality of the data we collect. Do you commit to providing your best and honest answers to every question in this survey? \\
    \textbf{Response Options:} ``I will provide my best answers'' OR ``I will not be able to provide my best answers'' \\
    \textbf{Comments:} Participants who answered ``I will not be able to provide my best answers'' were screened out.

    \item[Q3] \textbf{Question:} What is your year of birth? \\
    \textbf{Response Options:} A box for entering numerical values was provided. \\
    \textbf{Comments:} Participants who reported being younger than 18 years old were screened out. Non-numerical values could not be entered.

    \item[Q4] \textbf{Question:} Do you currently live in the United States? \\
    \textbf{Response Options:} ``Yes'' OR ``No'' \\
    \textbf{Comments:} Participants who answered ``No'' were screened out.

    \item[Q5] \textbf{Question:} What is your gender? \\
    \textbf{Response Options:} ``Male'' OR ``Female'' OR ``Other'' OR ``Prefer not to answer'' \\
    \textbf{Comments:} Participants who selected ``Other'' were provided with a box to fill.

    \item[Q6] \textbf{Question:} What is your racial or ethnic background? (Check all that apply) \\
    \textbf{Response Options:} ``Black or African American,'' ``American Indian or Alaska Native,'' ``Asian,'' ``Native Hawaiian or Pacific Islander,'' ``Hispanic or Latino/a,'' ``Other''  \\
    \textbf{Comments:} Participants who selected ``Other'' were provided with a box to fill.

    \item[Q7] \textbf{Question:} Please indicate the answer that includes your annual household income. \\
    \textbf{Response Options:} ``Less than \$10,000'' OR ``\$10,000 to \$14,999'' OR ``\$15,000 to \$24,999'' OR ``\$25,000 to \$49,999'' OR ``\$50,000 to \$99,999'' OR ``\$100,000 to \$149,999'' OR ``\$150,000 or more'' \\
    \textbf{Comments:} 

    \item[Q8] \textbf{Question:} In which state do you currently reside? \\
    \textbf{Response Options:} All 50 US states were provided as individual options, as well as ``District of Columbia,'' ``Puerto Rico,'' and ``I do not reside in the United States'' \\
    \textbf{Comments:} Participants who selected ``I do not reside in the United States'' were screened out.

    \item[Q9] \textbf{Question:} What is the highest level of education you have completed? \\
    \textbf{Response Options:} ``Less than high school'' OR ``High school or equivalent (diploma or GED)'' OR ``Some college but no degree'' OR ``Associate degree in college (2 years)'' OR ``Bachelor degree in college (4 years)'' OR ``Master's degree'' OR ``Doctoral degree'' OR ``Professional degree (JD, MD)'' \\
    \textbf{Comments:}

    \item[Q10] \textbf{Question:} Please tell us if you use any of the following social media sites. (Check all that apply). \\
    \textbf{Response Options:} ``Facebook,'' ``TikTok,'' ``WhatsApp,'' ``Twitter,'' ``Reddit,'' ``Telegram,'' ``Instagram,'' ``4chan,'' ``Truth Social,'' ``Snapchat,'' ``Pinterest,'' ``Rumble,'' ``Tumblr,'' ``Twitch,'' ``Parler,'' ``YouTube,'' ``LinkedIn,'' ``Gab'' \\
    \textbf{Comments:} 

    \item[Q11] \textbf{Question:} How frequently do you access the following sources to obtain news via the internet? \\
    \textbf{Sources:} ``Search engines (e.g. Google, Bing),'' ``Social media (e.g. Facebook, Twitter),'' ``News Aggregator (e.g., Google News, Flipboard),'' ``News websites (e.g., nyt.com, vox.com)'' \\
    \textbf{Response Options:} Seven point Likert Scale. Options: ``Never'' (1), ``About once every few months'' (2), ``About once a month'' (3), ``About once a week'' (4), ``A few times a week'' (5), ``About once a day'' (6), ``A few times a day'' (7). \\
    \textbf{Comments:} 

    \item[Q12] \textbf{Question:} Generally speaking, do you usually think of yourself as a Republican, a Democrat, an Independent, or what? \\
    \textbf{Response Options:} ``Republican,'' ``Democrat,'' ``Independent,'' ``Other,'' ``No preference,'' ``Don't know'' \\
    \textbf{Comments:} Participants who selected ``Other'' were provided with a box to fill. Participants who answered ``Republican'' or ``Democrat'' were then asked question 13. Those who provided other responses skipped Q13 and were directed to Q14.

    \item[Q13] \textbf{Question:} Would you call yourself a strong Republican (Democrat) or not a very strong Republican (Democrat)? \\
    \textbf{Response Options:} ``Strong'' OR ``Somewhat strong'' \\
    \textbf{Comments:} The words ``Republican'' and ``Democrat'' were not shown together in the question. Instead, one or the other was dynamically included to reflect the participant's response to Q12. Only asked if a participant answered ``Republican'' or ``Democrat'' for Q12.

    \item[Q14] \textbf{Question:} Do you think of yourself as closer to the Republican or Democratic Party? \\
    \textbf{Response Options:} ``Republican party'' OR ``Democratic party'' OR ``Neither'' OR ``Don't Know'' \\
    \textbf{Comments:} Only asked if a participant did not answer ``Republican'' or ``Democrat'' for Q12.

    \item[Q15] \textbf{Question:} To what extent do you agree with the following statements? \\
    \textbf{Statements:} ``I fear artificial intelligence,'' ``I trust artificial intelligence,'' ``Artificial intelligence will destroy humankind,'' ``Artificial intelligence will benefit humankind''\\
    \textbf{Response Options:} Seven point Likert Scale. Options: ``Strongly disagree'' (1), ``disagree'' (2), ``Somewhat disagree'' (3), ``Neither agree nor disagree'' (4), ``Somewhat agree'' (5), ``Agree'' (6), ``Strongly agree'' (7).  \\
    \textbf{Comments:} 

    \item[Q16] \textbf{Question:} In the past month, how often did you reference fact-checking websites (e.g., snopes.com or politifact.org) to check whether a headline you read is true? \\
    \textbf{Response Options:} ``A few times a week'' OR ``About once a week'' OR ``A few times every week'' OR ``At least once a day'' \\
    \textbf{Comments:} 

    \item[ChatGPT Introduction:] ChatGPT is an advanced language model developed by OpenAI. It is designed to generate human-like responses to questions and can be used for various purposes, including fact-checking. Simply ask ChatGPT a question, and it will provide you with an answer based on the information it was trained on. However, it's important to note that ChatGPT is not perfect and may not always provide accurate information. \\
    \textbf{Comments:}

    \item[Q17] \textbf{Question:} Have you used AI-powered tools such as ChatGPT before? \\
    \textbf{Response Options:} ``Yes'' OR ``No'' \\
    \textbf{Comments:} Participants who answered ``Yes'' were then asked questions Q18--Q21, otherwise these questions were skipped.

    \item[Q18] \textbf{Question:} In the past 30 days, how often have you used AI-powered tools such as ChatGPT? \\
    \textbf{Response Options:} ``About once,'' ``A couple of times,'' ``Several times,'' ``A few times every week,'' ``At least once every day'' \\
    \textbf{Comments:} Only asked if ``Yes'' was the answer to Q17.

    \item[Q19] \textbf{Question:} Have you ever used AI-powered tools such as ChatGPT to fact-check news reports before? \\
    \textbf{Response Options:} ``Yes'' OR ``No'' \\
    \textbf{Comments:} Only asked if ``Yes'' was the answer to question Q17.

    \item[Q20] \textbf{Question:} To what extent do you agree with the following statements? \\
    \textbf{Statements:} ``ChatGPT performs really well when fact-checking news reports,'' ``ChatGPT outperforms existing fact-checking services,'' ``Fact-checking answers provided by ChatGPT can change my mind,'' ``Fact-checking answers provided by ChatGPT are objective,'' ``Fact-checking answers provided by ChatGPT are trustworthy,'' ``Fact-checking answers provided by ChatGPT are informative.''\\
    \textbf{Response Options:} Seven point Likert Scale. Options: ``Strongly disagree'' (1), ``Disagree'' (2), ``Somewhat disagree'' (3), ``Neither agree nor disagree'' (4), ``Somewhat agree'' (5), ``Agree'' (6), ``Strongly agree'' (7).  \\
    \textbf{Comments:} Only asked if ``Yes'' was the answer to question Q17.

    \item[Q21] \textbf{Question:} To what extent do you agree with the following statements? \\
    \textbf{Statements:}  ``I would like to use ChatGPT to verify information in the future on a regular basis,'' ``I hope social media (e.g., Facebook, Twitter) incorporate ChatGPT fact-checking in their service,'' ``I hope search engines (e.g., Google, Bing) incorporate ChatGPT fact-checking in their service,'' ``I hope news aggregation apps (e.g., Apple News, Flipboard) incorporate ChatGPT fact-checking in their service,'' ``I will recommend ChatGPT fact-checking services to other people.''\\
    \textbf{Response Options:} Seven point Likert Scale. Options: ``Strongly disagree'' (1), ``Disagree'' (2), ``Somewhat disagree'' (3), ``Neither agree nor disagree'' (4), ``Somewhat agree'' (5), ``Agree'' (6), ``Strongly agree'' (7).  \\
    \textbf{Comments:} Only asked if ``Yes'' was the answer to question Q17.

    \item[Experiment Instructions:] Now we are going to show you approximately 40 news headlines that have appeared recently on the Internet and in media.
    \\\textit{Belief group only}: Please let us know if you think they are true or false.\\\textit{Sharing group only}: Please let us know whether you would consider sharing it.\\
    \textbf{Comments:} Participants in the fact-checking conditions were also provided with the following instructions directly below the above.\\\textit{LLM-forced}: You will also be provided with ChatGPT-generated fact-checking information for each headline.\\\textit{LLM-optional}: If you are unsure, you have the option to ask a ChatGPT fact-checker for help.\\\textit{Human fact check}: You will also be provided with fact-checking information for each headline.
    
    \item[Q22-Q43] \textbf{Question:} 41 randomly ordered headline stimuli (including 1 attention check item). \\\textit{Belief Question}: ``Do you believe the claim in the headline to be true?'' \\ \textit{Sharing Question}: ``Would you consider sharing this story online (for example, through Facebook or Twitter)?'' \\
    \textbf{Response Options:} ``Yes'' OR ``No''  \\
    \textbf{Comments:} Depending on one's experimental condition this question was accompanied by either no fact checks, human-generated fact checks, AI-generated fact checks that participants were forced to view, or the same AI-generated fact checks that participants were given the option to view.\\ \textit{Question to view optional AI fact checks}: ``Would you like ChatGPT to help you verify the headline?''\\ \textit{Options}: ``Yes'' OR ``No''

    \item[Q44] \textbf{Question:} Did you search the internet for more information about the headlines you were asked about? \\
    \textbf{Response Options:} ``Yes'' OR ``No'' \\
    \textbf{Comments:}

    \item[Q45] \textbf{Question:} To what extent do you still agree with the following statements? \\
    \textbf{Statements:} ``I fear artificial intelligence,'' ``I trust artificial intelligence,'' ``Artificial intelligence will destroy humankind,'' ``Artificial intelligence will benefit humankind.'' \\
    \textbf{Response Options:} Seven point Likert Scale. Options: ``Strongly disagree'' (1), ``Disagree'' (2), ``Somewhat disagree'' (3), ``Neither agree nor disagree'' (4), ``Somewhat agree'' (5), ``Agree'' (6), ``Strongly agree'' (7). \\
    \textbf{Comments:} Participants only saw this question in the LLM-optional and LLM-forced conditions.

    \item[Q46] \textbf{Question:} Based on your experience with ChatGPT in this study, to what extent do you agree with the following statements? \\
    \textbf{Statements:} ``ChatGPT performs really well when fact-checking news reports,'' ``ChatGPT outperforms existing fact-checking services,'' ``Fact-checking answers provided by ChatGPT have changed my mind,'' ``Fact-checking answers provided by ChatGPT are objective,'' ``Fact-checking answers provided by ChatGPT are trustworthy,'' ``Fact-checking answers provided by ChatGPT are informative.''
    \textbf{Response Options:} Seven point Likert Scale. Options: ``Strongly disagree'' (1), ``Disagree'' (2), ``Somewhat disagree'' (3), ``Neither agree nor disagree'' (4), ``Somewhat agree'' (5), ``Agree'' (6), ``Strongly agree'' (7).  \\
    \textbf{Comments:} Participants only saw this question if in the LLM-optional and LLM-forced conditions.

    \item[Q47] \textbf{Question:} Based on your experience with ChatGPT in this study, to what extent do you agree with the following statements? \\
    \textbf{Statements:} ``I would like to use ChatGPT to verify information in the future on a regular basis,'' ``I hope social media (e.g., Facebook, Twitter) incorporate ChatGPT fact-checking in their service,'' ``I hope search engines (e.g., Google, Bing) incorporate ChatGPT fact-checking in their service,'' ``I hope news aggregation apps (e.g., Apple News, Flipboard) incorporate ChatGPT fact-checking in their service,'' ``I will recommend ChatGPT fact-checking services to other people.'' \\
    \textbf{Response Options:} Seven point Likert Scale. Options: ``Strongly disagree'' (1), ``Disagree'' (2), ``Somewhat disagree'' (3), ``Neither agree nor disagree'' (4), ``Somewhat agree'' (5), ``Agree'' (6), ``Strongly agree'' (7). \\
    \textbf{Comments:} Participants only saw this question if in the LLM-optional and LLM-forced conditions.

    \item[Post-stimuli message:] Now we have just a few more questions about you.\\
    \textbf{Comments:}
    
    \item[Q48] \textbf{Question:} Did you vote in the 2020 Presidential election? \\
    \textbf{Response Options:} ``Yes'' OR ``No'' \\
    \textbf{Comments:} Participants who selected ``Yes'' were then asked Q49.

    \item[Q49] \textbf{Question:} Who did you vote for in the 2020 Presidential election? \\
    \textbf{Response Options:} ``Donald Trump/Mike Pence (Republican Party)'' OR ``Joe Biden/Kamala Harris (Democratic Party)'' OR ``Some other candidate'' \\
    \textbf{Comments:} Only asked if the answer to Q48 was ``Yes.''

    \item[Q50] \textbf{Question:} Did you vote in the 2022 midterm election? \\
    \textbf{Response Options:} ``Yes'' OR ``No'' \\
    \textbf{Comments:}

    \item[Q51-Q52] \textbf{Question:} Affective polarization feelings thermometer (voters and party). \\
    \textbf{Response Options:} Sliders allowed participants to select a value between 0--100 for four different items: ``Republican voters,'' ``Democrat voters,'' ``Republican party,'' ``Democrat party.'' \\
    \textbf{Comments:} Please reference the Qualtrics survey file in our preregistration\cite{DeVerna2022cgptOSF} for the exact wording of this question.

    \item[Debriefing:] All participants were informed about the study purpose in more detail, notified of the experimental group they participated within, and were again provided with contact information for the authors, should they have futher questions. \\
    \textbf{Comments:} Please reference the Qualtrics survey file in our preregistration\cite{DeVerna2022cgptOSF} for the exact wording in our debriefing.
\end{description}

\clearpage

\section{Headlines and fact checks}
\label{sec:headlines-fcs}

Here we present the text of each headline, along with its ideological leanings, and both AI- and human-generated fact-checking information. 
We also include the veracity of each headline and our coders' evaluations of the AI-generated fact checks, categorizing those judgments as true, false, or unsure.
AI judgments that are correct/incorrect are marked as \faCheckCircle/\faTimesCircle~for convenience (all ``Unsure'' responses are marked as incorrect).
Headlines ending in an ellipses (``...'') are truncated within the stimuli and the text presented here represents exactly the text that participants saw during the experiment.
Publication dates of stimuli articles were not provided by the developers of these stimuli (see the \nameref{sec:stimuli_curation} section for more details).
To gather them, we visit each article at the time of writing and extract the date directly from the article itself.
If the publication date was unavailable---due to inaccessible domains or articles, or because the date was not listed---we sought this information from internet archive sites or the webpage's metadata.
In instances where dates were obtained through these alternative means, we have marked them with an asterisk (*).

\begin{description}

    \item[Headline 1:] Blacks 4 Trump Group Caught Using Blackface To Pretend They Have Black Members
    \\
    \textbf{Publication Date:}  October 14, 2020
    \\
    \textbf{Source:} \url{halfwaypost.com}
    \\
    \textbf{Veracity:} False
    \\
    \textbf{Leaning:} Pro-Democrat
    \\
    \textbf{AI fact check:} I am unable to verify the credibility of this claim as it depends on the source of the information and the context surrounding it. However, it is important to note that blackface is a form of racist behavior that has been used historically to mock and stereotype black people, and it is generally considered unacceptable in modern society. If the claim is true, it is likely that the group's actions would be met with strong condemnation.
    \\
    \textbf{AI judgment label:} Unsure \faTimesCircle
    \\
    \textbf{Human fact check:} The above claim is False. This claim came from the Halfway Post, a website that publishes satire. This site states that it publishes ``daily jokes and satirical hot takes on the latest political events''. Therefore, this post was never intended to be interpreted as fact.

    \item[Headline 2:] Biden Buys Film Rights To New `MAD' Movie Just To Piss Off Trump
    \\
    \textbf{Publication Date:} N/A (domain unreachable)
    \\
    \textbf{Source:} \url{flake.news}
    \\
    \textbf{Veracity:} False
    \\
    \textbf{Leaning:} Pro-Democrat
    \\
    \textbf{AI fact check:} It is highly unlikely that this claim is true. As President of the United States, Joe Biden is a public figure, and any significant actions he takes, such as buying film rights, would be widely reported by reputable news sources. Additionally, it would be a highly unusual action for a President to take and would not align with the responsibilities and duties of the position. It is more likely that this claim is false, or a fabrication and it would be wise to verify the information from credible sources before sharing it.
    \\
    \textbf{AI judgment label:} False \faCheckCircle
    \\
    \textbf{Human fact check:} The above claim is False. This claim came from Flake News, a site that describes itself as a publisher of ``political satire and humor from the most witty writers across the United States.'' As such, the stories posted on this site are satirical and not intended to be interpreted as fact.

    \item[Headline 3:] Staring At Hard Times, Tucker Carlson May Be Forced To Sell Bow Tie Collection
    \\
    \textbf{Publication Date:} June 15, 2020
    \\
    \textbf{Source:} \url{dailykos.com}
    \\
    \textbf{Veracity:} False
    \\
    \textbf{Leaning:} Pro-Democrat
    \\
    \textbf{AI fact check:} It is unlikely that this claim is true. While I don't have the specific information about Tucker Carlson's financial situation, it is unlikely that he would be forced to sell his bow tie collection in order to make ends meet. The statement seems to be more of a joke or sarcastic comment rather than a serious claim. Furthermore, I'm not aware of any credible news sources that have reported on this topic. It would be wise to verify the information from credible sources before sharing it.
    \\
    \textbf{AI judgment label:} False \faCheckCircle
    \\
    \textbf{Human fact check:} The above claim is False. Tucker Carlson said that he had stopped wearing bow ties 10 years ago due to abuse. In a 2017 interview with the Los Angeles Times, he reported thinking, ``You know, this is pointlessly provocative...I’m happy to defend my views but I don’t want to have to defend my neckwear every day.''

    \item[Headline 4:] Steve Bannon Is Now Selling ``Freedom Yogurt'' Made Of White People's Semen
    \\
    \textbf{Publication Date:} October 9, 2021
    \\
    \textbf{Source:} \url{dailykos.com}
    \\
    \textbf{Veracity:} False
    \\
    \textbf{Leaning:} Pro-Democrat
    \\
    \textbf{AI fact check:} It is highly unlikely that this claim is true. Not only is the claim highly offensive and disrespectful, it is also illegal to sell food products made from human semen. Furthermore, this statement is not only highly offensive, but it is also illegal. This kind of claim is likely to be false or a fabrication, and it would be wise to verify the information from credible sources before sharing it. Additionally, spreading false or offensive information can cause harm to individuals and communities.
    \\
    \textbf{AI judgment label:} False \faCheckCircle
    \\
    \textbf{Human fact check:} The above claim is False This claim came from the Halfway Post, a website that publishes satire. This site states that it publishes ``daily jokes and satirical hot takes on the latest political events.'' Therefore, this post was never intended to be interpreted as fact.

    \item[Headline 5:] SUPREME COURT GRANTS BLACK MAN ``40 ACRES OF LAND AND A MULE''
    \\
    \textbf{Publication Date:} June 27, 2019*
    \\
    \textbf{Source:} \url{worldnewsdailyreport.com}
    \\
    \textbf{Veracity:} False
    \\
    \textbf{Leaning:} Pro-Democrat
    \\
    \textbf{AI fact check:} It is highly unlikely that this claim is true. The idea of granting ``40 acres of land and a mule'' to black Americans as reparations for slavery was proposed during the Civil War but was never implemented. The Supreme Court does not have the authority to grant land or property as it is not a legislative body, it is a judicial one, and its role is to interpret the law, not create it. Additionally, this would be a highly unusual action for the Supreme Court to take and would not align with the responsibilities and duties of the position. It is more likely that this claim is false, or a fabrication and it would be wise to verify the information from credible sources before sharing it.
    \\
    \textbf{AI judgment label:} False \faCheckCircle
    \\
    \textbf{Human fact check:} The above claim is False. This claim is from the World News Daily Report, a satirical website that publishes fictitious news for fun. This photograph depicts Ricky Dale Wyatt, who was exonerated thanks to DNA evidence after being wrongly convicted. His case had nothing to do with 40 acres of land.

    \item[Headline 6:] Trump Is Suing His Grandchildren For Violating NDA's He Made Them Sign As Infants
    \\
    \textbf{Publication Date:} May 5, 2021
    \\
    \textbf{Source:} \url{halfwaypost.com}
    \\
    \textbf{Veracity:} False
    \\
    \textbf{Leaning:} Pro-Democrat
    \\
    \textbf{AI fact check:} It is highly unlikely that this claim is true. First, it would be illegal to make infants sign a legal contract, as they are not of legal age to do so. Additionally, even if the grandchildren were of legal age, it would be highly unusual for a grandfather to sue his own grandchildren, especially for something as trivial as violating a non-disclosure agreement. Furthermore, I'm not aware of any credible news sources that have reported on this topic. It would be wise to verify the information from credible sources before sharing it.
    \\
    \textbf{AI judgment label:} False \faCheckCircle
    \\
    \textbf{Human fact check:} The above claim is False. This claim came from the Halfway Post, a website that publishes satire. This site states that it publishes ``daily jokes and satirical hot takes on the latest political events.'' Therefore, this post was never intended to be interpreted as fact.

    \item[Headline 7:] Trump Orders Americans To Stop Eating Chinese Food
    \\
    \textbf{Publication Date:} August 24, 2019
    \\
    \textbf{Source:} \url{dailykos.com}
    \\
    \textbf{Veracity:} False
    \\
    \textbf{Leaning:} Pro-Democrat
    \\
    \textbf{AI fact check:} It is unlikely that this claim is true. The President of the United States does not have the authority to order citizens to stop eating a certain type of food. Additionally, this would be a highly unusual action for the President to take and would not align with the responsibilities and duties of the position. Furthermore, I'm not aware of any credible news sources that have reported on this topic. It would be wise to verifv the information from credible sources before sharing it.
    \\
    \textbf{AI judgment label:} False \faCheckCircle
    \\
    \textbf{Human fact check:} The above claim is False. This claim originated from The Daily Noose under their ``satire'' tab, and then reposted on the Daily Kos website. This claim was not intended to be interpreted as fact. No other reputable news sources have reported Trump ordering Americans to stop eating Chinese food.

    \item[Headline 8:] Trump Routinely Forced Staffers To Shred And Eat White House Documents
    \\
    \textbf{Publication Date:} February 6, 2022
    \\
    \textbf{Source:} \url{halfwaypost.com}
    \\
    \textbf{Veracity:} False
    \\
    \textbf{Leaning:} Pro-Democrat
    \\
    \textbf{AI fact check:} It is highly unlikely that this claim is true. It would be illegal to destroy government documents and it would be highly unprofessional and unethical to force staff members to participate in such an act. Additionally, this would be a highly unusual action for the President to take and would not align with the responsibilities and duties of the position. Furthermore, I'm not aware of any credible news sources that have reported on this topic. It would be wise to verify the information from credible sources before sharing it.
    \\
    \textbf{AI judgment label:} False \faCheckCircle
    \\
    \textbf{Human fact check:} The above claim is False. This post came from the Halfway Post, a website that publishes satire. The website’s home page states that the site is ``A St. Louis gazette of satirical, halfway real news.'' As such, the content on the website not truthful and not intended to be interpreted as such.

    \item[Headline 9:] Trump Threatens To Sue Founding Fathers: Resorting to his most tried-and-true tactic, Donald Trump has...
    \\
    \textbf{Publication Date:} October 22, 2019 
    \\
    \textbf{Source:} \url{dailykos.com}
    \\
    \textbf{Veracity:} False
    \\
    \textbf{Leaning:} Pro-Democrat
    \\
    \textbf{AI fact check:} It is highly unlikely that this claim is true. The founding fathers of the United States are deceased, and it is not possible to sue them in a court of law. Additionally, this would be a highly unusual action for the President to take and would not align with the responsibilities and duties of the position. Furthermore, I'm not aware of any credible news sources that have reported on this topic. It would be wise to verify the information from credible sources before sharing it.
    \\
    \textbf{AI judgment label:} False \faCheckCircle
    \\
    \textbf{Human fact check:} The above claim is False. Donald Trump has never threatened to sue the founding fathers as it is not possible to sue individuals who have been dead for a long time. This article was originally posted by The Daily Nooze, a website that produces satirical content that is not intended to be interpreted as fact.

    \item[Headline 10:] Trump Voter Wants Taco Bell Shut Down For 'Being Too Mexican'
    \\
    \textbf{Publication Date:} N/A (domain unreachable)
    \\
    \textbf{Source:} \url{alternativelyfacts.com}
    \\
    \textbf{Veracity:} False
    \\
    \textbf{Leaning:} Pro-Democrat
    \\
    \textbf{AI fact check:} It is possible that an individual with racist views may want to shut down a business such as Taco Bell due to it being associated with Mexican culture, but it's hard to tell without more context and without knowing the source of the information. However, it is important to note that such a sentiment is discriminatory, and it would be unacceptable for a business to be shut down based on racist attitudes. Furthermore, it is worth noting that the president of the United States does not have the authority to shut down businesses and it would be a highly unusual action for an individual to take.It would be wise to verify the information from credible sources before sharing it.
    \\
    \textbf{AI judgment label:} Unsure \faTimesCircle
    \\
    \textbf{Human fact check:} The above claim is False. This post comes from Alternatively Facts, a website that publishes satire. The website states that it is ``satirical in scope and intent. It provides fake news and social criticism in a satirical setting.'' As such, this claim was not intended to be interpreted as fact.

    \item[Headline 11:] Biden Orders Arrest Of Black Leaders While His Elites Mock Appalachia Flood Victims
    \\
    \textbf{Publication Date:} N/A (domain unreachable)
    \\
    \textbf{Source:} \url{trusttheq.com}
    \\
    \textbf{Veracity:} False
    \\
    \textbf{Leaning:} Pro-Republican
    \\
    \textbf{AI fact check:} It is highly unlikely that this claim is true. The President of the United States does not have the power to arrest individuals, and it would be a highly unusual action for him to take. Additionally, the President would not be able to order the arrest of specific individuals based on their race or political beliefs as it would be unconstitutional and violative of the civil rights. Furthermore, it is not credible that the president or his administration would mock flood victims, as it would be highly disrespectful and unprofessional. Furthermore, I'm not aware of any credible news sources that have reported on this topic. It would be wise to verify the information from credible sources before sharing it.
    \\
    \textbf{AI judgment label:} False \faCheckCircle
    \\
    \textbf{Human fact check:} The above claim is False. This post appeared on the website Trust The Q which is associated with the QAnon conspiracy theory. The website reports that it was created in 2018 ``by URGAFIX (at my request, I’m a FAM, lol).'' According to Fox News, QAnon has been the source of frequent misinformation relating to politics.

    \item[Headline 12:] BREAKING: Hillary Clinton Found Dead Under Suspicious Circumstances
    \\
    \textbf{Publication Date:} February 13, 2022$^*$
    \\
    \textbf{Source:} \url{dunning-kruger-times.com}
    \\
    \textbf{Veracity:} False
    \\
    \textbf{Leaning:} Pro-Republican
    \\
    \textbf{AI fact check:} I have knowledge cut off at 2021, but as of that time, Hillary Clinton was alive and well. Additionally, any significant event such as the death of a high-profile individual like Hillary Clinton would be widely reported by reputable news sources and would be confirmed by credible sources. Therefore, it is likely that this claim is false or a fabrication and it would be wise to verify the information from credible sources before sharing it. Additionally, spreading false information can cause harm to individuals and communities.
    \\
    \textbf{AI judgment label:} False \faCheckCircle
    \\
    \textbf{Human fact check:} The above claim is False. Hillary Clinton is not dead. This article appeared on the Dunning Kruger Times, a website which states ``Everything on this website is fiction.'' Furthermore, Hillary Clinton gave a public speech two days after this article was posted.

    \item[Headline 13:] Delta Force Arrests Chelsea Clinton: Thanks for your support. The content we do is quite sensitive, s...
    \\
    \textbf{Publication Date:} N/A (article taken down)
    \\
    \textbf{Source:} \url{restoredrepublic.co}
    \\
    \textbf{Veracity:} False
    \\
    \textbf{Leaning:} Pro-Republican
    \\
    \textbf{AI fact check:} It is highly unlikely that this claim is true. The Delta Force is a specialized unit of the United States military that conducts sensitive operations and is not typically involved in civilian arrests. Additionally, it would be highly unusual for the military to arrest a civilian under such circumstances, and it would be a violation of the Posse Comitatus Act, which prohibits the use of military personnel to enforce domestic laws. Furthermore, I'm not aware of any credible news sources that have reported on this topic, and it would be wise to verify the information from credible sources before sharing it.
    \\
    \textbf{AI judgment label:} False \faCheckCircle
    \\
    \textbf{Human fact check:} The above claim is False. The claim that Chelsea Clinton was arrested for child trafficking is originally from Real Raw News, a website that publishes satirical content. Chelsea Clinton was tweeting the day after her supposed ``arrest,'' and no records indicate that she was incarcerated.

    \item[Headline 14:] Kamala: 'Our Military Are Soulless Cowards. Their Budget Must Be Cut.'
    \\
    \textbf{Publication Date:} N/A (domain unreachable)
    \\
    \textbf{Source:} \url{dailyworldupdate.us}
    \\
    \textbf{Veracity:} False
    \\
    \textbf{Leaning:} Pro-Republican
    \\
    \textbf{AI fact check:} It is highly unlikely that this claim is true. Kamala Harris, as the vice-president of the United States, has a strong commitment to the country and the military. She has shown her support for the men and women in the military and their families, and it would be highly unprofessional and disrespectful for her to make a statement like this. Additionally, this would be a highly unusual action for the vice-president to take and would not align with the responsibilities and duties of the position. Furthermore, I'm not aware of any credible news sources that have reported on this topic and it would be wise to verify the information from credible sources before sharing it.
    \\
    \textbf{AI judgment label:} False \faCheckCircle
    \\
    \textbf{Human fact check:} The above claim is False. This claim comes from the Daily World Update, a website that no longer exists. When the website was active, it had a disclaimer that it was intended to be satire and that ``Everything on this website is fiction.'' There is no other record of Kamala Harris saying this.

    \item[Headline 15:] Ilhan Omar Spits On The Tomb of the Unknown Soldier
    \\
    \textbf{Publication Date:} September 19, 2019$^*$
    \\
    \textbf{Source:} \url{tatersgonnatate.com}
    \\
    \textbf{Veracity:} False
    \\
    \textbf{Leaning:} Pro-Republican
    \\
    \textbf{AI fact check:} It is highly unlikely that this claim is true. Ilhan Omar is a U.S Congress woman, and asan elected official, she would have a strong commitment to her country and its military. Additionally, spitting on the Tomb of the Unknown Soldier would be a highly disrespectful and unprofessional act, and it would be unlikely for a member of Congress to do such a thing. Furthermore, I'm not aware of any credible news sources that have reported on this topic and it would be wise to verify the information from credible sources before sharing it. Additionally, spreading false information can cause harm to individuals and communities.
    \\
    \textbf{AI judgment label:} False \faCheckCircle
    \\
    \textbf{Human fact check:} The above claim is False. This claim is from a website that no longer exists, Taters Gonna Tate. When the site was active, it had a disclaimer that it published satire: “Everything on this website is fiction”. This post was categorized in the ``Conservative Fan Fiction'' category and therefore was not intended to be interpreted as fact.

    \item[Headline 16:] Malia Obama Arrested Again, Complete Media Blackout: The truth is right in front of you.
    \\
    \textbf{Publication Date:} June 7, 2019$^*$
    \\
    \textbf{Source:} \url{dailyworldupdate.us}
    \\
    \textbf{Veracity:} False
    \\
    \textbf{Leaning:} Pro-Republican
    \\
    \textbf{AI fact check:} I have knowledge cut off at 2021, but as far as I know, there is no credible information that Malia Obama has been arrested again and there is no evidence of a media blackout. Furthermore, it would be highly unusual for a media blackout to occur insuch a high-profile case, and it would be reported by reputable news sources. Additionally, spreading false information can cause harm to individuals and communities. It would be wise to verify the information from credible sources before sharing it.
    \\
    \textbf{AI judgment label:} False \faCheckCircle
    \\
    \textbf{Human fact check:} The above claim is False. The claim that Malia Obama was arrested comes from Daily World Update, a website that publishes satire. The disclaimer on this site states ``Everything on this website is fiction... If you believe that it is real, you should have your head examined.''

    \item[Headline 17:] QAnon Wants You To Pee On Your Kids So Democrats Can't Smell Their Youthful Blood
    \\
    \textbf{Publication Date:} February 1, 2021
    \\
    \textbf{Source:} \url{halfwaypost.com}
    \\
    \textbf{Veracity:} False
    \\
    \textbf{Leaning:} Pro-Republican
    \\
    \textbf{AI fact check:} This claim is not only highly offensive and disrespectful but also completely unfounded and untrue. It is not associated with the QAnon movement, which is a far-right conspiracy theory that has been associated with false and dangerous information. Additionally, it is illegal and highly inappropriate to engage in such activities with children and would be considered abuse. It is important to be critical of information from sources that are not credible and to verify information from reputable sources before sharing it. This claim should be disregarded as it is patently false and offensive.
    \\
    \textbf{AI judgment label:} False \faCheckCircle
    \\
    \textbf{Human fact check:} The above claim is False. This claim came from the Halfway Post, a website that publishes satire. This site states that it publishes ``daily jokes and satirical hot takes on the latest political events.'' Therefore, this post was never intended to be interpreted as fact.

    \item[Headline 18:] Special Forces Arrest Deep State Dr. Anthony Fauci: US Special Forces on Saturday scored a major victory in the war...
    \\
    \textbf{Publication Date:} April 10, 2022
    \\
    \textbf{Source:} \url{realrawnews.com}
    \\
    \textbf{Veracity:} False
    \\
    \textbf{Leaning:} Pro-Republican
    \\
    \textbf{AI fact check:} It is highly unlikely that this claim is true. Dr. Anthony Fauci is a prominent public health official and the director of the National Institute of Allergy and Infectious Diseases. He is a respected figure in the field of medicine and has been a key advisor to multiple presidents on public health matters, especially during the COVID-19 pandemic. The notion that special forces would arrest him, especially for something related to the ``Deep State'' is baseless and not supported by any credible sources. Furthermore, I'm not aware of any credible news sources that have reported on this topic. It would be wise to verify the information from credible sources before sharing it.
    \\
    \textbf{AI judgment label:} False \faCheckCircle
    \\
    \textbf{Human fact check:} The above claim is False. The director at the National Institutes of Health confirmed that Fauci was not arrested. Indeed, he appeared live on ABC news the day after the arrest supposedly occurred. The claim is from Real Raw News, a website that publishes false content and has a disclaimer that it ``contains humor, parody, and satire.''

    \item[Headline 19:] WOW! Biden Caught Fake Driving - Someone Else Is Steering Vehicle - It Was All a Stunt! - VIDEO and PICS
    \\
    \textbf{Publication Date:} May 19, 2021
    \\
    \textbf{Source:} \url{thegatewaypundit.com}
    \\
    \textbf{Veracity:} False
    \\
    \textbf{Leaning:} Pro-Republican
    \\
    \textbf{AI fact check:} It is highly unlikely that this claim is true. The President of the United States is a public figure and his movements are closely monitored by the Secret Service and the press. It would be highly unusual for the President to participate in a staged event of this nature and it would be reported by credible news sources. Additionally, the President of the United States is not legally allowed to drive due to security reasons, and he is always driven by Secret Service agents. Furthermore, I'm not aware of any credible news sources that have reported on this topic. It would be wise to verify the information from credible sources before sharing it.
    \\
    \textbf{AI judgment label:} False \faCheckCircle
    \\
    \textbf{Human fact check:} The above claim is False. A frame-by-frame analysis of this video shows Biden’s passenger was not steering the truck, but adjusting a camera that was resting on the dashboard. According to Melissa Miller, a spokesperson at Ford, the vehicle Biden drove had only one steering wheel.

    \item[Headline 20:] YOKO ONO: ``I HAD AN AFFAIR WITH HILLARY CLINTON IN THE '70S''
    \\
    \textbf{Publication Date:} N/A (domain unreachable)
    \\
    \textbf{Source:} \url{worldnewsdailyreport.com}
    \\
    \textbf{Veracity:} False
    \\
    \textbf{Leaning:} Pro-Republican
    \\
    \textbf{AI fact check:} It is highly unlikely that this claim is true. Yoko Ono and Hillary Clinton were both involved in different fields and it is unlikely that they would have had the opportunity to meet, let alone have an affair in the 1970s. Furthermore, I'm not aware of any credible news sources that have reported on this topic, and Yoko Ono has not made any statements or claims of this nature. It would be wise to verify the information from credible sources before sharing it. Additionally spreading false or offensive information can cause harm to individuals and communities.
    \\
    \textbf{AI judgment label:} False \faCheckCircle
    \\
    \textbf{Human fact check:} The above claim is False. This claims is from the World News Faily Report, is a satirical website that publishes fictitious news for fun. The website itself has a disclaimer that it ``assumes all responsibility for the satirical nature of its articles and for the fictional nature of their content.''

    \item[Headline 21:] Biden administration cancels another \$3.9 billion in student loan debt for former for-profit college students
    \\
    \textbf{Publication Date:} August 16, 2022
    \\
    \textbf{Source:} \url{cnn.com}
    \\
    \textbf{Veracity:} True
    \\
    \textbf{Leaning:} Pro-Democrat
    \\
    \textbf{AI fact check:} As of my knowledge cut off, I am not aware of any recent announcements from the Biden administration cancelling \$3.9 billion in student loan debt for former for-profit college students. However, it is important to note that the Biden administration has made student loan forgiveness and cancelling a priority in their agenda, and they have taken steps to cancel student loan debt for certain groups of borrowers such as those who attended defrauded for-profit colleges. Additionally, there have been several proposed bills on student loan forgiveness and cancelling that have been introduced in Congress, and it is possible that some of these bills call for cancelling \$3.9 billion in student loan debt for former for-profit college students. However, it is important to verify the information from credible sources before sharing it, and to understand that the process of cancelling student loan debt is a complex one and it may take some time for the administration to implement their plans.
    \\
    \textbf{AI judgment label:} Unsure \faTimesCircle
    \\
    \textbf{Human fact check:} The above claim is True. CNN is a reputable and trustworthy news organization. Addressing disparities and financial challenges within the education system has been a priority of the Biden administration.

    \item[Headline 22:] Biden has tamped down talk of a primary challenge, for now
    \\
    \textbf{Publication Date:} September 2, 2022
    \\
    \textbf{Source:} \url{washingtonpost.com}
    \\
    \textbf{Veracity:} True
    \\
    \textbf{Leaning:} Pro-Democrat
    \\
    \textbf{AI fact check:} As of my knowledge cut off, I am not aware of any recent announcements from the Biden administration regarding primary challenges, but it is important to note that during a presidential term, there is always a possibility of a primary challenge, particularly as the mid-term elections approach. However, the probability of a primary challenge is influenced by various factors such as the President's approval rating, the state of the economy and the political climate, and the support of the President within his party. Additionally, it is important to note that the primary process is a democratic one, and the decision to run or not to run is ultimately up to the individual. It would be wise to verify the information from credible sources before sharing it, and to understand that the political landscape is constantly changing and it's hard to predict the future.
    \\
    \textbf{AI judgment label:} Unsure \faTimesCircle
    \\
    \textbf{Human fact check:} The above claim is True. The Washington Post is a reputable and trustworthy news organization. At the time of reporting, despite rumors of Democrats challenging President Joe Biden's second presidential campaign circulating in private, it appeared that these discussions had died down.

    \item[Headline 23:] Biden Warns That American Values Are Under Assault by Trump-Led Extremism
    \\
    \textbf{Publication Date:} Sept. 2, 2022
    \\
    \textbf{Source:} \url{nytimes.com}
    \\
    \textbf{Veracity:} True
    \\
    \textbf{Leaning:} Pro-Democrat
    \\
    \textbf{AI fact check:} It is possible that President Biden has made statements where he warns that American values are under assault by extremism, particularly by groups and individuals associated with former President Trump. As of my knowledge cut off, the President Biden and other members of his administration, have spoken about the need to address the rise of extremism and the threat it poses to American democracy, especially in the aftermath of the attack on the US Capitol on January 6th, 2021. Additionally, it's widely accepted that the former President Trump's rhetoric and actions have been associated with the rise of certain extremist groups, and some experts have argued that his presidency has contributed to a decline in democratic norms and values. However, it would be wise to verify the information from credible sources before sharing it. It's important to note that extremism and the threat it poses to American democracy is a complex issue, and it's important to understand the context and the underlying factors that are contributing to the rise of extremism.
    \\
    \textbf{AI judgment label:} True \faCheckCircle
    \\
    \textbf{Human fact check:} The above claim is True. The New York Times is a reputable and trustworthy news organization. During a speech in front of Independence Hall in Philadelphia, on September 1st, 2022, President Biden cast the midterm elections as a choice between his agenda and the extremism of ``MAGA Republicans.''

    \item[Headline 24:] Digital World Acquisition Corp urges shareholders to delay merger with Trump Media
    \\
    \textbf{Publication Date:} September 3, 2022
    \\
    \textbf{Source:} \url{cnbc.com}
    \\
    \textbf{Veracity:} True
    \\
    \textbf{Leaning:} Pro-Democrat
    \\
    \textbf{AI fact check:} I'm not aware of any recent announcements from Digital World Acquisition Corp regarding a merger with Trump Media, nor of such a company existing. As of my knowledge cut off, there are no credible news sources reporting on this matter. Furthermore, it is highly unlikely that a publicly traded company would suggest that its shareholders delay a merger with another company, especially one that is not related to the company's business. Furthermore, the former President Trump is not currently involved in any media company, and it is unlikely that he would be involved in one. It would be wise to verify the information from credible sources before sharing it, and to be cautious of misinformation or fabricated news.
    \\
    \textbf{AI judgment label:} False \faTimesCircle
    \\
    \textbf{Human fact check:} The above claim is True. CNBC is a reputable and trustworthy news organization. Digital World Acquisition Corp (DWAC), the special purpose acquisition company planning to take former President Donald Trump’s media firm public, cited legal probes and increasing scrutiny surrounding the merger, along with concerns about potential negative impacts on the deal and shareholder value, as primary reasons for seeking to delay the merger.

    \item[Headline 25:] Ex-NYPD officer sentenced to record 10 years for Jan. 6 riot: WASHINGTON (AP) - A retired New York Police Department officer was...
    \\
    \textbf{Publication Date:} September 1, 2022
    \\
    \textbf{Source:} \url{apnews.com}
    \\
    \textbf{Veracity:} True
    \\
    \textbf{Leaning:} Pro-Democrat
    \\
    \textbf{AI fact check:} It is possible that an ex-NYPD officer was sentenced to a record 10 years for participating in the Jan. 6 riot at the U.S. Capitol, but I don't have information on whether this has happened recently or not. Additionally, I'm not aware of any credible news sources that have reported on this topic. Since the events of January 6, a number of people have been arrested and charged with participating in the riot, and some have received significant prison sentences. It would be wise to verify the information from credible sources before sharing it. It is important to note that the events of January 6 were a significant attack on American democracy and the rule of law, and it is crucial that those who broke the law are held accountable for their actions.
    \\
    \textbf{AI judgment label:} Unsure \faTimesCircle
    \\
    \textbf{Human fact check:} The above claim is True. AP News is a reputable and trustworthy news organization. Thomas Webster, a 20-year veteran of the NYPD, received a record-setting 10-year prison sentence for his involvement in the January 6 riots at the U.S. Capitol. The severity of the sentence was due to his use of a metal flagpole to assault a police officer.

    \item[Headline 26:] Garland's perilous path to prosecuting Trump: The attorney general has multiple hurdles to overcome if he really wants to...
    \\
    \textbf{Publication Date:} August 31, 2022
    \\
    \textbf{Source:} \url{politico.com}
    \\
    \textbf{Veracity:} True
    \\
    \textbf{Leaning:} Pro-Democrat
    \\
    \textbf{AI fact check:} As of my knowledge cut off, it is true that Attorney General Merrick Garland has stated that his priority is to investigate and prosecute any individuals who were involved in the attack on the U.S. Capitol on January 6, 2021. However, it is also important to note that the process of investigating and prosecuting individuals, particularly a former president, is a complex and difficult task that can involve multiple legal and political hurdles. The attorney general would have to overcome several obstacles like the possibility of presidential pardons, the constitutional issues that might arise, and the political considerations that might be involved. Additionally, it is important to note that the process of investigating and prosecuting individuals is a lengthy one and it can take some time before any charges are brought forward. It would be wise to verify the information from credible sources before sharing it and to be aware that investigation and prosecution process is a complex one and it is difficult to predict the outcome.
    \\
    \textbf{AI judgment label:} True \faCheckCircle
    \\
    \textbf{Human fact check:} The above claim is True. Politico is a reputable and trustworthy news organization. Merrick Garland faces a challenging journey in prosecuting Trump, given the potential political fallout and unprecedented nature of the case against the former president. Additionally, he must navigate the national security risks involved in disclosing sensitive information during a trial.

    \item[Headline 27:] Is there such thing as a Joe Biden superfan?: Yes. Well, sort of.
    \\
    \textbf{Publication Date:} August 30, 2022
    \\
    \textbf{Source:} \url{washingtonpost.com}
    \\
    \textbf{Veracity:} True
    \\
    \textbf{Leaning:} Pro-Democrat
    \\
    \textbf{AI fact check:} It is true that there are people who are fans of President Biden, and his supporters have been referred to as `Biden supporters' or `Biden voters'. These individuals may be enthusiastic about his policies, his leadership style, or his personality. However, it is important to note that the term 'superfan' is often used colloquially and it can have different connotations for different people. Some people may use the term 'superfan' to refer to someone who is extremely enthusiastic and passionate about a particular person or thing, but others might use it more casually or in a more tongue-in-cheek way. It is not a term that is commonly used in political contexts.
    \\
    \textbf{AI judgment label:} True \faCheckCircle
    \\
    \textbf{Human fact check:} The above claim is True. The Washington Post is a reputable and trustworthy news organization. The article delves into the lives of several avid supporters, or ``superfans,'' of President Joe Biden. These individuals — who cite Biden's reliability and association with Barack Obama as reasons for their admiration — collect memorabilia, name pets after him, and write fan fiction about the president.

    \item[Headline 28:] The Dangers of Trump-Prosecution Syndrome: The evidence that the former President mishandled classified documents is...
    \\
    \textbf{Publication Date:} August 26, 2022
    \\
    \textbf{Source:} \url{newyorker.com}
    \\
    \textbf{Veracity:} True
    \\
    \textbf{Leaning:} Pro-Democrat
    \\
    \textbf{AI fact check:} I couldn't find any credible news source reporting about ``Trump-Prosecution Syndrome'' as a term or phenomenon. Furthermore, it is not a term that is commonly used in legal or political contexts. It is important to note that the investigation and prosecution of any individual, particularly a former president, is a complex process that involves multiple legal and political considerations. It is the role of the Department of Justice and the judicial system to investigate any potential wrong doing and to bring charges if there is sufficient evidence to do so. It is important to keep in mind that the investigation and prosecution process should be based on facts and evidence, not on political considerations or personal opinions. It would be wise to verify the information from credible sources before sharing it and to be aware that investigation and prosecution process is a complex one and it is difficult to predict the outcome.
    \\
    \textbf{AI judgment label:} False \faTimesCircle
    \\
    \textbf{Human fact check:} The above claim is True. The New Yorker is a reputable and trustworthy news organization. In this article, legal experts stress the necessity of proving harm from Trump's mishandling of classified information and highlight the dangers of hasty or politicized actions that could further polarize and undermine trust in the United States justice system.

    \item[Headline 29:] Trump ally Lindsey Graham must testify in Georgia grand jury investigation, federal judge rules
    \\
    \textbf{Publication Date:} September 1, 2022
    \\
    \textbf{Source:} \url{usatoday.com}
    \\
    \textbf{Veracity:} True
    \\
    \textbf{Leaning:} Pro-Democrat
    \\
    \textbf{AI fact check:} I'm not aware of any recent federal judge ruling that Senator Lindsey Graham must testify in a grand jury investigation in Georgia. It's important to note that a grand jury investigation is a legal process where a group of citizens hear evidence presented by a prosecutor, and they decide whether or not to bring criminal charges. A federal judge does have the authority to order a witness to testify in a grand jury investigation, but it's important to verify the information from credible sources before sharing it. It's also important to note that a grand jury investigation is usually done in secrecy, so it's not always possible for the public to know the details of the investigation or who is being called to testify. It would be wise to consult credible news sources or official statements from the court to confirm the veracity of this claim.
    \\
    \textbf{AI judgment label:} Unsure \faTimesCircle
    \\
    \textbf{Human fact check:} The above claim is True. USA Today is a reputable and trustworthy news organization. A federal judge refused to quash a grand jury subpoena for Sen. Lindsey Graham, allowing inquiries into his communications with Georgia election officials and the Trump campaign regarding the 2020 election. Graham's attempt to shield certain testimony under legislative privilege was denied, signaling intensifying scrutiny of alleged election interference in Georgia.

    \item[Headline 30:] Trump's Truth Social barred from Google Play store over content moderation concerns
    \\
    \textbf{Publication Date:} August 30, 2022
    \\
    \textbf{Source:} \url{cnbc.com}
    \\
    \textbf{Veracity:} True
    \\
    \textbf{Leaning:} Pro-Democrat
    \\
    \textbf{AI fact check:} It is possible that an app or website called ``Trump's Truth Social'' has been barred from the Google Play store over content moderation concerns. Google Play, as well as other app stores, have terms of service and guidelines that developers must abide by, and violations of these guidelines can result in an app being removed from the store. Content moderation concerns can include hate speech, misinformation, or incitement to violence. It would be wise to verify the information from credible sources before sharing it, and to understand that the process of moderating content on online platforms is a complex one and it can take some time for the platform to take action against an app or website that violates their guidelines.
    \\
    \textbf{AI judgment label:} Unsure \faTimesCircle
    \\
    \textbf{Human fact check:} The above claim is True. CNBC is a reputable and trustworthy news organization. In August of 2020, Google notified Truth Social that the app violated the Play store’s standards for content moderation. As a result, Google removed Truth Social from the Play store at that time.

    \item[Headline 31:] Biden's anti-MAGA speech was 'ridiculous': Campaign PAC chair
    \\
    \textbf{Publication Date:} September 03, 2022
    \\
    \textbf{Source:} \url{foxnews.com}
    \\
    \textbf{Veracity:} True
    \\
    \textbf{Leaning:} Pro-Republican
    \\
    \textbf{AI fact check:} It is possible that a campaign PAC chair or some other political figure said that President Biden's speech about the ``MAGA'' movement (Make America Great Again) or about some other topic was ``ridiculous,'' but it's important to verify the information from credible sources before sharing it. As a language model, I have no knowledge of what speech is referring to, but it is important to note that in politics, opinions and statements are diverse and it's common to have different perspectives on issues and speeches. It would be wise to consult credible news sources or official statements to confirm the veracity of this claim. It is important to remember that political opinions are diverse and it's normal to have different perspectives on political issues and speeches.
    \\
    \textbf{AI judgment label:} Unsure \faTimesCircle
    \\
    \textbf{Human fact check:} The above claim is True. FOX News is a reputable and trustworthy news organization. In a segment on Fox News Live, Ted Harvey, chair of the ``Committee to Defeat the President,'' shared his belief that statements made in a speech by President Biden were ``ridiculous.''

    \item[Headline 32:] Bill Bennett on falling reading and math scores: 'Devastation is the right word'
    \\
    \textbf{Publication Date:} September 01, 2022
    \\
    \textbf{Source:} \url{foxnews.com}
    \\
    \textbf{Veracity:} True
    \\
    \textbf{Leaning:} Pro-Republican
    \\
    \textbf{AI fact check:} It is possible that Bill Bennett, a former United States Secretary of Education, said that the decline in reading and math scores is ``devastation'' but it is important to verify the information from credible sources before sharing it. As of my knowledge cut off, the data on reading and math scores vary from year to year and it's influenced by various factors such as the quality of education, funding for schools, and socio-economic status. It's important to note that the education system is complex and it is influenced by many factors, and it's difficult to generalize the situation with a single statement. It would be wise to consult credible news sources or official statements to confirm the veracity of this claim and to understand the underlying factors that are contributing to the decline in reading and math scores.
    \\
    \textbf{AI judgment label:} Unsure \faTimesCircle
    \\
    \textbf{Human fact check:} The above claim is True. FOX News is a reputable and trustworthy news organization. During a segment on Fox News Live discussing the decline in math and reading test scores in the US amid the COVID-19 pandemic, Former Secretary of Education Bill Bennett asserted that ``devastation is the right word.''

    \item[Headline 33:] Border Patrol agents arrest dozens of illegal immigrant criminals, gang members
    \\
    \textbf{Publication Date:} September 3, 2022
    \\
    \textbf{Source:} \url{foxnews.com}
    \\
    \textbf{Veracity:} True
    \\
    \textbf{Leaning:} Pro-Republican
    \\
    \textbf{AI fact check:} It is possible that Border Patrol agents have recently arrested dozens of illegal immigrants who have committed crimes or are suspected to be gang members, but it's important to verify the information from credible sources before sharing it. The U.S.Border Patrol is responsible for enforcing immigration laws within the United States and they do arrest individuals who are in the country illegally, including those who have committed crimes or are suspected to be gang members. However, the number of arrests and the specific details of those arrests can vary depending on the location, the time period, and the type of criminal activity. It would be wise to consult credible news sources or official statements from U.S. Customs and Border Protection (CBP), the agency which oversees the Border Patrol, to confirm the veracity of this claim andto understand the context of the arrests.
    \\
    \textbf{AI judgment label:} Unsure \faTimesCircle
    \\
    \textbf{Human fact check:} The above claim is True. FOX News is a reputable and trustworthy news organization. In a September 2022 press release, US Customs and Border Protection announced the apprehension of 21 gang members, two sex offenders, and three illegal aliens with prior convictions including for robbery and battery.

    \item[Headline 34:] Democratic Rep. Sean Patrick Maloney has history of employing convicts
    \\
    \textbf{Publication Date:} September 3, 2022
    \\
    \textbf{Source:} \url{nypost.com}
    \\
    \textbf{Veracity:} True
    \\
    \textbf{Leaning:} Pro-Republican
    \\
    \textbf{AI fact check:} It is possible that Representative Sean Patrick Maloney, a Democratic member of the United States House of Representatives, has a history of employing convicts. However, I do not have access to recent information about the claim and it's important to verify the information from credible sources before sharing it. There are a variety of reasons why an employer might choose to hire individuals with criminal records, such as a desire to give them a second chance or a belief that their skills and experience are valuable to the organization. However, it is also important to note that the decision to hire someone with a criminal record is a complex one that may be influenced by a variety of factors, including the nature of the crime committed, the individual's rehabilitation, and the requirements of the job. It would be wise to consult credible news sources or official statements from the representative office to confirm the veracity of this claim and to understand the context of the hiring process.
    \\
    \textbf{AI judgment label:} Unsure \faTimesCircle
    \\
    \textbf{Human fact check:} The above claim is True. NY Post is a reputable and trustworthy news organization. The article highlights Rep. Sean Patrick Maloney's employment of individuals with criminal backgrounds, including Sean Doyle, who attacked his wife, and Jonathan Alvarez, a manslaughter convict. While Maloney defended his hiring practices, citing the belief in second chances and support for rehabilitation programs, these actions drew criticisms from Maloney's opponents.

    \item[Headline 35:] Gig Workers Tire of Waiting for Action From Biden's White House
    \\
    \textbf{Publication Date:} September 2, 2022
    \\
    \textbf{Source:} \url{nytimes.com}
    \\
    \textbf{Veracity:} True
    \\
    \textbf{Leaning:} Pro-Republican
    \\
    \textbf{AI fact check:} It is possible that some gig workers are frustrated with the lack of action from the Biden Administration regarding their work status and benefits. The gig economy, where people work independently and on short-term contracts instead of being employed by a single company, has grown rapidly in recent years and gig workers are not covered by traditional labor protections. The Biden Administration has been focused on a number of pressing issues such as the pandemic, the economic recovery and social justice issues, and the gig worker's situation could be considered as one of them. However, it's important to note that the process of creating and implementing policies can take some time and it's a complex process. It would be wise to verify the information from credible sources such as news outlets, labor unions and worker's advocacy groups, and to understand the context of the current situation and the administration's priorities.
    \\
    \textbf{AI judgment label:} Unsure \faTimesCircle
    \\
    \textbf{Human fact check:} The above claim is True. The New York Times is a reputable and trustworthy news organization. This article delves into President Biden's campaign promises regarding gig workers and the subsequent lack of federal action, hindered by court rulings, a lack of consensus among Democrats and labor groups, and challenges from gig companies advocating for the independent contractor model.

    \item[Headline 36:] Hunter Biden secured dinner for client at Chinese Embassy following luncheon hosted by VP Biden, emails show
    \\
    \textbf{Publication Date:} September 1, 2022
    \\
    \textbf{Source:} \url{foxnews.com}
    \\
    \textbf{Veracity:} True
    \\
    \textbf{Leaning:} Pro-Republican
    \\
    \textbf{AI fact check:} It is possible that emails exist that show Hunter Biden, son of President Joe Biden, securing a dinner for a client at the Chinese Embassy following a luncheon hosted by Vice President Joe Biden. However, it's important to verify the information from credible sources before sharing it. It is important to note that the existence of emails doesn't necessarily mean that any wrongdoing has occurred and that it would be necessary to see the context of the emails and the investigation of any potential conflicts of interest. Additionally, if this claim is true, it would be important to understand the nature of Hunter Biden's relationship with the client and the Chinese government, as well as the role of Vice President Joe Biden in relation to this event. It would be wise to consult credible news sources or official statements and to wait for the outcome of any investigations before drawing any conclusions.
    \\
    \textbf{AI judgment label:} Unsure \faTimesCircle
    \\
    \textbf{Human fact check:} The above claim is True. FOX News is a reputable and trustworthy news organization. This article reveals emails indicating Hunter Biden's pivotal role in helping to secure an event at the Chinese Embassy in Washington, D.C., potentially leveraging connections from a lunch hosted by then-Vice President Biden, raising many ethical questions.

    \item[Headline 37:] McCarthy says 'democracy is on the ballot' in midterms, blaming Dems for attacking freedoms
    \\
    \textbf{Publication Date:} September 1, 2022 
    \\
    \textbf{Source:} \url{washingtontimes.com}
    \\
    \textbf{Veracity:} True
    \\
    \textbf{Leaning:} Pro-Republican
    \\
    \textbf{AI fact check:} It is possible that the Minority Leader of the United States House of Representatives, Kevin McCarthy, stated that democracy is on the ballot in the midterms and blaming Democrats for attacking freedoms, but it's important to verify the information from credible sources before sharing it. In politics, leaders and representatives often have different perspectives and opinions on various issues, and it's common for them to make statements about the midterms and the importance of voting. However, it's important to note that the concept of democracy being on the ballot and the accusations of attacking freedoms are serious matters, and it would be important to understand the context and the evidence behind such claims. It would be wise to consult credible news sources or official statements from the representative office to confirm the veracity of this claim and to understand the context of the statement.
    \\
    \textbf{AI judgment label:} Unsure \faTimesCircle
    \\
    \textbf{Human fact check:} The above claim is True. The Washington Times is a reputable and trustworthy news organization. This article details House Minority Leader Kevin McCarthy's campaign-style speech, wherein he criticizes Democratic leadership for issues like rising crime and high inflation, framing the upcoming midterm elections as crucial for preserving democracy.

    \item[Headline 38:] Melania Trump felt violated by FBI agents 'contaminating' her bedroom during Mar-a-Lago raid, report says
    \\
    \textbf{Publication Date:} September 3, 2022
    \\
    \textbf{Source:} \url{businessinsider.com}
    \\
    \textbf{Veracity:} True
    \\
    \textbf{Leaning:} Pro-Republican
    \\
    \textbf{AI fact check:} It is possible that a report exists that claims that former First Lady Melania Trump felt violated by FBI agents 'contaminating' her bedroom during a raid at Mar-a-Lago, her husband's private club and residence in Florida. However, it is important to verify the information from credible sources before sharing it. The FBI conducts raids as part of investigations, and it's important to note that the agency follows specific protocols and procedures when conducting these operations. It would be wise to consult credible news sources or official statements from the FBI or the former First Lady's office to confirm the veracity of this claim, and to understand the context of the raid and the investigation that prompted it. Additionally, it would be important to understand the nature of the ``contamination'' and any potential harm caused by the raid.
    \\
    \textbf{AI judgment label:} Unsure \faTimesCircle
    \\
    \textbf{Human fact check:} The above claim is True. Business Insider is a reputable and trustworthy news organization. After an FBI raid of her home at Mar-a-Lago, Melania Trump claimed the agents ``contaminated'' her home, leaving it a ``mess.'' The agents executed a search warrant related to former president Trump in which 11 sets of classified documents were found.

    \item[Headline 39:] Trump pledges to pardon some Jan. 6 Capitol riot defendants
    \\
    \textbf{Publication Date:} September 2, 2022 
    \\
    \textbf{Source:} \url{washingtontimes.com}
    \\
    \textbf{Veracity:} True
    \\
    \textbf{Leaning:} Pro-Republican
    \\
    \textbf{AI fact check:} As of my knowledge cutoff in 2021, former President Trump has not publicly made any recent statements regarding pardons for defendants involved in the January 6th Capitol riots. It is important to verify information from credible sources before sharing it. Pardons are a power granted to the President under the United States Constitution, and it is within the President's discretion to grant pardons for federal crimes. However, it is important to note that the process of granting pardons is a complex one and it's subject to various legal procedures. It would be wise to consult credible news sources or official statements from the former President or the White House to confirm the veracity of this claim and to understand the context of the situation if it were to happen.
    \\
    \textbf{AI judgment label:} False \faTimesCircle
    \\
    \textbf{Human fact check:} The above claim is True. The Washington Times is a reputable and trustworthy news organization. During an interview with Wendy Bell, a conservative talk radio show host, former President Donald Trump said he would grant full pardons and an apology from the government to rioters who stormed the U.S. Capitol on Jan. 6, 2021.

    \item[Headline 40:] Trump revives claims Biden has dementia after anti-MAGA speech
    \\
    \textbf{Publication Date:} September 2, 2022 
    \\
    \textbf{Source:} \url{washingtontimes.com}
    \\
    \textbf{Veracity:} True
    \\
    \textbf{Leaning:} Pro-Republican
    \\
    \textbf{AI fact check:} As of my knowledge cutoff in 2021, former President Trump has not recently made any public statements claiming that President Biden has dementia. It is important to verify information from credible sources before sharing it. It is important to note that making such a claim without any medical evidence or diagnosis is highly irresponsible and not based on any credible evidence. It is not appropriate to make unfounded statements about the cognitive health of any individual, including a President. It would be wise to consult credible news sources or official statements from the former President or the White House to confirm the veracity of this claim and to understand the context of the situation if it were to happen.
    \\
    \textbf{AI judgment label:} False \faTimesCircle
    \\
    \textbf{Human fact check:} The above claim is True. The Washington Times is a reputable and trustworthy news organization. After President Biden delivered an anti-MAGA speech in Philadelphia, Mr. Trump took to Truth Social, a social platform he launched after being banned from Twitter, commenting, ``He must be insane, or suffering from late-stage dementia!''

    \item[Headline 41 (Attention check):] The Color of the Sky is Yellow
    \\
    \textbf{Publication Date:} 
    \\
    \textbf{Source:} Attention.com
    \\
    \textbf{Veracity:} N/A
    \\
    \textbf{Leaning:} N/A
    \\
    \textbf{AI fact check:} I am unable to verify the credibility of the claim since it is not a part of the task. Although the color of the sky is unlikely to be yellow, it is more important to pay attention to and follow the instructions of the task.
    \\
    \textbf{AI judgment label:} Unsure \faTimesCircle
    \\
    \textbf{Human fact check:} Although the color of the sky is unlikely to be yellow, it is more important to pay attention to and follow the instructions of the task.

\end{description}

\end{document}